\documentclass[prb,twocolumn,superscriptaddress]{revtex4-1}
\usepackage{graphicx}
\usepackage{color}
\usepackage{amsmath}
\usepackage{amssymb}
\usepackage{hyperref}
\usepackage{anysize}
\usepackage{natbib}

\begin{document}

\title{Plectoneme tip bubbles: Coupled denaturation and writhing in supercoiled DNA}
\author{Christian Matek}
\author{Thomas E. Ouldridge}
\affiliation{Rudolf Peierls Centre for Theoretical Physics, University of Oxford, 1 Keble Road, Oxford, OX1 3NP, United Kingdom}
\author{Jonathan P. K. Doye}
\affiliation{Physical and Theoretical Chemistry Laboratory, Department of Chemistry, University of Oxford, South Parks Road, Oxford, OX1 3QZ, United Kingdom}
\author{Ard A. Louis}
\affiliation{Rudolf Peierls Centre for Theoretical Physics, University of Oxford, 1 Keble Road, Oxford, OX1 3NP, United Kingdom}

\begin{abstract}
  Biological information is not only stored  in the digital chemical sequence of double helical DNA, but is also encoded in the mechanical properties of the DNA strands,  which can  influence biochemical processes involving its readout~\cite{Wang2013}.   For example, loop formation in the Lac operon~\cite{Zhang2006} can regulate the expression of key genes, and  DNA supercoiling is closely correlated to rhythmic circardian gene expression in cyanobacteria~\cite{Vijayan2009}.   Supercoiling is also important for large scale organisation of the genome in both eukaryotic and prokaryotic cells.   
  DNA can respond to torsional stress by writhing to form looped structures called plectonemes, thus  transferring energy stored as twist into energy stored in bending.  
Denaturation bubbles can also relax torsional stress, with the enthalpic cost of breaking bonds being compensated by their ability to absorb undertwist.
Here we predict a novel regime where bubbles form at the tips of plectonemes, and study its properties using coarse-grained simulations.   These tip bubbles can occur for both positive and negative supercoiling and greatly reduce plectoneme diffusion by a pinning mechanism.   They can cause plectonemes to preferentially localise to AT rich regions, because bubbles more easily form there.   The tip-bubble regime occurs for supercoiling densities and forces that are typically encountered for DNA {\em in vivo}, and may be exploited for biological control of genomic processes.

\end{abstract}

\maketitle

Much remains to be understood about the physical mechanisms by which DNA supercoiling affects cellular control in biology.  For this reason, the rich mechanical properties of DNA have been intensively studied by single-molecule techniques such as as magnetic and optical tweezers \cite{Strick1996,Forth2008, Kapanidis2009,Mosconi2009,Brutzer2010,Schoepflin2012,Janssen2012,Loenhout2012,Salerno2012a,Tempestini2013}, 
and by various theoretical techniques ranging from continuum models of DNA to atomistic simulations~\cite{Marko2012, Daniels2011, Emanuel2012,Klenin1991, Vologodskii1992, Schoepflin2012,Mitchell2011,Fye1999,Mielke2005}
(see also Section \ref{sec:other_models} of the Supplementary Material).
\begin{figure*}
 \includegraphics[width=1.95\columnwidth]{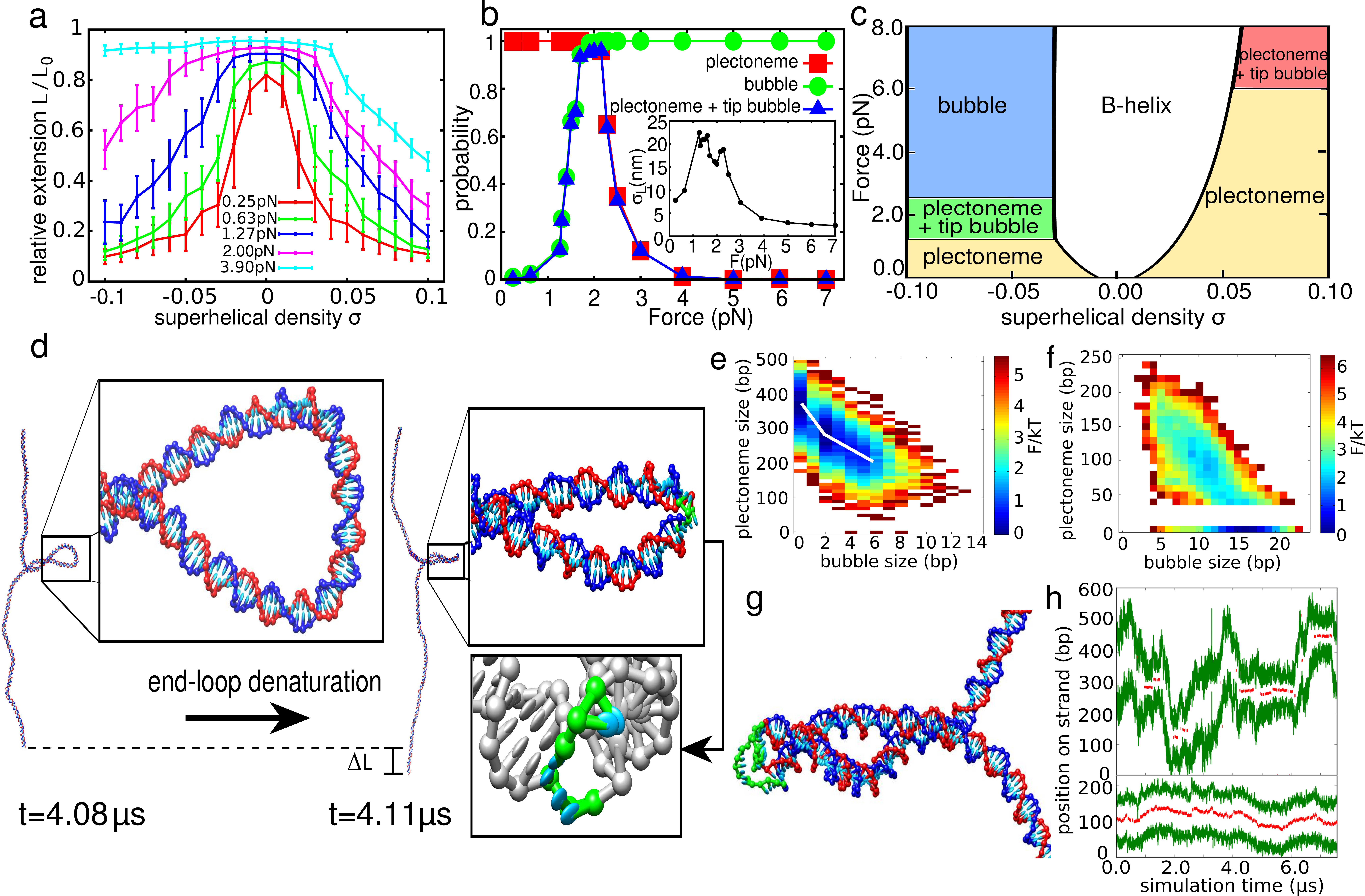}
  \caption{{\bf Plectoneme tip-bubble regime: a}, ``Hat-curves'' show the  mean relative extension of a $600$-bp duplex against superhelical density $\sigma$ for various applied forces $F$. Error bars indicate thermal fluctuations in the end-to-end distance, rather than sampling uncertainties. {\bf b}, Mean fraction of plectonemes, bubbles and  tip-bubble plectonemes,  as a function of force for $\sigma = -0.08$. {\bf Inset}:  Fluctuations (standard deviation $\sigma_L$) of end-to-end distances as a function of force for  $\sigma = -0.08$ show two maxima, the first at the point when tip bubbles form in plectonemes, the second at the transition from tip-bubble plectonemes to bubbles only. Results for $L$=1500\,bp as well as for other values of $\sigma$ can be found in Supplementary Sections III and VII. {\bf c}, State diagram of structures. Tip-bubble regions indicate points with at least a 40\% probability of a 
plectoneme with a tip bubble. {\bf d}, The 600-bp plectoneme system for  $\sigma=-0.05$ and $F=1.27$\,pN. Enlarged structures show the end loops with and without denatured bases (coloured green). The formation of a 3-bp tip bubble leads to a smaller plectoneme, because the tip can bend more easily and absorb extra undertwist, leading to an increased extension of the full strand by $\Delta L$. {\bf e}, A free-energy landscape for $\sigma = -0.08$ and $F=1.5$\,pN. The white line schematically shows the variation of the most likely plectoneme size with bubble size, illustrating the initial size reduction due to end-loop kinking. {\bf f}, A free-energy landscape for $\sigma = -0.08$ and $F=2.3$\,pN illustrating how the growth of bubbles leads to shrinking of the plectoneme. Tip-bubble plectonemes with small size  ($\lesssim 40$\,bp) are hard to detect or distinguish from writhed bubbles (see Methods) and so are classed here as bubbles. See Supplementary Section V for further information on the free energy landscape of tip bubbles. {\bf g}, Structure of a tip-bubble plectoneme at $\sigma=-0.08$ and $F=2.3$\,pN, posessing a 12-bp tip bubble and a 134-bp plectoneme. Denatured nucleotides are coloured green.
 {\bf h}, Plectoneme kinetics depicted by kymographs of the plectoneme boundaries (green 
lines).  Red denotes the centre of denatured base-pair stretches (bubbles), which pin the plectoneme and slow diffusion. The upper panel shows a simulation at $\sigma=-0.05$ and $F=1.27$\,pN, from which the structures in ({\bf d}) are taken. The lower panel shows a simulation for a fully pinned state at positive supercoiling, $\sigma=+0.08$, $F=7.9$\,pN, exhibiting much slower effective diffusion.
 } 
 \label{fig:hatcurves_torques}
\end{figure*}
\begin{figure*}
 \includegraphics[width=\linewidth]{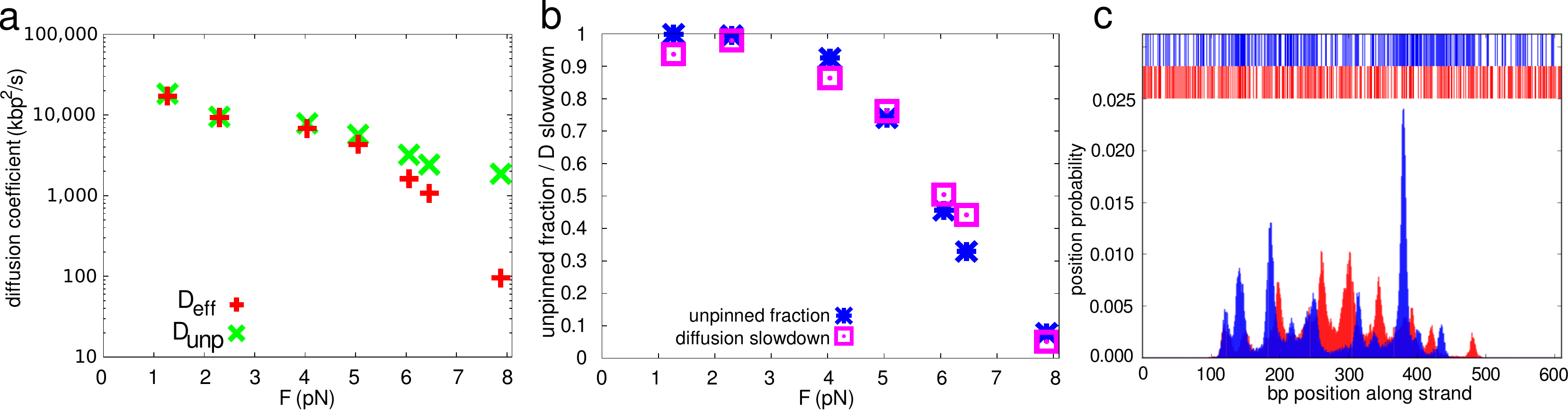}
 \caption{{\bf Plectoneme diffusion and sequence-dependent localization: a}, Diffusion coefficients for plectonemes at positive  $\sigma$ as a function of stretching force. $\sigma$ is chosen as in Ref.~\onlinecite{Loenhout2012} such that approximately 25\% of the strand length is in the plectonemic state on average. The diffusion constant $D_{\rm unp}$ for unpinned plectonemes (green) is compared to the observed effective diffusion constant $D_{\rm eff}$ (red). {\bf b}, Fraction of time plectonemes are found without a tip bubble (blue) compared to the relative slowdown of diffusion $D_{\rm eff}/D_{\rm unp}$ (magenta).  The close agreement suggests that the slowing down of the observed $D_{\rm eff}$ compared to $D_{\rm unp}$ is mainly due to pinning.
{\bf c}, Position distribution of plectonemes at $\sigma = -0.06$ and $F=1.27$\,pN. Results are shown in red for a random sequence, and in blue for a block-random sequence, as explained in the text. Plectoneme formation is suppressed near strand ends because these are clamped. For each sequence, the upper part of the figure shows coloured positions for AT basepairs and white for CG basepairs.  Plectonemes in the tip-bubble regime strongly localise to AT-rich regions.
 }
 \label{fig:pinning-diffusion}
\end{figure*}

Here we study the interplay between plectonemes and bubbles by employing a recently developed model, oxDNA \cite{Ouldridge2011,Sulc2012}, that treats nucleotides as rigid bodies possessing three interaction sites that mediate short-ranged backbone, stacking, excluded volume and hydrogen bonding interactions. 
This resolution allows us, in contrast to standard continuum models,  to study the effect of  strand denaturations on the behaviour of supercoiled DNA.   The model is simple enough to explore the  time and length scales relevant to the  formation and dynamics of plectonemes, something currently beyond the purview of atomistic simulations. 
OxDNA has successfully captured a range of systems where base pairs break and form, including nanotechnological devices \cite{Doye2013}, as well as biophysical processes such as overstretching~\cite{Romano2013} and cruciform formation~\cite{Matek2012}, suggesting it is well suited for studying the interplay between writhing and denaturation in supercoiled DNA.

To begin, we test our model by calculating canonical ``hat-curves" for the strand extension over a range of torsions and forces  similar to those used in single molecule assays \cite{Strick1996,Forth2008, Kapanidis2009,Mosconi2009,Brutzer2010,Schoepflin2012,Janssen2012,Loenhout2012,Salerno2012a,Tempestini2013} and  found {\it in vivo}.   Torsion is quantified  using the length-independent superhelical density $\sigma$, defined such that $\sigma=+(-)1$ for one full positive (negative)  imposed turn per pitch length.
We studied a strand of length 600 base-pairs (bp) using an ``average-base" parameterisation that treats each base-pair as having the same strength, allowing us to focus on generic DNA behaviour~\cite{Ouldridge2011}. Simulations are for one salt concentration, $500$\,mM NaCl, and further details are described in \mbox{\bf Methods}. 
   Fig.~\ref{fig:hatcurves_torques}a shows that, at a fixed force, and  for increasing positive $\sigma$,  the extension $L$ (here normalised by maximum extension $L_0$) does not change appreciably until $\sigma = \sigma_b$ at which point a   buckling transition occurs to a writhed plectonemic structure. For $\sigma \gtrsim \sigma_b$ the overall length of the strand decreases linearly for increasing $\sigma$ because the extra writhe is stored in the growing plectoneme.  For $F \lesssim 1$\,pN DNA strand extension is symmetric for $\sigma \rightarrow -\sigma$.  
But for $F \gtrsim 2.5$\,pN increasing negative supercoiling causes the formation of bubbles that absorb the twist, leading to little or  no  shortening of the strand.      
In all simulations, we observed at most one plectoneme, as is expected at high salt concentration and short strand length \cite{Loenhout2012,Emanuel2012}.
 We show in Supplementary Section~\ref{sec:experiment_comparison} that good agreement can be achieved with single-molecule experiments~\cite{Brutzer2010,Salerno2012a,Tempestini2013}, both for the hat curves and for direct torque response curves~\cite{Janssen2012}.  This agreement strengthens our confidence in the ability of oxDNA to predict behaviour for DNA under torsion and tension.

Here we address the unresolved question of how the system
transitions between the plectoneme-dominated and bubble-dominated regimes. 
Instead of a simple competition between spatially separated bubbles and plectonemes, 
we observe a ``tip-bubble'' regime where states with a co-localised
bubble/plectoneme pair are dominant.  This novel regime can be seen in the population diagram
Fig.~\ref{fig:hatcurves_torques}b, and in our overall schematic
state-diagram shown in Fig.~\ref{fig:hatcurves_torques}c. We now explore
these tip-bubble states in more detail.

At the plectoneme tip the DNA must bend back on itself relatively sharply. At
low forces this is achieved by a homogeneously bent loop.  However, as
recognised in studies on DNA bending \cite{Vologodskii13}, when the curvature
becomes too large an alternative is for the bending to be localized at a kink
defect where typically a few base pairs are broken. Similarly, as the force is
increased and the plectoneme becomes more tightly wound, a transition to a
tip-bubble state can occur, as illustrated in Fig.\
\ref{fig:hatcurves_torques}d. Importantly, the sharper bending at the kink
allows the same amount of writhe to be achieved by a smaller plectoneme, and
thus the tip-bubble state is stabilized by an increase in the extension $\Delta L$ along the force.
Additionally, for negative supercoiling the tip bubble is also able to absorb
some of the negative twist, allowing the plectoneme to shrink further. The
latter is the reason why, although tip bubble formation is seen for positive and
negative supercoiling (Fig.\ \ref{fig:hatcurves_torques}c) the transition
occurs at a significantly higher force for positive $\sigma$. 

For negative supercoiling, larger bubbles absorb additional twist, allowing plectonemes to shrink further. 
However, small kinks are stable against bubble growth at forces around 1-1.5\,pN, as shown in Fig.~\ref{fig:hatcurves_torques}e, because larger bubbles cause a smaller contraction in plectoneme size per base pair disrupted than the initial kink. At large enough forces, the tip bubble does grow, eventually eliminating the plectoneme. 

Fig.~\ref{fig:hatcurves_torques}f shows a free-energy landscape in the vicinity of the transition from the tip-bubble regime to the bubble dominated regime. The landscape is relatively flat along the diagonal, showing that plectoneme size and bubble size can be easily interchanged, and that states with both bubbles and plectonemes of intermediate size are common.  It is instructive to compare this scenario to what might be expected if bubbles and plectonemes did not co-localise.  The free-energy landscape would then be bi-stable, with plectoneme and bubble states separated by a substantial free-energy barrier due to the significant nucleation costs of both. 
Here, instead, each state helps lower the nucleation cost of the other; namely, bubble growth occurs from the plectoneme tip,
and the enhanced flexibility of a bubble allows the DNA to more easily writhe (Fig.\ \ref{fig:hatcurves_torques}g).
Compared to separate bubbles and plectonemes, co-localisation thus leads to enhanced fluctuations
in the extension (see inset of Fig.~\ref{fig:hatcurves_torques}b) over a broader range of force. 
It is interesting to note that in the experiments of Refs.\
\onlinecite{Salerno2012a,Tempestini2013} fluctuations were observed over a
significantly wider range of force than was expected from their simple theory (see also Supplementary Sec.~\ref{sec:length_comparison}).
 
Fig.~\ref{fig:hatcurves_torques}h shows kymographs for the diffusion of the plectonemes. 
The upper panel is in a regime where tip bubbles form about 34\% of the time.  When there is no tip bubble, the plectoneme diffuses by a strand-slithering mechanism where the strand reptates through the fluctuating plectonemic structure.  However, when there is a tip bubble, the plectonemes are effectively pinned because their diffusion requires the coupled motion of the plectoneme and the writhed bubble at its tip.
For example, the plectoneme formed by positive supercoiling in the lower panel of 
Fig.~\ref{fig:hatcurves_torques}h nearly always has a tip bubble, and so is effectively pinned.  

To further quantify the dynamics of these plectonemes we calculated their diffusion coefficients (see also Supplementary Section~\ref{sec:plectoneme_displacement}).  Fig.~\ref{fig:pinning-diffusion}a shows diffusion coefficients for positive supercoiling.   It is generally expected that plectoneme diffusion will slow down with increasing tension~\cite{Gennes1984}, an effect we observe by plotting the diffusion coefficient only for configurations with no tip bubble.  However, the full diffusion coefficient $D_{\rm eff}$ exhibits a marked further decrease as the fraction of time that the system is pinned increases. Fig.~\ref{fig:pinning-diffusion}b shows that the additional slow-down reflects the unpinned fraction.    

 A similar slowing down of plectoneme diffusion with increasing force for positive supercoiling was observed in the pioneering experiments of Ref.~\onlinecite{Loenhout2012}.  Here we provide a microscopic mechanism for part of this rapid slowing of the diffusion, namely the creation of tip bubbles with increasing force.   A further discussion of the comparison of our (much smaller) system to the experiments of  Ref.~\onlinecite{Loenhout2012} can be found in the Supplementary Section~\ref{sec:plectoneme_displacement}.
 
 Bubble formation is known to be highly sequence dependent \cite{Fye1999,Alberts2007}. To study how sequence affects the tip-bubble regime we performed additional simulations at $\sigma=-0.06$ and $F=1.27$\,pN using a sequence-dependent parametrization of oxDNA for a fully random sequence with a GC content of 49$\%$, and a block-random sequence of the form 5'-SWSWS-3', where S and W are 120\,bp stretches with a GC content of approximately 70$\%$ and 30$\%$ respectively, and an overall average GC content of $52\%$ (sequences are given in Supplementary Section~\ref{sec:sequences}).  
 We first note that for these parameters, tip bubble prevalence is $59 \%$ for the randomised sequence and $72\%$ for the block-random sequence (in contrast to $19 \%$ for the average-base model), even though each has nearly the same number of GC as AT base pairs.   These differences are unsurprising, as bubbles can form more easily in weaker AT-rich regions.   
This sequence dependence carries over into the 
hat-curves for negative $\sigma$; for example, the random sequence deviates from the $\sigma\rightarrow -\sigma$ symmetry at lower forces than the average-base model does (see Supplementary Section~\ref{sec:sequences}).

 We show the distribution of plectoneme locations for the random and block-random sequences  in Fig.~\ref{fig:pinning-diffusion}c. Both sequences show (in contrast to the average-base model - see Supplementary Section~\ref{sec:sequences}) strong preferential localisation of the plectonemes within AT-rich regions of the strand. 
Denaturations in tip-bubble plectonemes possess an average AT-content of $84\%$ and $91\%$ for the random and block-random sequence.     
Hence, local sequence properties can both influence the large-scale structure of DNA and stabilize denaturations in weak sequences by guiding plectoneme position.
As relevant forces and torsions lie within the regime thought to be relevant for DNA {\em in vivo}~\cite{Wang2013}, we speculate that plectoneme localisation can be used by the cell to regulate access to weak parts of the sequence, 
which are known to be important in several key biological processes including transcription and replication~\cite{Alberts2007}.

\bigskip

\noindent {\bf \normalsize Methods}

 We performed dynamical simulations of oxDNA, using an Andersen-like thermostat described in Ref.~\onlinecite{Russo2009} at $T=300$\,K.  Trajectories were generated using a time step of 12.1\,fs, and production runs were started from pre-thermalized configurations and run for $5 \times 10^{8}$ time steps. We studied DNA molecules of length 600\,bp, subject to torsional stress, quantified using the length-independent superhelical density $\sigma$ to facilitate comparison with different strand lengths $L$.
Torsionally relaxed states were chosen by demanding vanishing torque $\Gamma$ on unwound molecules, $\Gamma(F,\sigma=0)=0$.
During simulation runs, $\sigma$ was fixed to values in the range $-0.1\leq \sigma \leq +0.1$ by trapping the strand ends.
A stretching force $F$ with $0.25$\,pN$\leq F\leq7.9$\,pN was applied to the molecule ends. Further details of the set-up,  boundary conditions, and simulation techniques are described in Supplementary Section~\ref{sec:boundaries}.\\
Plectoneme structures were detected using an algorithm that compares the spatial distance between parts of the double strand to their proximity along the strand, allowing plectoneme position and size to be measured (see Supplementary Section \ref{sec:plecto_pos_size_pinn}).\\
To study the effect of strand length on our results, we also performed simulations of 1500-bp strands, which are described in more detail in Supplementary Section \ref{sec:length_comparison}. 
For these simulations, we used a GPU version of the oxDNA code.

\bigskip
\noindent {\bf \normalsize Acknowledgements } 

The authors are grateful to the Engineering and Physical Sciences Research Council.
C.M. acknowledges financial and material support from German Academic Exchange Service (DAAD) and Studienstiftung des deutschen Volkes. T.E.O. acknowledges funding from University College, Oxford.
\bigskip 
\newpage
\bibliographystyle{apsrev}
\bibliography{plectonemes_refs_nourl.bib}

\begin{thebibliography}{51}
\expandafter\ifx\csname natexlab\endcsname\relax\def\natexlab#1{#1}\fi
\expandafter\ifx\csname bibnamefont\endcsname\relax
  \def\bibnamefont#1{#1}\fi
\expandafter\ifx\csname bibfnamefont\endcsname\relax
  \def\bibfnamefont#1{#1}\fi
\expandafter\ifx\csname citenamefont\endcsname\relax
  \def\citenamefont#1{#1}\fi
\expandafter\ifx\csname url\endcsname\relax
  \def\url#1{\texttt{#1}}\fi
\expandafter\ifx\csname urlprefix\endcsname\relax\def\urlprefix{URL }\fi
\providecommand{\bibinfo}[2]{#2}
\providecommand{\eprint}[2][]{\url{#2}}

\bibitem[{\citenamefont{Wang et~al.}(2013)\citenamefont{Wang, Llopis, and
  Rudner}}]{Wang2013}
\bibinfo{author}{\bibfnamefont{X.}~\bibnamefont{Wang}},
  \bibinfo{author}{\bibfnamefont{P.~M.} \bibnamefont{Llopis}},
  \bibnamefont{and} \bibinfo{author}{\bibfnamefont{D.~Z.}
  \bibnamefont{Rudner}}, \bibinfo{journal}{Nat. Rev. Genet.}
  \textbf{\bibinfo{volume}{14}}, \bibinfo{pages}{191} (\bibinfo{year}{2013}).

\bibitem[{\citenamefont{Zhang et~al.}(2006)\citenamefont{Zhang, McEwen,
  Crothers, and Levene}}]{Zhang2006}
\bibinfo{author}{\bibfnamefont{Y.}~\bibnamefont{Zhang}},
  \bibinfo{author}{\bibfnamefont{A.~E.} \bibnamefont{McEwen}},
  \bibinfo{author}{\bibfnamefont{D.~M.} \bibnamefont{Crothers}},
  \bibnamefont{and} \bibinfo{author}{\bibfnamefont{S.~D.}
  \bibnamefont{Levene}}, \bibinfo{journal}{PLoS One}
  \textbf{\bibinfo{volume}{1}}, \bibinfo{pages}{e136} (\bibinfo{year}{2006}).

\bibitem[{\citenamefont{Vijayan et~al.}(2009)\citenamefont{Vijayan, Zuzow, and
  O'Shea}}]{Vijayan2009}
\bibinfo{author}{\bibfnamefont{V.}~\bibnamefont{Vijayan}},
  \bibinfo{author}{\bibfnamefont{R.}~\bibnamefont{Zuzow}}, \bibnamefont{and}
  \bibinfo{author}{\bibfnamefont{E.~K.} \bibnamefont{O'Shea}},
  \bibinfo{journal}{Proc. Natl. Acad. Sci. USA} \textbf{\bibinfo{volume}{106}},
  \bibinfo{pages}{22564} (\bibinfo{year}{2009}).

\bibitem[{\citenamefont{Strick et~al.}(1996)\citenamefont{Strick, Allemand,
  Bensimon, Bensimon, and Croquette}}]{Strick1996}
\bibinfo{author}{\bibfnamefont{T.~R.} \bibnamefont{Strick}},
  \bibinfo{author}{\bibfnamefont{J.~F.} \bibnamefont{Allemand}},
  \bibinfo{author}{\bibfnamefont{D.}~\bibnamefont{Bensimon}},
  \bibinfo{author}{\bibfnamefont{A.}~\bibnamefont{Bensimon}}, \bibnamefont{and}
  \bibinfo{author}{\bibfnamefont{V.}~\bibnamefont{Croquette}},
  \bibinfo{journal}{Science} \textbf{\bibinfo{volume}{271}},
  \bibinfo{pages}{1835} (\bibinfo{year}{1996}).

\bibitem[{\citenamefont{Forth et~al.}(2008)\citenamefont{Forth, Deufel,
  Sheinin, Daniels, Sethna, and Wang}}]{Forth2008}
\bibinfo{author}{\bibfnamefont{S.}~\bibnamefont{Forth}},
  \bibinfo{author}{\bibfnamefont{C.}~\bibnamefont{Deufel}},
  \bibinfo{author}{\bibfnamefont{M.~Y.} \bibnamefont{Sheinin}},
  \bibinfo{author}{\bibfnamefont{B.}~\bibnamefont{Daniels}},
  \bibinfo{author}{\bibfnamefont{J.~P.} \bibnamefont{Sethna}},
  \bibnamefont{and} \bibinfo{author}{\bibfnamefont{M.~D.} \bibnamefont{Wang}},
  \bibinfo{journal}{Phys. Rev. Lett.} \textbf{\bibinfo{volume}{100}},
  \bibinfo{pages}{148301} (\bibinfo{year}{2008}).

\bibitem[{\citenamefont{Kapanidis and Strick}(2009)}]{Kapanidis2009}
\bibinfo{author}{\bibfnamefont{A.~N.} \bibnamefont{Kapanidis}}
  \bibnamefont{and} \bibinfo{author}{\bibfnamefont{T.~R.}
  \bibnamefont{Strick}}, \bibinfo{journal}{Trends Biochem. Sci.}
  \textbf{\bibinfo{volume}{34}}, \bibinfo{pages}{234} (\bibinfo{year}{2009}).

\bibitem[{\citenamefont{Mosconi et~al.}(2009)\citenamefont{Mosconi, Allemand,
  Bensimon, and Croquette}}]{Mosconi2009}
\bibinfo{author}{\bibfnamefont{F.}~\bibnamefont{Mosconi}},
  \bibinfo{author}{\bibfnamefont{J.~F.} \bibnamefont{Allemand}},
  \bibinfo{author}{\bibfnamefont{D.}~\bibnamefont{Bensimon}}, \bibnamefont{and}
  \bibinfo{author}{\bibfnamefont{V.}~\bibnamefont{Croquette}},
  \bibinfo{journal}{Phys. Rev. Lett.} \textbf{\bibinfo{volume}{102}},
  \bibinfo{pages}{078301} (\bibinfo{year}{2009}).

\bibitem[{\citenamefont{Brutzer et~al.}(2010)\citenamefont{Brutzer, Luzzietti,
  Klaue, and Seidel}}]{Brutzer2010}
\bibinfo{author}{\bibfnamefont{H.}~\bibnamefont{Brutzer}},
  \bibinfo{author}{\bibfnamefont{N.}~\bibnamefont{Luzzietti}},
  \bibinfo{author}{\bibfnamefont{D.}~\bibnamefont{Klaue}}, \bibnamefont{and}
  \bibinfo{author}{\bibfnamefont{R.}~\bibnamefont{Seidel}},
  \bibinfo{journal}{Biophys. J.} \textbf{\bibinfo{volume}{98}},
  \bibinfo{pages}{1267} (\bibinfo{year}{2010}).

\bibitem[{\citenamefont{Sch\"{o}pflin et~al.}(2012)\citenamefont{Sch\"{o}pflin,
  Brutzer, M\"{u}ller, Seidel, and Wedemann}}]{Schoepflin2012}
\bibinfo{author}{\bibfnamefont{R.}~\bibnamefont{Sch\"{o}pflin}},
  \bibinfo{author}{\bibfnamefont{H.}~\bibnamefont{Brutzer}},
  \bibinfo{author}{\bibfnamefont{O.}~\bibnamefont{M\"{u}ller}},
  \bibinfo{author}{\bibfnamefont{R.}~\bibnamefont{Seidel}}, \bibnamefont{and}
  \bibinfo{author}{\bibfnamefont{G.}~\bibnamefont{Wedemann}},
  \bibinfo{journal}{Biophys. J.} \textbf{\bibinfo{volume}{103}},
  \bibinfo{pages}{323} (\bibinfo{year}{2012}).

\bibitem[{\citenamefont{Janssen et~al.}(2012)\citenamefont{Janssen, Lipfert,
  Jager, Daudey, Beekman, and Dekker}}]{Janssen2012}
\bibinfo{author}{\bibfnamefont{X.~J.~A.} \bibnamefont{Janssen}},
  \bibinfo{author}{\bibfnamefont{J.}~\bibnamefont{Lipfert}},
  \bibinfo{author}{\bibfnamefont{T.}~\bibnamefont{Jager}},
  \bibinfo{author}{\bibfnamefont{R.}~\bibnamefont{Daudey}},
  \bibinfo{author}{\bibfnamefont{J.}~\bibnamefont{Beekman}}, \bibnamefont{and}
  \bibinfo{author}{\bibfnamefont{N.~H.} \bibnamefont{Dekker}},
  \bibinfo{journal}{Nano Lett.} \textbf{\bibinfo{volume}{12}},
  \bibinfo{pages}{3634} (\bibinfo{year}{2012}).

\bibitem[{\citenamefont{van Loenhout et~al.}(2012)\citenamefont{van Loenhout,
  de~Grunt, and Dekker}}]{Loenhout2012}
\bibinfo{author}{\bibfnamefont{M.~T.~J.} \bibnamefont{van Loenhout}},
  \bibinfo{author}{\bibfnamefont{M.~V.} \bibnamefont{de~Grunt}},
  \bibnamefont{and} \bibinfo{author}{\bibfnamefont{C.}~\bibnamefont{Dekker}},
  \bibinfo{journal}{Science} \textbf{\bibinfo{volume}{338}},
  \bibinfo{pages}{94} (\bibinfo{year}{2012}).

\bibitem[{\citenamefont{Salerno et~al.}(2012)\citenamefont{Salerno, Tempestini,
  Mai, Brogioli, Ziano, Cassina, and Mantegazza}}]{Salerno2012a}
\bibinfo{author}{\bibfnamefont{D.}~\bibnamefont{Salerno}},
  \bibinfo{author}{\bibfnamefont{A.}~\bibnamefont{Tempestini}},
  \bibinfo{author}{\bibfnamefont{I.}~\bibnamefont{Mai}},
  \bibinfo{author}{\bibfnamefont{D.}~\bibnamefont{Brogioli}},
  \bibinfo{author}{\bibfnamefont{R.}~\bibnamefont{Ziano}},
  \bibinfo{author}{\bibfnamefont{V.}~\bibnamefont{Cassina}}, \bibnamefont{and}
  \bibinfo{author}{\bibfnamefont{F.}~\bibnamefont{Mantegazza}},
  \bibinfo{journal}{Phys. Rev. Lett.} \textbf{\bibinfo{volume}{109}},
  \bibinfo{pages}{118303} (\bibinfo{year}{2012}).

\bibitem[{\citenamefont{Tempestini et~al.}(2013)\citenamefont{Tempestini,
  Cassina, Brogioli, Ziano, Erba, Giovannoni, Cerrito, Salerno, and
  Mantegazza}}]{Tempestini2013}
\bibinfo{author}{\bibfnamefont{A.}~\bibnamefont{Tempestini}},
  \bibinfo{author}{\bibfnamefont{V.}~\bibnamefont{Cassina}},
  \bibinfo{author}{\bibfnamefont{D.}~\bibnamefont{Brogioli}},
  \bibinfo{author}{\bibfnamefont{R.}~\bibnamefont{Ziano}},
  \bibinfo{author}{\bibfnamefont{S.}~\bibnamefont{Erba}},
  \bibinfo{author}{\bibfnamefont{R.}~\bibnamefont{Giovannoni}},
  \bibinfo{author}{\bibfnamefont{M.~G.} \bibnamefont{Cerrito}},
  \bibinfo{author}{\bibfnamefont{D.}~\bibnamefont{Salerno}}, \bibnamefont{and}
  \bibinfo{author}{\bibfnamefont{F.}~\bibnamefont{Mantegazza}},
  \bibinfo{journal}{Nucleic Acids Res.} \textbf{\bibinfo{volume}{41}},
  \bibinfo{pages}{2009} (\bibinfo{year}{2013}).

\bibitem[{\citenamefont{Marko and Neukirch}(2012)}]{Marko2012}
\bibinfo{author}{\bibfnamefont{J.~F.} \bibnamefont{Marko}} \bibnamefont{and}
  \bibinfo{author}{\bibfnamefont{S.}~\bibnamefont{Neukirch}},
  \bibinfo{journal}{Phys. Rev. E} \textbf{\bibinfo{volume}{85}},
  \bibinfo{pages}{011908} (\bibinfo{year}{2012}).

\bibitem[{\citenamefont{Daniels and Sethna}(2011)}]{Daniels2011}
\bibinfo{author}{\bibfnamefont{B.~C.} \bibnamefont{Daniels}} \bibnamefont{and}
  \bibinfo{author}{\bibfnamefont{J.~P.} \bibnamefont{Sethna}},
  \bibinfo{journal}{Phys. Rev. E} \textbf{\bibinfo{volume}{83}},
  \bibinfo{pages}{041924} (\bibinfo{year}{2011}).

\bibitem[{\citenamefont{Emanuel et~al.}(2013)\citenamefont{Emanuel, Lanzani,
  and Schiessel}}]{Emanuel2012}
\bibinfo{author}{\bibfnamefont{M.}~\bibnamefont{Emanuel}},
  \bibinfo{author}{\bibfnamefont{G.}~\bibnamefont{Lanzani}}, \bibnamefont{and}
  \bibinfo{author}{\bibfnamefont{H.}~\bibnamefont{Schiessel}},
  \bibinfo{journal}{Phys. Rev. E} \textbf{\bibinfo{volume}{88}},
  \bibinfo{pages}{022706} (\bibinfo{year}{2013}).

\bibitem[{\citenamefont{Klenin et~al.}(1991)\citenamefont{Klenin, Vologodskii,
  Anshelevich, Dykhne, and Frank-Kamenetskii}}]{Klenin1991}
\bibinfo{author}{\bibfnamefont{K.~V.} \bibnamefont{Klenin}},
  \bibinfo{author}{\bibfnamefont{A.~V.} \bibnamefont{Vologodskii}},
  \bibinfo{author}{\bibfnamefont{V.~V.} \bibnamefont{Anshelevich}},
  \bibinfo{author}{\bibfnamefont{A.~M.} \bibnamefont{Dykhne}},
  \bibnamefont{and} \bibinfo{author}{\bibfnamefont{M.~D.}
  \bibnamefont{Frank-Kamenetskii}}, \bibinfo{journal}{J. Mol. Biol.}
  \textbf{\bibinfo{volume}{217}}, \bibinfo{pages}{413} (\bibinfo{year}{1991}).

\bibitem[{\citenamefont{Vologodskii et~al.}(1992)\citenamefont{Vologodskii,
  Levene, Klenin, Frank-Kamenetskii, and Cozzarelli}}]{Vologodskii1992}
\bibinfo{author}{\bibfnamefont{A.~V.} \bibnamefont{Vologodskii}},
  \bibinfo{author}{\bibfnamefont{S.~D.} \bibnamefont{Levene}},
  \bibinfo{author}{\bibfnamefont{K.~V.} \bibnamefont{Klenin}},
  \bibinfo{author}{\bibfnamefont{M.}~\bibnamefont{Frank-Kamenetskii}},
  \bibnamefont{and} \bibinfo{author}{\bibfnamefont{N.~R.}
  \bibnamefont{Cozzarelli}}, \bibinfo{journal}{J. Mol. Biol.}
  \textbf{\bibinfo{volume}{227}}, \bibinfo{pages}{1224} (\bibinfo{year}{1992}).

\bibitem[{\citenamefont{Mitchell et~al.}(2011)\citenamefont{Mitchell, Laughton,
  and Harris}}]{Mitchell2011}
\bibinfo{author}{\bibfnamefont{J.~S.} \bibnamefont{Mitchell}},
  \bibinfo{author}{\bibfnamefont{C.~A.} \bibnamefont{Laughton}},
  \bibnamefont{and} \bibinfo{author}{\bibfnamefont{S.~A.}
  \bibnamefont{Harris}}, \bibinfo{journal}{Nucleic Acids Res.}
  \textbf{\bibinfo{volume}{39}}, \bibinfo{pages}{3928} (\bibinfo{year}{2011}).

\bibitem[{\citenamefont{Mielke et~al.}(2005)\citenamefont{Mielke,
  Gr{\o}nbech-Jensen, Krishnan, Fink, and Benham}}]{Mielke2005}
\bibinfo{author}{\bibfnamefont{S.~P.} \bibnamefont{Mielke}},
  \bibinfo{author}{\bibfnamefont{N.}~\bibnamefont{Gr{\o}nbech-Jensen}},
  \bibinfo{author}{\bibfnamefont{V.~V.} \bibnamefont{Krishnan}},
  \bibinfo{author}{\bibfnamefont{W.~H.} \bibnamefont{Fink}}, \bibnamefont{and}
  \bibinfo{author}{\bibfnamefont{C.~J.} \bibnamefont{Benham}},
  \bibinfo{journal}{J. Chem. Phys.} \textbf{\bibinfo{volume}{123}},
  \bibinfo{pages}{124911} (\bibinfo{year}{2005}).

\bibitem[{\citenamefont{Ouldridge et~al.}(2011)\citenamefont{Ouldridge, Louis,
  and Doye}}]{Ouldridge2011}
\bibinfo{author}{\bibfnamefont{T.~E.} \bibnamefont{Ouldridge}},
  \bibinfo{author}{\bibfnamefont{A.~A.} \bibnamefont{Louis}}, \bibnamefont{and}
  \bibinfo{author}{\bibfnamefont{J.~P.~K.} \bibnamefont{Doye}},
  \bibinfo{journal}{J. Chem. Phys.} \textbf{\bibinfo{volume}{134}},
  \bibinfo{pages}{085101} (\bibinfo{year}{2011}).

\bibitem[{\citenamefont{\v{S}ulc et~al.}(2012)\citenamefont{\v{S}ulc, Romano,
  Ouldridge, Rovigatti, Doye, and Louis}}]{Sulc2012}
\bibinfo{author}{\bibfnamefont{P.}~\bibnamefont{\v{S}ulc}},
  \bibinfo{author}{\bibfnamefont{F.}~\bibnamefont{Romano}},
  \bibinfo{author}{\bibfnamefont{T.~E.} \bibnamefont{Ouldridge}},
  \bibinfo{author}{\bibfnamefont{L.}~\bibnamefont{Rovigatti}},
  \bibinfo{author}{\bibfnamefont{J.~P.~K.} \bibnamefont{Doye}},
  \bibnamefont{and} \bibinfo{author}{\bibfnamefont{A.~A.} \bibnamefont{Louis}},
  \bibinfo{journal}{J. Chem. Phys.} \textbf{\bibinfo{volume}{137}},
  \bibinfo{pages}{135101} (\bibinfo{year}{2012}).

\bibitem[{\citenamefont{Doye et~al.}(2013)\citenamefont{Doye, Ouldridge, Louis,
  Romano, {\v{S}}ulc, Matek, Snodin, Rovigatti, Schreck, Harrison
  et~al.}}]{Doye2013}
\bibinfo{author}{\bibfnamefont{J.~P.~K.} \bibnamefont{Doye}},
  \bibinfo{author}{\bibfnamefont{T.~E.} \bibnamefont{Ouldridge}},
  \bibinfo{author}{\bibfnamefont{A.~A.} \bibnamefont{Louis}},
  \bibinfo{author}{\bibfnamefont{F.}~\bibnamefont{Romano}},
  \bibinfo{author}{\bibfnamefont{P.}~\bibnamefont{{\v{S}}ulc}},
  \bibinfo{author}{\bibfnamefont{C.}~\bibnamefont{Matek}},
  \bibinfo{author}{\bibfnamefont{B.~E.} \bibnamefont{Snodin}},
  \bibinfo{author}{\bibfnamefont{L.}~\bibnamefont{Rovigatti}},
  \bibinfo{author}{\bibfnamefont{J.~S.} \bibnamefont{Schreck}},
  \bibinfo{author}{\bibfnamefont{R.~M.} \bibnamefont{Harrison}},
  \bibnamefont{et~al.}, \bibinfo{journal}{Phys. Chem. Chem. Phys.}
  \textbf{\bibinfo{volume}{15}}, \bibinfo{pages}{20395} (\bibinfo{year}{2013}).

\bibitem[{\citenamefont{Romano et~al.}(2013)\citenamefont{Romano, Chakraborty,
  Doye, Ouldridge, and Louis}}]{Romano2013}
\bibinfo{author}{\bibfnamefont{F.}~\bibnamefont{Romano}},
  \bibinfo{author}{\bibfnamefont{D.}~\bibnamefont{Chakraborty}},
  \bibinfo{author}{\bibfnamefont{J.~P.~K.} \bibnamefont{Doye}},
  \bibinfo{author}{\bibfnamefont{T.~E.} \bibnamefont{Ouldridge}},
  \bibnamefont{and} \bibinfo{author}{\bibfnamefont{A.~A.} \bibnamefont{Louis}},
  \bibinfo{journal}{J. Chem. Phys.} \textbf{\bibinfo{volume}{138}},
  \bibinfo{pages}{085101} (\bibinfo{year}{2013}).

\bibitem[{\citenamefont{Matek et~al.}(2012)\citenamefont{Matek, Ouldridge,
  Levy, Doye, and Louis}}]{Matek2012}
\bibinfo{author}{\bibfnamefont{C.}~\bibnamefont{Matek}},
  \bibinfo{author}{\bibfnamefont{T.~E.} \bibnamefont{Ouldridge}},
  \bibinfo{author}{\bibfnamefont{A.}~\bibnamefont{Levy}},
  \bibinfo{author}{\bibfnamefont{J.~P.~K.} \bibnamefont{Doye}},
  \bibnamefont{and} \bibinfo{author}{\bibfnamefont{A.~A.} \bibnamefont{Louis}},
  \bibinfo{journal}{J. Phys. Chem. B} \textbf{\bibinfo{volume}{116}},
  \bibinfo{pages}{11616} (\bibinfo{year}{2012}).

\bibitem[{\citenamefont{Vologodskii and
  Frank-Kamenetskii}(2013)}]{Vologodskii13}
\bibinfo{author}{\bibfnamefont{A.}~\bibnamefont{Vologodskii}} \bibnamefont{and}
  \bibinfo{author}{\bibfnamefont{M.~D.} \bibnamefont{Frank-Kamenetskii}},
  \bibinfo{journal}{Nucleic Acids Res.} \textbf{\bibinfo{volume}{41}},
  \bibinfo{pages}{6785} (\bibinfo{year}{2013}).

\bibitem[{\citenamefont{de~Gennes}(1984)}]{Gennes1984}
\bibinfo{author}{\bibfnamefont{P.~G.} \bibnamefont{de~Gennes}},
  \bibinfo{journal}{Macromolecules} \textbf{\bibinfo{volume}{17}},
  \bibinfo{pages}{703} (\bibinfo{year}{1984}).

\bibitem[{\citenamefont{Fye and Benham}(1999)}]{Fye1999}
\bibinfo{author}{\bibfnamefont{R.~M.} \bibnamefont{Fye}} \bibnamefont{and}
  \bibinfo{author}{\bibfnamefont{C.~J.} \bibnamefont{Benham}},
  \bibinfo{journal}{Phys. Rev. E} \textbf{\bibinfo{volume}{59}},
  \bibinfo{pages}{3408} (\bibinfo{year}{1999}).

\bibitem[{\citenamefont{Alberts et~al.}(2007)\citenamefont{Alberts, Johnson,
  Walter, Lewis, Raff, Roberts, and Orme}}]{Alberts2007}
\bibinfo{author}{\bibfnamefont{B.}~\bibnamefont{Alberts}},
  \bibinfo{author}{\bibfnamefont{A.}~\bibnamefont{Johnson}},
  \bibinfo{author}{\bibfnamefont{P.}~\bibnamefont{Walter}},
  \bibinfo{author}{\bibfnamefont{J.}~\bibnamefont{Lewis}},
  \bibinfo{author}{\bibfnamefont{M.}~\bibnamefont{Raff}},
  \bibinfo{author}{\bibfnamefont{K.}~\bibnamefont{Roberts}}, \bibnamefont{and}
  \bibinfo{author}{\bibfnamefont{N.}~\bibnamefont{Orme}},
  \emph{\bibinfo{title}{Molecular Biology of the Cell.}}
  (\bibinfo{publisher}{Taylor \& Francis}, \bibinfo{year}{2007}),
  \bibinfo{edition}{5th} ed.

\bibitem[{\citenamefont{Russo et~al.}(2009)\citenamefont{Russo, Tartaglia, and
  Sciortino}}]{Russo2009}
\bibinfo{author}{\bibfnamefont{J.}~\bibnamefont{Russo}},
  \bibinfo{author}{\bibfnamefont{P.}~\bibnamefont{Tartaglia}},
  \bibnamefont{and}
  \bibinfo{author}{\bibfnamefont{F.}~\bibnamefont{Sciortino}},
  \bibinfo{journal}{J. Chem. Phys.} \textbf{\bibinfo{volume}{131}},
  \bibinfo{pages}{014504} (\bibinfo{year}{2009}).

\bibitem[{\citenamefont{Padding and Louis}(2006)}]{Padding2006}
\bibinfo{author}{\bibfnamefont{J.~T.} \bibnamefont{Padding}} \bibnamefont{and}
  \bibinfo{author}{\bibfnamefont{A.~A.} \bibnamefont{Louis}},
  \bibinfo{journal}{Phys. Rev. E} \textbf{\bibinfo{volume}{74}},
  \bibinfo{pages}{031402} (\bibinfo{year}{2006}).

\bibitem[{\citenamefont{Louis}(2010)}]{Louis2010}
\bibinfo{author}{\bibfnamefont{A.~A.} \bibnamefont{Louis}},
  \bibinfo{journal}{Faraday Discuss.} \textbf{\bibinfo{volume}{133}},
  \bibinfo{pages}{323} (\bibinfo{year}{2010}).

\bibitem[{\citenamefont{Murtola et~al.}(2009)\citenamefont{Murtola, Bunker,
  Vattulainen, Deserno, and Karttunen}}]{Murtola2009}
\bibinfo{author}{\bibfnamefont{T.}~\bibnamefont{Murtola}},
  \bibinfo{author}{\bibfnamefont{A.}~\bibnamefont{Bunker}},
  \bibinfo{author}{\bibfnamefont{I.}~\bibnamefont{Vattulainen}},
  \bibinfo{author}{\bibfnamefont{M.}~\bibnamefont{Deserno}}, \bibnamefont{and}
  \bibinfo{author}{\bibfnamefont{M.}~\bibnamefont{Karttunen}},
  \bibinfo{journal}{Phys. Chem. Chem. Phys.} \textbf{\bibinfo{volume}{11}},
  \bibinfo{pages}{1869} (\bibinfo{year}{2009}).

\bibitem[{\citenamefont{Ouldridge et~al.}(2013)\citenamefont{Ouldridge,
  \v{S}ulc, Romano, Doye, and Louis}}]{Ouldridge2013a}
\bibinfo{author}{\bibfnamefont{T.~E.} \bibnamefont{Ouldridge}},
  \bibinfo{author}{\bibfnamefont{P.}~\bibnamefont{\v{S}ulc}},
  \bibinfo{author}{\bibfnamefont{F.}~\bibnamefont{Romano}},
  \bibinfo{author}{\bibfnamefont{J.~P.~K.} \bibnamefont{Doye}},
  \bibnamefont{and} \bibinfo{author}{\bibfnamefont{A.~A.} \bibnamefont{Louis}},
  \bibinfo{journal}{Nucleic Acids Res.} \textbf{\bibinfo{volume}{41}},
  \bibinfo{pages}{8886} (\bibinfo{year}{2013}).

\bibitem[{\citenamefont{C\u{a}lug\u{a}reanu}(1959)}]{Calufareanu1959}
\bibinfo{author}{\bibfnamefont{G.}~\bibnamefont{C\u{a}lug\u{a}reanu}},
  \bibinfo{journal}{Rev. Roum. Math. Pures et Appl.}
  \textbf{\bibinfo{volume}{4}}, \bibinfo{pages}{5} (\bibinfo{year}{1959}).

\bibitem[{\citenamefont{White}(1969)}]{White1969}
\bibinfo{author}{\bibfnamefont{J.~H.} \bibnamefont{White}},
  \bibinfo{journal}{Am. J. Math.} \textbf{\bibinfo{volume}{91}},
  \bibinfo{pages}{693} (\bibinfo{year}{1969}).

\bibitem[{\citenamefont{Fuller}(1971)}]{Fuller1971o}
\bibinfo{author}{\bibfnamefont{F.~B.} \bibnamefont{Fuller}},
  \bibinfo{journal}{Proc. Natl. Acad. Sci. USA} \textbf{\bibinfo{volume}{68}},
  \bibinfo{pages}{815} (\bibinfo{year}{1971}).

\bibitem[{\citenamefont{Maffeo et~al.}(2010)\citenamefont{Maffeo,
  Sch\"{o}pflin, Brutzer, Stehr, Aksimentiev, Wedemann, and
  Seidel}}]{Maffeo2010}
\bibinfo{author}{\bibfnamefont{C.}~\bibnamefont{Maffeo}},
  \bibinfo{author}{\bibfnamefont{R.}~\bibnamefont{Sch\"{o}pflin}},
  \bibinfo{author}{\bibfnamefont{H.}~\bibnamefont{Brutzer}},
  \bibinfo{author}{\bibfnamefont{R.}~\bibnamefont{Stehr}},
  \bibinfo{author}{\bibfnamefont{A.}~\bibnamefont{Aksimentiev}},
  \bibinfo{author}{\bibfnamefont{G.}~\bibnamefont{Wedemann}}, \bibnamefont{and}
  \bibinfo{author}{\bibfnamefont{R.}~\bibnamefont{Seidel}},
  \bibinfo{journal}{Phys. Rev. Lett.} \textbf{\bibinfo{volume}{105}},
  \bibinfo{pages}{158101} (\bibinfo{year}{2010}).

\bibitem[{\citenamefont{Neukirch and Marko}(2011)}]{Neukirch2011}
\bibinfo{author}{\bibfnamefont{S.}~\bibnamefont{Neukirch}} \bibnamefont{and}
  \bibinfo{author}{\bibfnamefont{J.~F.} \bibnamefont{Marko}},
  \bibinfo{journal}{Phys. Rev. Lett.} \textbf{\bibinfo{volume}{106}},
  \bibinfo{pages}{138104} (\bibinfo{year}{2011}).

\bibitem[{\citenamefont{Moroz and Nelson}(1997)}]{Moroz1997}
\bibinfo{author}{\bibfnamefont{J.~D.} \bibnamefont{Moroz}} \bibnamefont{and}
  \bibinfo{author}{\bibfnamefont{P.}~\bibnamefont{Nelson}},
  \bibinfo{journal}{Proc. Natl. Acad. Sci. USA} \textbf{\bibinfo{volume}{94}},
  \bibinfo{pages}{14418} (\bibinfo{year}{1997}).

\bibitem[{\citenamefont{Hills et~al.}(2010)\citenamefont{Hills, Lu, and
  Voth}}]{Hills2010}
\bibinfo{author}{\bibfnamefont{R.~D.} \bibnamefont{Hills}},
  \bibinfo{author}{\bibfnamefont{L.}~\bibnamefont{Lu}}, \bibnamefont{and}
  \bibinfo{author}{\bibfnamefont{G.~A.} \bibnamefont{Voth}},
  \bibinfo{journal}{PLoS Comput. Biol.} \textbf{\bibinfo{volume}{6}},
  \bibinfo{pages}{e1000827} (\bibinfo{year}{2010}).

\bibitem[{\citenamefont{Marko and Siggia}(1994)}]{Marko1994}
\bibinfo{author}{\bibfnamefont{J.~F.} \bibnamefont{Marko}} \bibnamefont{and}
  \bibinfo{author}{\bibfnamefont{E.~D.} \bibnamefont{Siggia}},
  \bibinfo{journal}{Science} \textbf{\bibinfo{volume}{265}},
  \bibinfo{pages}{506} (\bibinfo{year}{1994}).

\bibitem[{\citenamefont{Marko and Neukirch}(2013)}]{Marko2013}
\bibinfo{author}{\bibfnamefont{J.~F.} \bibnamefont{Marko}} \bibnamefont{and}
  \bibinfo{author}{\bibfnamefont{S.}~\bibnamefont{Neukirch}},
  \bibinfo{journal}{Phys. Rev. E} \textbf{\bibinfo{volume}{88}},
  \bibinfo{pages}{062722} (\bibinfo{year}{2013}).

\bibitem[{\citenamefont{Benham}(1992)}]{Benham1992}
\bibinfo{author}{\bibfnamefont{C.~J.} \bibnamefont{Benham}},
  \bibinfo{journal}{J. Mol. Biol.} \textbf{\bibinfo{volume}{225}},
  \bibinfo{pages}{835} (\bibinfo{year}{1992}).

\bibitem[{\citenamefont{Bauer and Benham}(1993)}]{Bauer1993}
\bibinfo{author}{\bibfnamefont{W.~R.} \bibnamefont{Bauer}} \bibnamefont{and}
  \bibinfo{author}{\bibfnamefont{C.~J.} \bibnamefont{Benham}},
  \bibinfo{journal}{J. Mol. Biol.} \textbf{\bibinfo{volume}{234}},
  \bibinfo{pages}{1184} (\bibinfo{year}{1993}).

\bibitem[{\citenamefont{Strawbridge et~al.}(2010)\citenamefont{Strawbridge,
  Benson, Gelfand, and Benham}}]{Strawbridge2010}
\bibinfo{author}{\bibfnamefont{E.~M.} \bibnamefont{Strawbridge}},
  \bibinfo{author}{\bibfnamefont{G.}~\bibnamefont{Benson}},
  \bibinfo{author}{\bibfnamefont{Y.}~\bibnamefont{Gelfand}}, \bibnamefont{and}
  \bibinfo{author}{\bibfnamefont{C.~J.} \bibnamefont{Benham}},
  \bibinfo{journal}{Curr. Genet.} \textbf{\bibinfo{volume}{56}},
  \bibinfo{pages}{321} (\bibinfo{year}{2010}).

\bibitem[{\citenamefont{Jeon et~al.}(2010)\citenamefont{Jeon, Adamcik, Dietler,
  and Metzler}}]{Jeon2010}
\bibinfo{author}{\bibfnamefont{J.-H.} \bibnamefont{Jeon}},
  \bibinfo{author}{\bibfnamefont{J.}~\bibnamefont{Adamcik}},
  \bibinfo{author}{\bibfnamefont{G.}~\bibnamefont{Dietler}}, \bibnamefont{and}
  \bibinfo{author}{\bibfnamefont{R.}~\bibnamefont{Metzler}},
  \bibinfo{journal}{Phys. Rev. Lett.} \textbf{\bibinfo{volume}{105}},
  \bibinfo{pages}{208101} (\bibinfo{year}{2010}).

\bibitem[{\citenamefont{Randall et~al.}(2009)\citenamefont{Randall,
  Zechiedrich, and Pettitt}}]{Randall2009}
\bibinfo{author}{\bibfnamefont{G.~L.} \bibnamefont{Randall}},
  \bibinfo{author}{\bibfnamefont{L.}~\bibnamefont{Zechiedrich}},
  \bibnamefont{and} \bibinfo{author}{\bibfnamefont{B.~M.}
  \bibnamefont{Pettitt}}, \bibinfo{journal}{Nucleic Acids Res.}
  \textbf{\bibinfo{volume}{37}}, \bibinfo{pages}{5568} (\bibinfo{year}{2009}).

\bibitem[{\citenamefont{Kannan and Zacharias}(2009)}]{Kannan2009}
\bibinfo{author}{\bibfnamefont{S.}~\bibnamefont{Kannan}} \bibnamefont{and}
  \bibinfo{author}{\bibfnamefont{M.}~\bibnamefont{Zacharias}},
  \bibinfo{journal}{Phys. Chem. Chem. Phys.} \textbf{\bibinfo{volume}{11}},
  \bibinfo{pages}{10589} (\bibinfo{year}{2009}).

\bibitem[{\citenamefont{Harris}(2006)}]{Harris2006}
\bibinfo{author}{\bibfnamefont{S.~A.} \bibnamefont{Harris}},
  \bibinfo{journal}{Phil. Trans. A} \textbf{\bibinfo{volume}{364}},
  \bibinfo{pages}{3319} (\bibinfo{year}{2006}).

\bibitem[{\citenamefont{Liverpool et~al.}(2008)\citenamefont{Liverpool, Harris,
  and Laughton}}]{Liverpool2008}
\bibinfo{author}{\bibfnamefont{T.~B.} \bibnamefont{Liverpool}},
  \bibinfo{author}{\bibfnamefont{S.~A.} \bibnamefont{Harris}},
  \bibnamefont{and} \bibinfo{author}{\bibfnamefont{C.~A.}
  \bibnamefont{Laughton}}, \bibinfo{journal}{Phys. Rev. Lett.}
  \textbf{\bibinfo{volume}{100}}, \bibinfo{pages}{238103}
  (\bibinfo{year}{2008}).

\end{thebibliography}

\newpage
\setcounter{figure}{0}
\setcounter{table}{0} 
\setcounter{section}{0} 
\renewcommand{\thefigure}{S\arabic{figure}}
\renewcommand{\thetable}{S\arabic{table}}
\renewcommand{\thesection}{\Roman{section}}
\onecolumngrid

\begin{center}
 {\Large \bf Supplementary Material}
\end{center}

\section{Simulation methods and boundary conditions}\label{sec:boundaries}
\noindent \underline{Simulation algorithm}\\

As oxDNA is an implicit solvent model, it is inappropriate to directly integrate the equations of motion resulting from its interaction potentials, which would lead to ballistic particle motion.
Instead, we used the Andersen-like thermostat that is described in more detail in Ref.~\onlinecite{Russo2009}. It propagates the system for a set number of time steps $N_{\rm Newt}$ according to Newton's equations, using the common Verlet integrator with a time step of 12.2\,fs.
Linear and angular velocities in the system are then assigned new values drawn from a Maxwell-Boltzmann distribution at $T=300$\,K with probabilities $p_{\rm lin}$ and $p_{\rm ang}$.
For all simulations performed for this work, we chose $N_{\rm Newt}=103$, $p_{\rm lin}=0.0204$ and $p_{\rm ang}=0.0068$. 
The time-scales reported in this paper were set by mass, energy and length scales used in the integrator. 
However, making direct comparisons of time-scales between coarse-grained simulations and experiments is complex, see for example the discussions in Refs.~\onlinecite{Padding2006,Louis2010,Murtola2009}, and in Supplementary Sec.~\ref{sec:plectoneme_displacement}, where we discuss diffusion.
Similar simulations settings have been successfully used in other applications of oxDNA (see e.g. Refs.~\onlinecite{Romano2013,Ouldridge2013a,Doye2013}).\\

\noindent \underline{Boundary conditions}\\

In this work, we study the response of DNA to superhelical stress.
A common measure of superhelical stress is the linking number $Lk$, a topological quantity which equals the number of times the single strands wrap around each other.
In topologically constrained systems, $Lk$ decomposes according to the Fuller-White-C\u{a}lug\u{a}reanu relation $Lk=Tw+Wr$ \cite{Calufareanu1959,White1969,Fuller1971o}, 
where $Tw$ is the number of single-strand crossings, and $Wr$ the number of self-crossings of the double strand axis of the system.
To facilitate comparison to systems of different strand length, we quantify torsion using the length independent superhelical density defined as $\sigma = (Lk-Lk_0)/Lk_0$, 
where $Lk_0$ is the number of single-strand crossings in a torsionally relaxed, linear double strand.

\begin{figure}[h!]
 \includegraphics[width=.9\columnwidth]{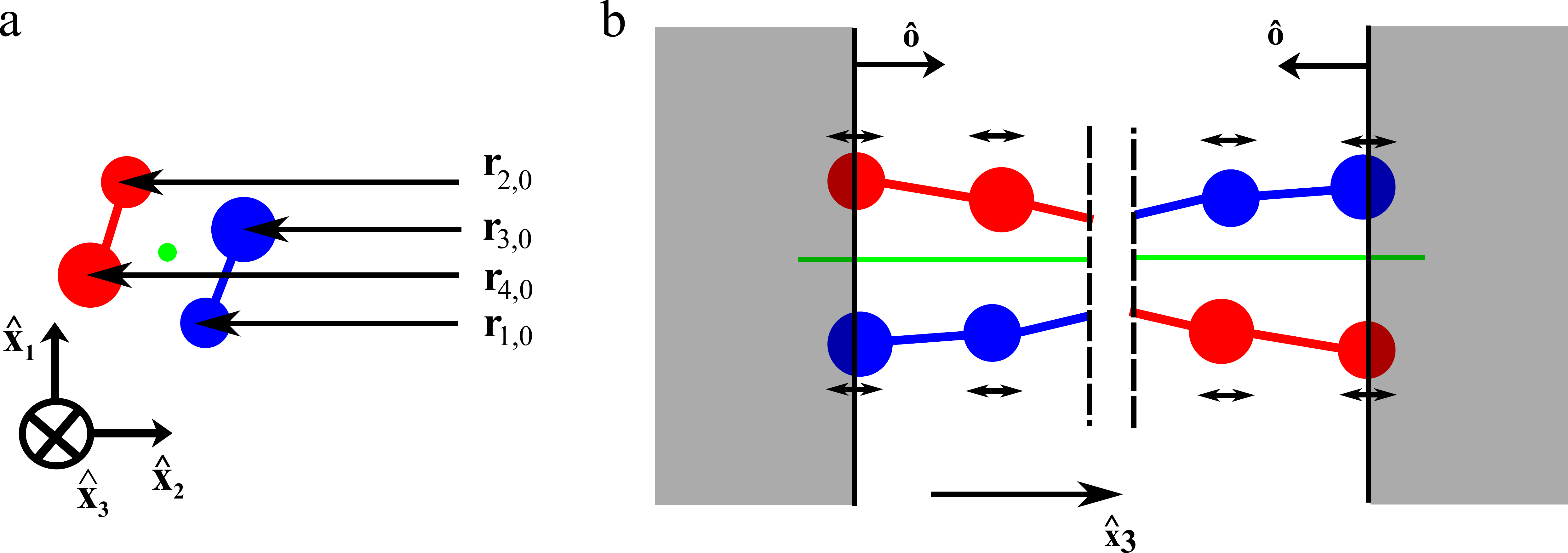}
 \caption{Schematic depiction of the applied boundary conditions illustrated for the last 2\,bp at each end of the strand: (a) View along the strand axis. 5 nucleotides at each strand end are constrained by 2-dimensional harmonic traps, which fix boundary nucleotides to positions $\textbf r_{n,0}$ in planes perpendicular to the strand axis (green). (b) View perpendicular to the double strand. Due to the 2-dimensional traps, nucleotides are unconstrained only in the strand-axis direction. A repulsion plane perpendicular to the strand axis is tagged to the last base pair. Movement of nucleotides into the area below the end base pair (shaded grey) is excluded. In order to allow unconstrained strand extensibility, the repulsion plane does not act on the first two base pairs along the strand.}
 \label{fig:boundary_schematic}
\end{figure}

To impose superhelical stress in a simulation, we constrain the ends of a double strand, and disallow passing of the double strand across the strand ends.
Similar constraining boundary conditions have been successfully used before in simulations of cruciform extrusion \cite{Matek2012}.
A schematic overview of the boundary conditions applied is shown in Fig.~\ref{fig:boundary_schematic}.

Fixation is implemented by adding five boundary base pairs to the strand ends, which are trapped in harmonic potentials. 
These potentials acting on the $n$-th trapped nucleotide have the form
\begin{equation}
 V_{\rm trap}({\textbf r_{n}} ; {\textbf r_{n,0}}) = \frac{1}{2} \sum_{i=1}^3 k_{\rm trap}^i (r^i_{n}-r_{n,0}^i)^2,
\end{equation}
where ${\textbf r_n = (r^1_n, r^2_n, r^3_n)}$ is the centre-of-mass position of the $n$-th trapped nucleotide and the corresponding trap position is ${ \textbf r_{n,0}=(r^1_{n,0}, r^2_{n,0}, r^3_{n,0})}$, chosen initially such as to fix a given twist angle of the strand.
To keep the overall twist angle on the system fixed, but ensure free extensibility of the strand along the setup axis $\mathbf{\hat{x}_3}$, we choose $k_1^{\rm trap}=k_2^{\rm trap}=57.1$N/m and $k_3^{\rm trap}=0$.
The traps defined in this way only constrain the end nucleotides in a co-moving 2-dimensional plane perpendicular to the strand setup axis, while not hindering the strand in the $\mathbf{\hat{x}_3}$ direction.
High trap stiffness in that plane ensures that fluctuations of the linking number $Lk$ in the course of a simulation are negligible.

Strands have finite lengths, which means that more distant parts of the system can pass around the strand ends. 
Such a process would change the linking number $Lk$ and thus the superhelical density $\sigma$ of the system.
We therefore prevent this process by introducing repulsion planes oriented perpendicular to the setup axis $\mathbf{\hat{x}_3}$ which co-move with the first boundary nucleotide of the two single strands in the system.
Repulsion planes generate a potential 
\begin{equation}
 V_{\rm plane}({\textbf r};{\textbf R}) = \frac{1}{2} k^{\rm plane} \left( \left( {\textbf r}-{\textbf R} \right) \cdot \mathbf{\hat{o}} \right)^2 \theta(-\left( {\textbf r}-{\textbf R} \right) \cdot \mathbf{\hat{o}}),
\end{equation}
where {\bf r} is the centre-of-mass position of an affected particle, {\bf R} and $\mathbf{\hat{o}}$ are anchor point and orientation of the plane, and $\theta$ is the Heaviside step function.
We choose $\mathbf{\hat{o}}=\mathbf{\hat{x}_3}$ and $\mathbf{\hat{o}}=-\mathbf{\hat{x}_3}$ for the lower and upper repulsion planes respectively, and set ${\textbf R}$ to the instantaneous positions of the  the first and last double strand boundary base pair.
To avoid hindering free strand extensibility in the $\mathbf{\hat{x}_3}$ direction, the repulsion planes do not interact with the next-to-last boundary base pairs at both strand ends.
In all simulations, we chose $k^{\rm plane}=28.5$\,pN/nm, which prevented the duplex from passing over its ends.

\noindent \underline{Microscopic determination of base-pair breaking}\\

In several sections of this work, we determine if a given base pair is formed or broken. 
As in previous applications of oxDNA~\cite{Ouldridge2011,Sulc2012}, a base-pair was counted as formed if the energy contribution from hydrogen-bonding was below $-4.13 \times 10^{-21}$\,J, corresponding to approximately $15\%$ of the typical energy of a fully formed hydrogen bond.\\

\noindent \underline{Sequences and symmetry of hydrogen bonding}\\

\begin{figure}[h]
 \includegraphics[width=0.4\columnwidth]{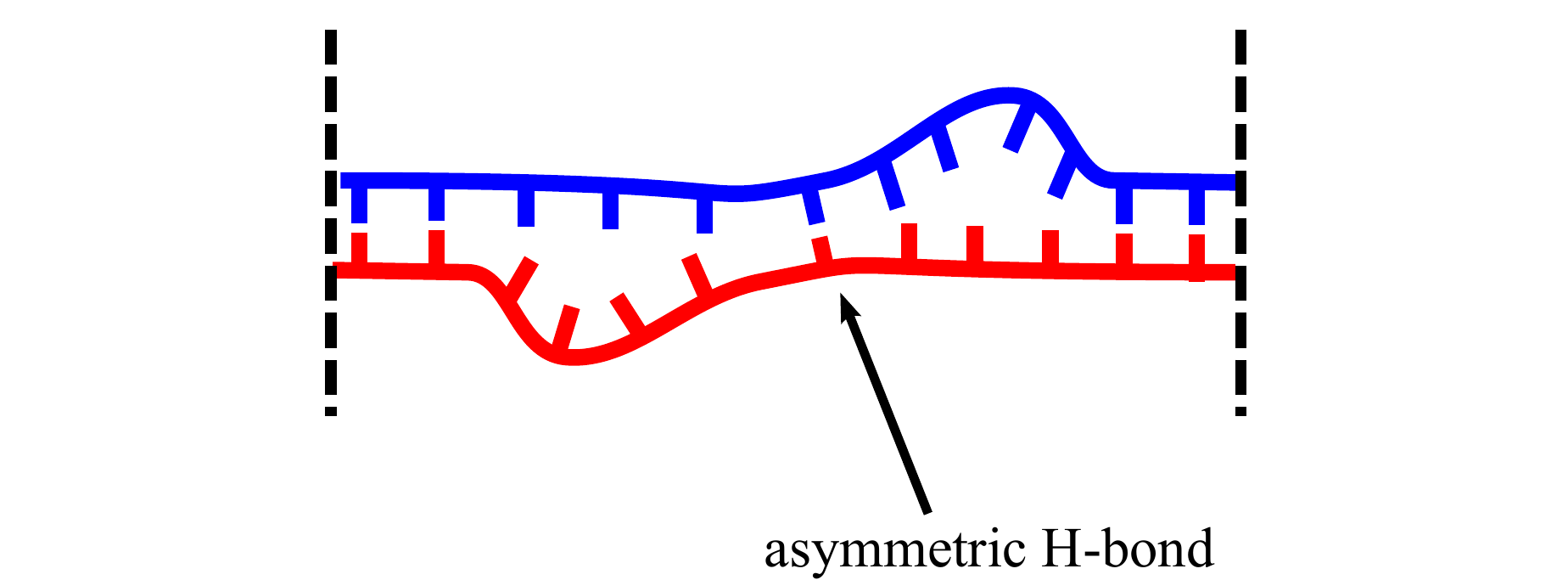}
 \caption{Asymmetric hydrogen bonds were excluded in simulations.}
 \label{fig:asymBonds}
\end{figure}

The double strands studied in this work possess fully complementary sequences. 
Hydrogen bonding was only permitted between each nucleotide and its direct counterpart on the other single strand, excluding shifted bonds as shown in Fig. \ref{fig:asymBonds}. 
This restriction was imposed to avoid non-generic secondary structure effects, and increase simulation efficiency. It has been successfully used before for other studies with oxDNA~\cite{Matek2012,Ouldridge2013a,Romano2013,Doye2013}.
The sequences used within the sequence-dependent parametrization are given in Supplementary Sec.~\ref{sec:sequences}.

\newpage
\section{Comparison of mechanical behaviour to experimental data}\label{sec:experiment_comparison}
The mechanical behaviour we observe can be compared directly to experimental data from single molecule assays.
Most available data has been obtained using magnetic \cite{Strick1996,Brutzer2010,Janssen2012,Mosconi2009} or optical tweezers \cite{Forth2008}, which can measure the end-to-end extension and torque response of a single DNA molecule.
Available datasets differ in molecule length $L$, ionic strength of the buffer, and sequences properties of the DNA strands used in the respective experiments.

Here, we relate our results to experimental findings and explain the influence of different parameters on the quality of agreement.
We note that oxDNA is limited to an ionic strength of $[{\rm Na}^+]=500$\,mM, as the effective potentials of the model were parametrized at that salt concentration. 
Furthermore, for reasons of computational efficiency, we simulated double strands of length  $L=600$\,bp, for which equilibrium data for end-to-end extension can be obtained within about two weeks of runtime on a current CPU.
This is shorter than the strand lengths typically used in experiment, which will affect the comparison of results.

In order to facilitate comparison to experimental data, we report our results in terms of the relative strand extension $l=L/L_0$ and the superhelical density $\sigma$, which are length-independent up to finite-size corrections vanishing in the long-strand limit~\cite{Marko2012}.
Some uncertainty exists about the relaxed length $L_0$, because in typical experimental assays, measurements of end-to-end distance take place on molecules which are attached to a cover slip and a bead via chemically functionalized ends.
Attachment may slightly modify the effective free length of the double strand. 
The simulated system only uses a very short piece of strand to constrain the duplex ends (see Supplementary Sec.~\ref{sec:boundaries}) and therefore does not include these chemical attachment effects.
Uncertainty about the effective free length of attachment may therefore introduce small systematic differences when comparing experiments to our simulations, especially when measuring the absolute length $L$.
By contrast, experimental measurements of the buckling superhelical densities $\sigma_b$ and postbuckling slopes $dl / d\sigma$ should be largely unaffected by the length of attachment, and may therefore be easier to compare with our simulations.

The buckling superhelical density $\sigma_b$ is determined primarily by the values of bending persistence length and twist persistence length~\cite{Brutzer2010}, which in oxDNA take the values $B_0=42.5$~nm and $C_0=114.7$~nm respectively~\cite{Ouldridge2011}. 
As these values lie in the range of values reported for DNA~\cite{Mosconi2009}, oxDNA might be expected to reproduce experimental values for $\sigma_b$. 
Differences with experiment may simply stem from small differences in the value of these constants, and would occur for any elastic model using these values for $B_0$ and $C_0$.

The post-buckling slopes are thought to mainly depend on the radius of the plectoneme stem, which is set by the physics of twisting and bending, as well as by screened electrostatic interactions between the double strands~\cite{Maffeo2010, Neukirch2011}.
Good agreement of post-buckling slopes may thus be an indication for the consistency of oxDNA's treatment of strongly screened electrostatics by excluded volume interactions.

In addition to reproducing elastic properties of the double strand, oxDNA provides a good representation of the melting curves of DNA, as well as several other systems where breaking of base pairs plays a role~\cite{Ouldridge2011,Romano2013,Doye2013}.
This gives us confidence to apply the model to study plectonemes, in which both effects of strand elasticity and double strand denaturation might be expected to be relevant.

All simulation results presented in this section were obtained using the average-base parametrization of oxDNA. 
Details of the mechanical response of DNA to imposed twist may however depend on the specific sequence used.
In particular, the precise value of the crossover force $F_{\rm char}$ between bubbles and plectonemic structures might be expected to depend on sequence properties of the DNA strand, as the enthalpic cost of bubble formation decreases with increasing AT content. 
Therefore, AT-rich stretches represent preferred nucleation sites for denaturation bubbles, which can facilitate denaturation and therefore decrease $F_{\rm char}$.
The effect of different sequences is described in more detail in Supplementary Sec.~\ref{sec:sequences}.\\\\\\

\newpage
\noindent \underline{Molecular extension ``hat curves''}\\

\begin{figure}[h!]
 \includegraphics[scale=.7]{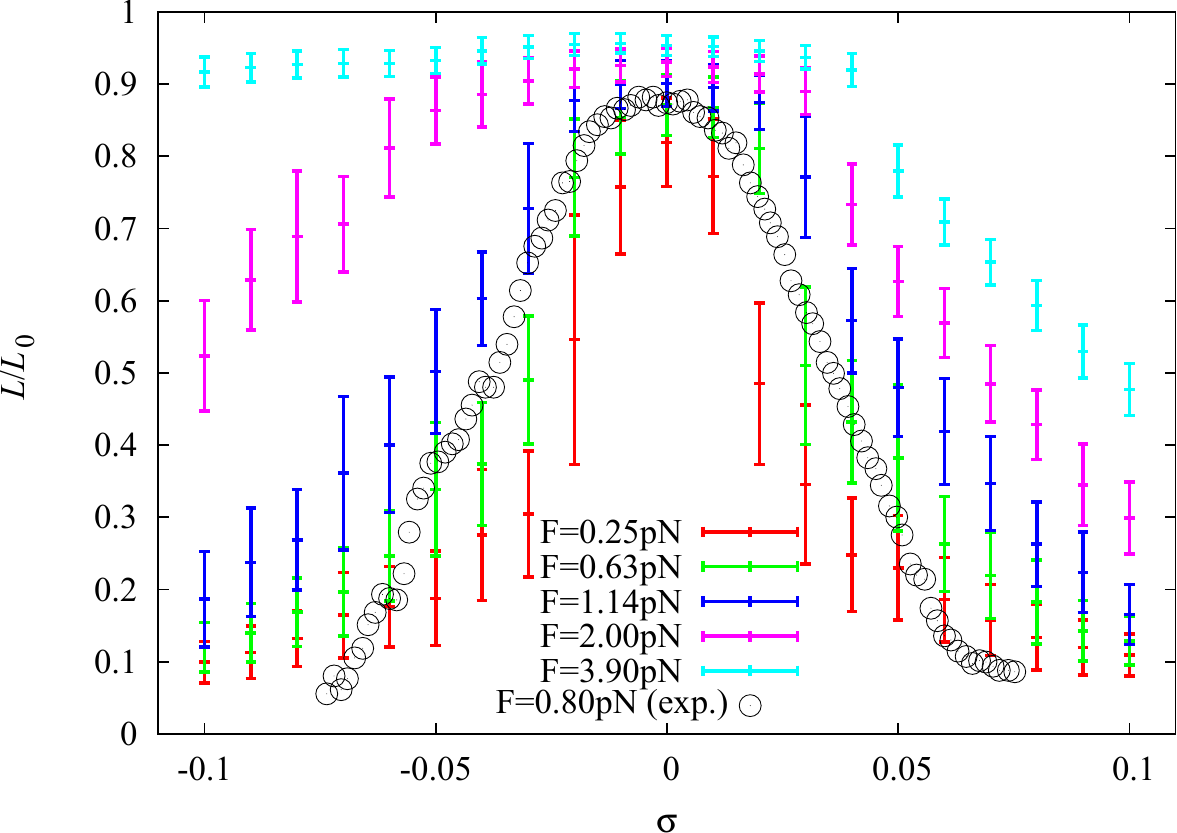}
 \caption{Experimental strand extension data of Ref.~\onlinecite{Tempestini2013} at $[{\rm Na}^+]=500$\,mM and $L \approx 5.8$\,kbp for $F=0.8$\,pN, compared to data from simulations using oxDNA ($[{\rm Na}^+]=500$\,mM and $L=600$\,bp). Error bars on simulation results indicate thermal fluctuations in the end-to-end distance, rather than sampling uncertainties. Good agreement is observed, while nonlinear effects in the post-buckling slopes for high $|\sigma|$, caused by end effects, are somewhat more pronounced for the shorter simulated system.}
 \label{fig:hatcurves_tempestini}
\end{figure}

\begin{figure}[h!]
 \includegraphics[scale=.7]{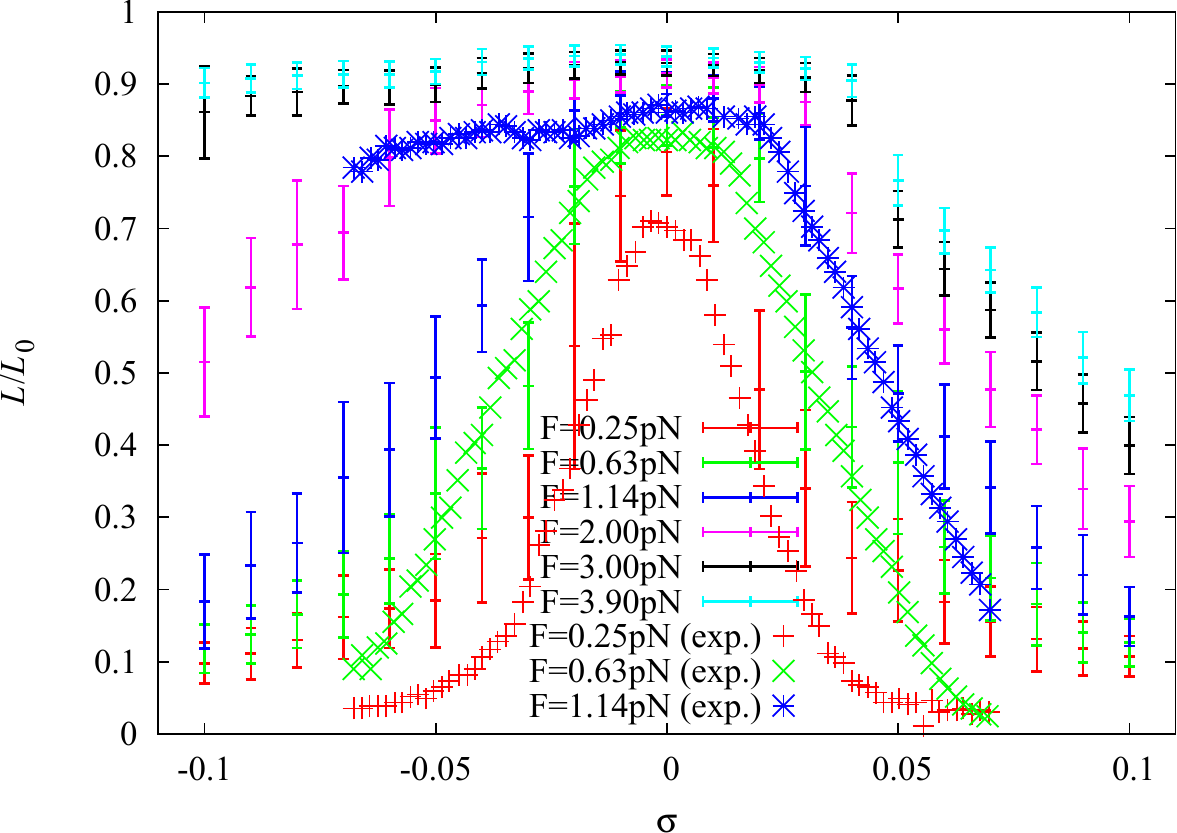}
 \caption{Experimental strand extension data of Ref.~\onlinecite{Salerno2012a} at $[{\rm Na}^+]=150$\,mM and $L \approx 6$\,kbp for $F=0.25$\,pN, $0.63$\,pN and $1.14$\,pN compared to data from simulations using oxDNA ($[{\rm Na}^+]=500$\,mM and $L=600$\,bp). Data for equal stretching force is represented by the same color. Error bars on simulation results indicate thermal fluctuations in the end-to-end distance, rather than sampling uncertainties. At lower salt concentration, plectoneme structures are enthalpically penalized as electrostatic interactions are screened less, making them less favourable compared to bubble configurations. Lower salt concentration also reduces the free-energy cost of base pair breaking. Hence, the crossover from tip-bubble plectonemes to extended bubble states occurs at lower forces than would be the case for $[{\rm Na}^{+}]=500$\,mM. Sequence dependence may additionally decrease the crossover force $F_{\rm char}$ by providing AT-rich bubble nucleation sites, as discussed in Supplementary Sec.~\ref{sec:sequences}. Non-linear regions in the post-buckling slope are somewhat more pronounced for the shorter simulated system.}
 \label{fig:hatcurves_salerno}
\end{figure}

\begin{figure}[h!]
 \includegraphics[scale=.7]{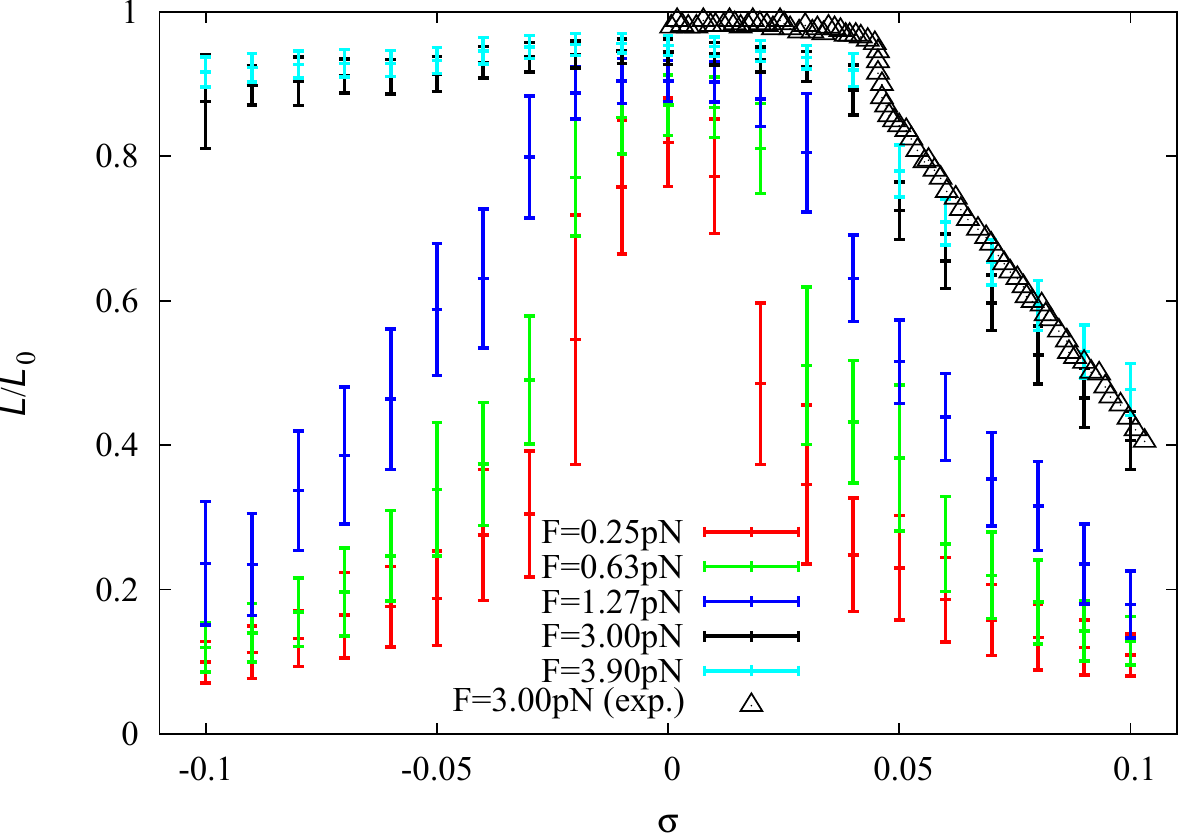}
 \caption{Experimental strand extension data of Ref.~\onlinecite{Brutzer2010}(black triangles) for $[{\rm Na}^+]=320$\,mM and $L=1.9$\,kbp at $F=3.0$\,pN compared to data from simulations using oxDNA ($[{\rm Na}^+]=500$\,mM and $L=600$\,bp). For simulation data, error bars indicate standard deviations due to thermal fluctuations, rather than sampling uncertainties. Good agreement of buckling point and post-buckling slope is observed. Note also the abrupt length reduction upon buckling, present in both experiment and simulations.}
 \label{fig:hatcurves_brutzer}
\end{figure}

Measurements of strand extension as a function of applied force $F$ and superhelical density $\sigma$ are very reproducible, and have been measured in many setups for different ambient conditions~\cite{Brutzer2010,Mosconi2009,Tempestini2013,Salerno2012a,Strick1996,Janssen2012}.
Here, we compare to a set of experiments performed at conditions close to those used in the parametrization of oxDNA.
Our strand extension results are plotted together with experimental data in Figs.~\ref{fig:hatcurves_tempestini} -~\ref{fig:hatcurves_brutzer}. 

Fig.~\ref{fig:hatcurves_tempestini} shows a comparison to experimental data of Tempestini \textit{et al.} \cite{Tempestini2013} on a stand of length 7~kbp, with an effective free length of $5.8$\,kbp. 
This effective length is inferred from the length of the attachment described in Ref.~\onlinecite{Tempestini2013}.
The salt concentration used in these experiments is $[{\rm Na}^+]=500$\,mM, the value at which oxDNA was parametrized.
Post-buckling slopes are in good agreement with simulation data, whereas buckling occurs at slightly lower experimental superhelical density $\sigma_b$.
This may be due to additive finite-size corrections $O(L^{-1/2})$ predicted in continuum models of the buckling transition \cite{Marko2012}.

Fig. \ref{fig:hatcurves_salerno} shows a comparison to data by Salerno \textit{et al.}~\cite{Salerno2012a}, obtained at $L=6$\,kbp and a lower salt concentration $[{\rm Na}^+]=150$\,mM.
Lower salt is expected to destabilize plectonemes and make bubbles more favourable, as electrostatic interactions of the backbone are less screened.
In particular, this effect is expected to become more significant for $\sigma<0$, as has been shown experimentally~\cite{Tempestini2013}. 
Comparing to simulation results, we still observe good agreement in the post-buckling slopes for all forces when $\sigma>0$.
In contrast, the crossover to the extended bubble state for $\sigma<0$ happens at higher forces for oxDNA. 
That the force is higher is expected because of the difference in the salt concentration between simulation and experiment.
Although appropriate hat curves are unavailable at higher salt for $\sigma<0$, data on the fluctuation of end-to-end lengths from Ref.~\onlinecite{Tempestini2013} suggests that the transition to the extended bubble state occurs at roughly 1\,pN, about 1\,pN below our estimate for $[{\rm Na}^+]=500$\,mM. 
We discuss these differences in more detail in Supplementary Sec.~\ref{sec:length_comparison} on end-to-end fluctuations, and Supplementary Sec.~\ref{sec:sequences} on 
sequence dependence, where we show that since AT rich regions are more likely to form bubbles, taking this into account with a sequence dependent model leads to lower forces for the crossover to bubbles than using an average base model for oxDNA does. 
Nevertheless, even if these effects don't explain the full difference with experiment because oxDNA underestimates the stability of internal bubbles, this is unlikely to affect our qualitative results. 

In the data of Refs.~\onlinecite{Salerno2012a} and~\onlinecite{Tempestini2013}, deviations of the post-buckling slopes from linearity were observed at low stretching forces and high values of $|\sigma|$.
Similar, more pronounced deviations are observed in our results due to finite-size effects arising from the interaction of the coiled DNA strand with the system boundaries (cf. Supplementary Sec.~\ref{sec:boundaries}).
Boundary interactions become relevant when the diameter of the plectonemic region is comparable to the end-to-end extension of the double strand.
This is expected to be the case at somewhat lower values of $|\sigma|$ in the 600-bp system studied here, as compared to the roughly 10 times longer experimental systems of Refs.~\onlinecite{Salerno2012a} and~\onlinecite{Tempestini2013}, shown in Figs.~\ref{fig:hatcurves_tempestini} and~\ref{fig:hatcurves_salerno}.

Fig.~\ref{fig:hatcurves_brutzer} depicts data obtained by Brutzer \textit{et al.}~\cite{Brutzer2010} at $[{\rm Na}^+]=320$\,mM, $L=1.9$\,kbp and $\sigma>0$.
Close agreement is observed concerning both the buckling superhelical densities $\sigma_b$ and the post-buckling slopes $dl/d \sigma$ (Fig.~\ref{fig:hatcurves_brutzer}).
We note that in this study, abrupt strand shortening at the buckling point was observed and ascribed to the energetic cost of forming the plectoneme end-loop. 
Abrupt strand shortening was shown to become more pronounced with increasing ionic strength. 
In oxDNA, we consistently observe a marked shortening of the strand at similar force around the buckling point (see Fig.~\ref{fig:hatcurves_brutzer}; for an example with the sequence-dependent parametrization of oxDNA, see Fig.~\ref{fig:hatcurves_avg_ran}).\\\\\\

\noindent \underline{Torque response curves}\\

\begin{figure}[h!]
 \includegraphics[scale=.7]{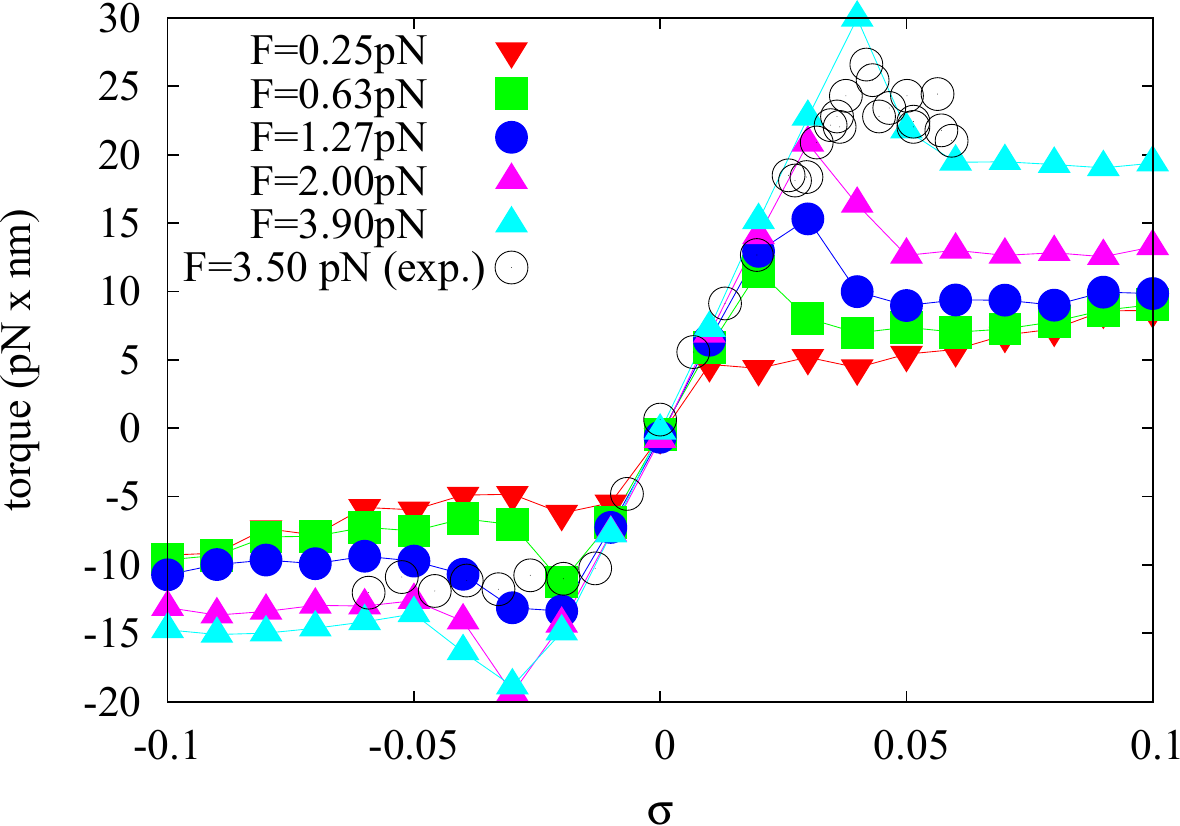}
 \caption{Experimental torque response data of Ref.~\onlinecite{Janssen2012} at $[{\rm Na}^+]=550$\,mM and $L=7.9$\,kbp for $F=3.50$\,pN compared to data from simulations using oxDNA ($[{\rm Na}^+]=500$\,mM and $L=600$\,bp). Excellent agreement is observed both in the linear regime (see also Fig.~\ref{fig:MorozNelson}) and for the location and size of the torque overshoot.}
 \label{fig:torques_janssen}
\end{figure}

\begin{figure}[h!]
 \includegraphics[width = 0.7\columnwidth]{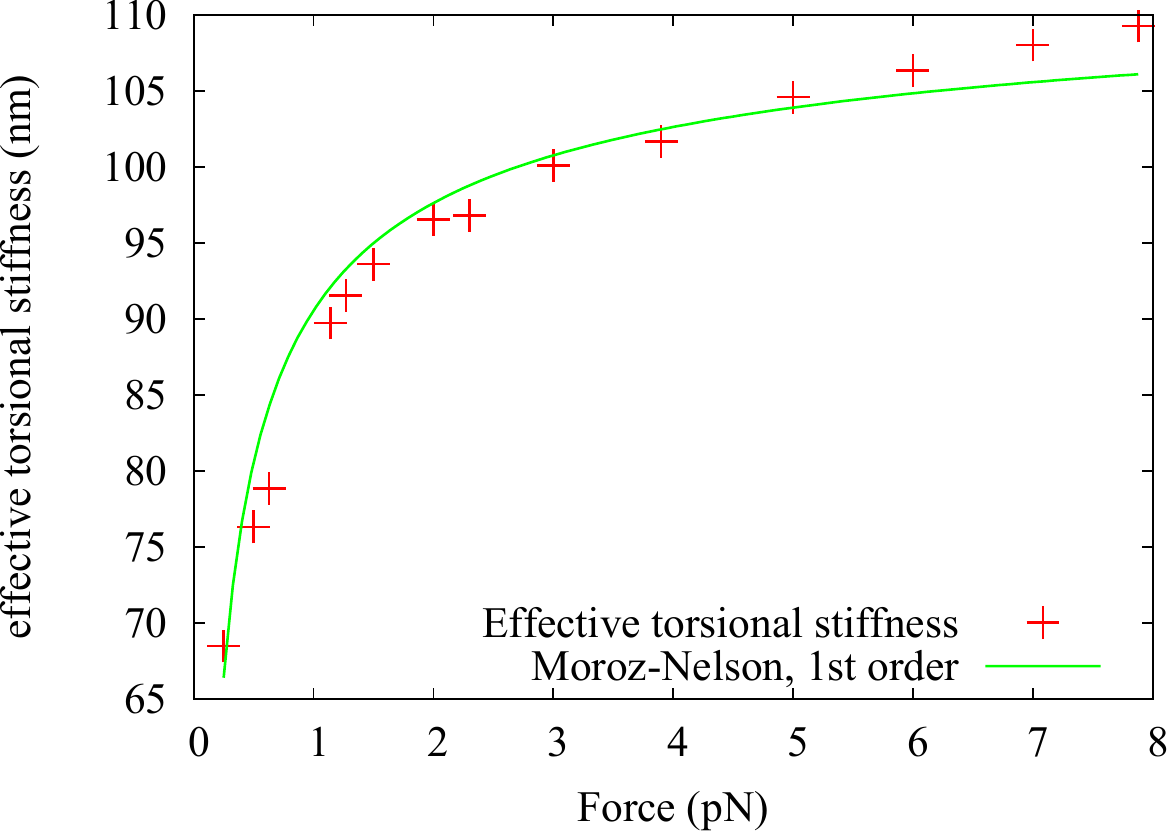}
 \caption{Prediction of the effective twist persistence length from microscopic values according to Moroz-Nelson theory (green lines show first-order expansion; no free parameters) and values determined from simulations (red crosses).}
 \label{fig:MorozNelson}
\end{figure}

The torque response of DNA to imposed twist has been studied for systems of different length and salt concentration~\cite{Forth2008,Mosconi2009,Janssen2012}.
A common feature of the torque response is a linear regime for low $|\sigma|$, followed by a torque overshoot at buckling and a constant post-buckling torque.
The published dataset which is closest to the conditions used in oxDNA is the one recently measured by Janssen \textit{et al.}~\cite{Janssen2012} at $[{\rm Na}^+]=550$\,mM and $L=7.9$\,kbp using a variant of the magnetic tweezer setup.
A comparison to the torque response data is shown in Fig.~\ref{fig:torques_janssen}, exhibiting very good agreement for both the size and location of the torque overshoot, as well as the slope of the torque response in the linear regime and the magnitude of the constant post-buckling torque.
\clearpage
\noindent \underline{Correspondence to Moroz-Nelson theory} \\\\
The effective twist persistence length of a DNA strand is renormalized because of thermal fluctuations~\cite{Moroz1997}.
By using a torsional directed walk model, Moroz and Nelson estimated the effect of fluctuations to lead to an effective twist persistence length of
\begin{equation} 
 C_{\rm eff} = C_0 \left[1-\frac{C_0}{4B_0} \sqrt{\frac{k_B T}{B_0F}}\right],
 \label{eq:moroznelson}
\end{equation}
where $B_0$ and $C_0$ are the microscopic bending and twist persistence lengths of the double strand respectively. In oxDNA, these values have previously been determined as $B_0=42.5$\,nm and $C_0=114.7$\,nm~\cite{Ouldridge2011}.\\
We determined the force-dependent effective twist persistence length $C_{\rm eff}$ of the simulated 600-bp system by fitting the slope of the linear regime of the torque response curve (see e.g. Fig.~\ref{fig:torques_janssen}).
Good agreement of the measured torsional moduli with the theoretical prediction is observed (Fig.~\ref{fig:MorozNelson}). 
If the functional form of Eq.~\ref{eq:moroznelson} was fitted to the data with $B_0$ and $C_0$ as free parameters, we obtain $B_0=40.8$\,nm and $C_0=115.6$\,nm, which are very similar to the previously determined values.
Note that the agreement shown in Fig.~\ref{fig:MorozNelson} is obtained without free fit parameters.
\newpage
\section{Simulations of plectonemes of length 1500\,bp}\label{sec:length_comparison}
For reasons of computational efficiency, the majority of the data presented in the main paper of this work was taken from systems at a strand length of $L=600$\,bp.
In order to study the influence of differences in strand length and for comparison with experimental work we also performed simulations at strand length $L=1500$\,bp, for which a version of oxDNA for CUDA-enabled GPUs was used.

At forces between 1 and 3\,pN, the 600-bp fluctuation spectrum exhibits enhanced fluctuations, and shows a two-peak substructure (see inset of Fig.~\ref{fig:hatcurves_torques}b in the main text). 
The first peak may be explained by fluctuations arising from the initial opening of the tip bubble, while the second peak is due to the competition between a tip-bubble plectoneme and the extended bubble state.
From this consideration, the first peak of the fluctuation spectrum might be expected to be approximately independent of system size, kink formation being a local effect associated with the plectoneme end-loop.
On the other hand, the high-force peak of the enhanced fluctuation region is expected to grow linearly with system size for constant $\sigma$, as it is related to the plectoneme size, and the maximum size of a plectoneme is set by the absolute linking difference $\Delta Lk=  \sigma (N-1)/p$, where $N$ is the number of base pairs in the strand, and $ p \approx 10.4$ is the average number of base pairs per turn. 

\begin{figure}[bh!]
\centering
\begin{minipage}{.5\textwidth}
  \centering
  \includegraphics[width=1.\linewidth]{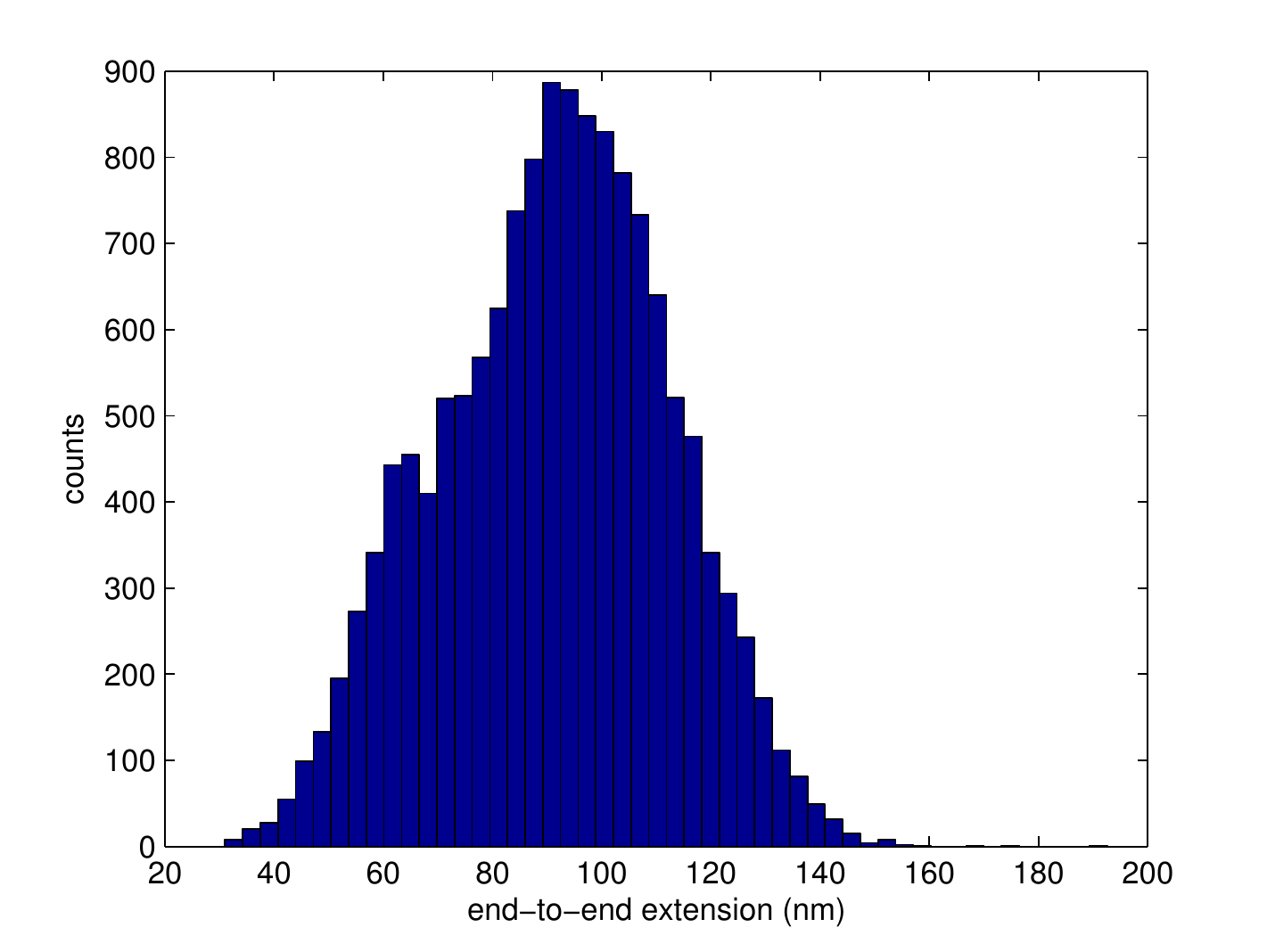}
\end{minipage}%
\begin{minipage}{.5\textwidth}
  \centering
  \includegraphics[width=1.\linewidth]{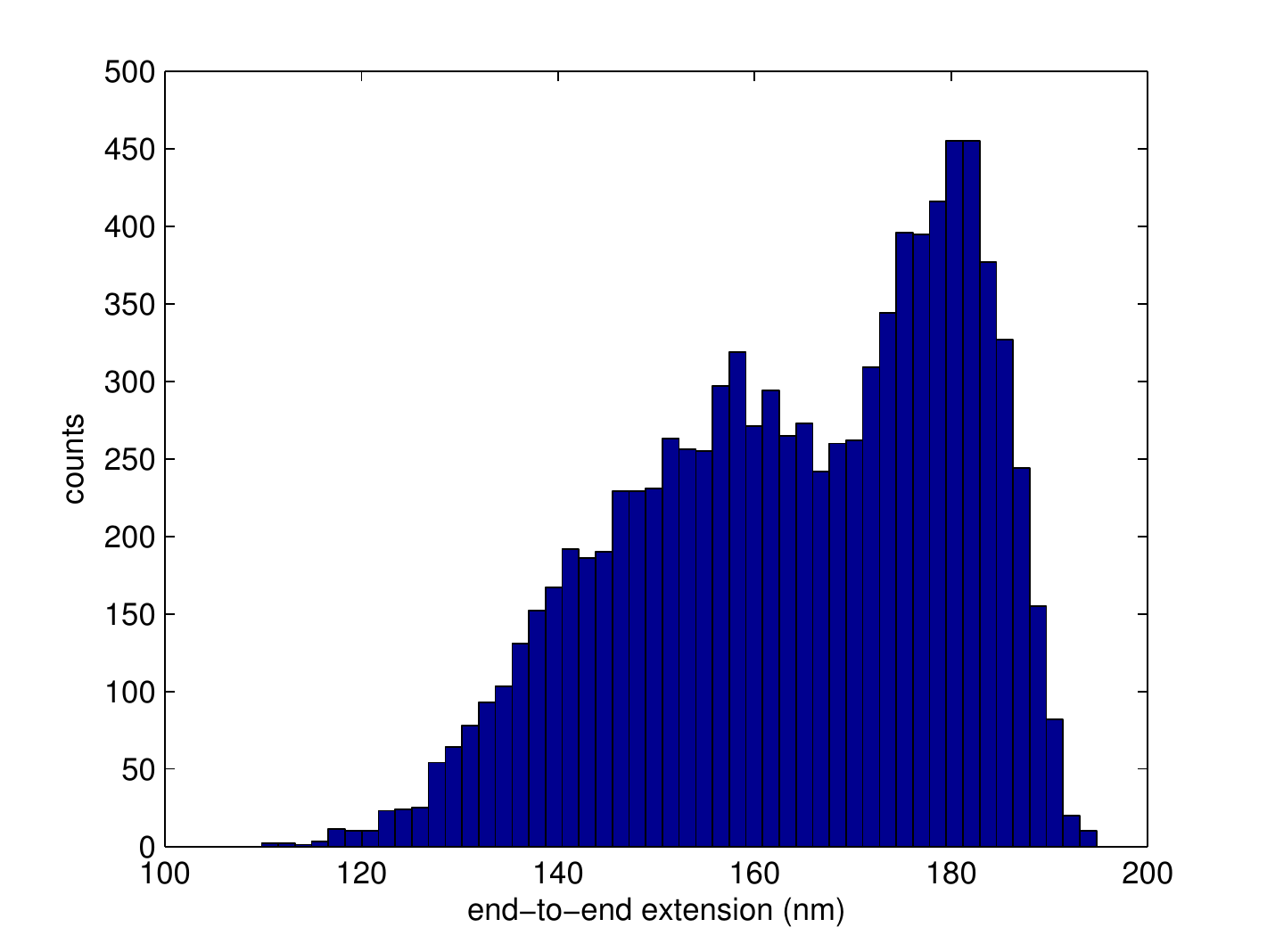}
\end{minipage}
\caption{Extension distributions for $L=600$\,bp, at superhelical density $\sigma=-0.08$, with $F=1.5$\,pN (left) and $F=2.3$\,pN (right). Note the weak bimodality in the right histogram, due to interconversion between bubble and tip-bubble plectoneme states.}
\label{fig:600bp_end2end_hist}
\end{figure}

\begin{figure}[bh!]
\centering
\begin{minipage}{.5\textwidth}
  \centering
  \includegraphics[width=1.\linewidth]{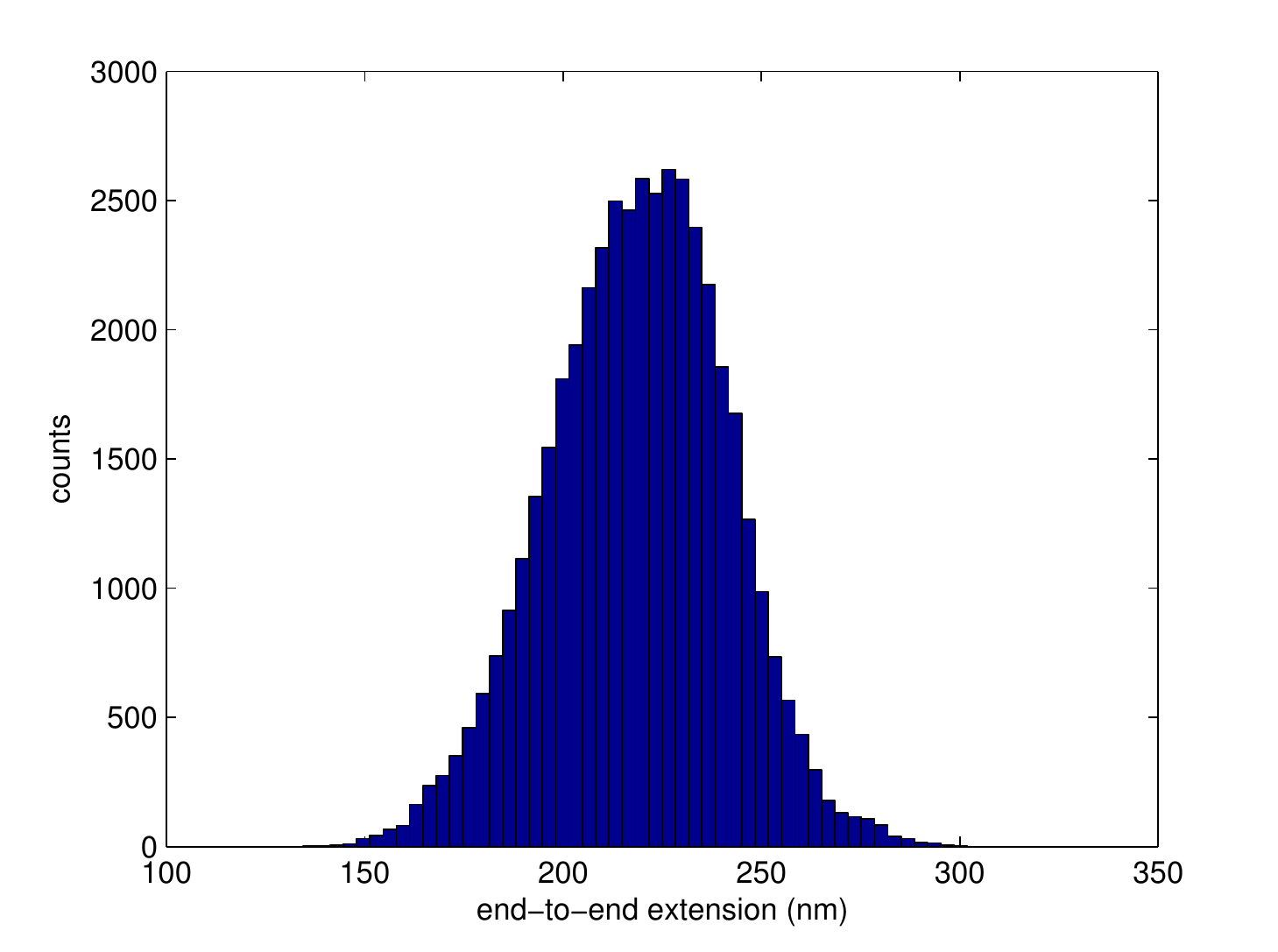}
\end{minipage}%
\begin{minipage}{.5\textwidth}
  \centering
  \includegraphics[width=1.\linewidth]{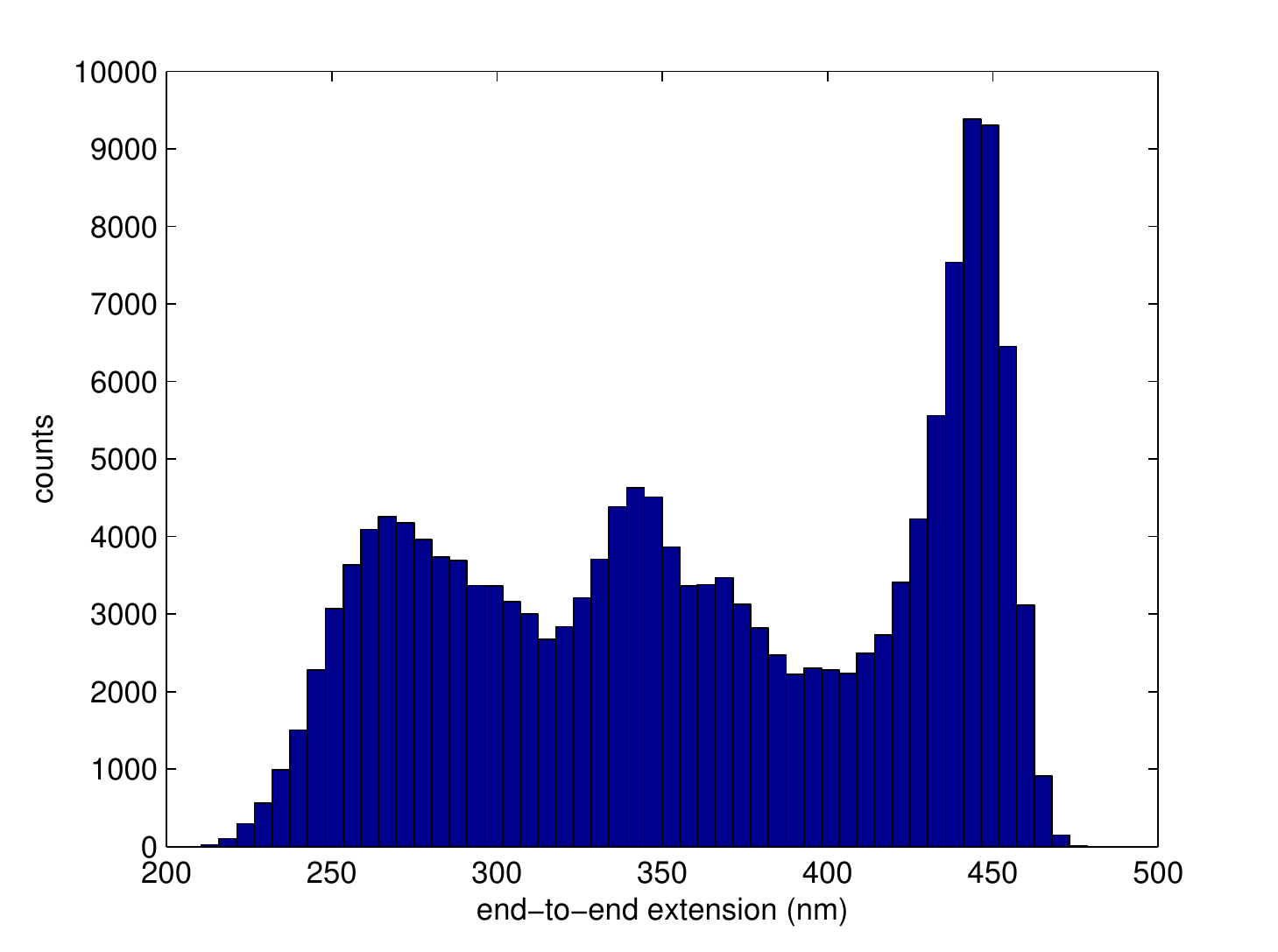}
\end{minipage}
\caption{Extension distributions for $L=1500$\,bp, at superhelical density $\sigma=-0.08$, with $F=1.5$\,pN (left) and $F=2.3$\,pN (right). The results are qualitatively similar to those for the $L=600$\,bp system shown in Fig.~\ref{fig:600bp_end2end_hist}.}
\label{fig:1500bp_end2end_hist}
\end{figure}

To test this expectation, we ran simulations of the 1500-bp system at $\sigma=-0.08$ and $F=1.5$\,pN and $F=2.3$\,pN, near the expected peaks in the fluctuation spectrum. 
In particular the second point is difficult to sample, as it involves a global redistribution of $\Delta Lk$, which is known to be a relatively slow process~\cite{Brutzer2010}.
We ran four independent simulations of a total simulation time of approximately $128$\,$\mu$s at $F=1.5$\,pN and ten independent simulations of a total simulation time of approximately $443$\,$\mu$s at $F=2.3$\,pN.
The end-to-end distributions obtained for these state points at $L=600$\,bp and $L=1500$\,bp are shown in Figs.~\ref{fig:600bp_end2end_hist} and \ref{fig:1500bp_end2end_hist}. At $F=2.3$\,pN, clear bimodal behaviour is observed, where a high-extension population represents the extended bubble state, and a low-extension population the tip-bubble plectoneme state.
We note that a somewhat similar distribution has been reported from experiments in Ref.~\onlinecite{Brutzer2010} for positive supercoiling, where the low-extension state is the unwrithed structure without bubbles and the high-extension case is expected to be plectonemes without tip-bubbles. 
In our case the low-extension population consists of plectonemes with tip bubbles and is much broader because plectonemes with tip bubbles of different sizes have comparable stabilities. 
We observe further substructure in the low-extension distribution for $L=1500$\,bp, but would caution that the sampling of large supercoiled structures converges slowly and so we have not been able to verify that this feature is robust.  At any rate,   
 we expect its influence on the standard deviation calculated for the distribution to be small.

\begin{figure}[h!]
 \includegraphics[width = 0.7\columnwidth]{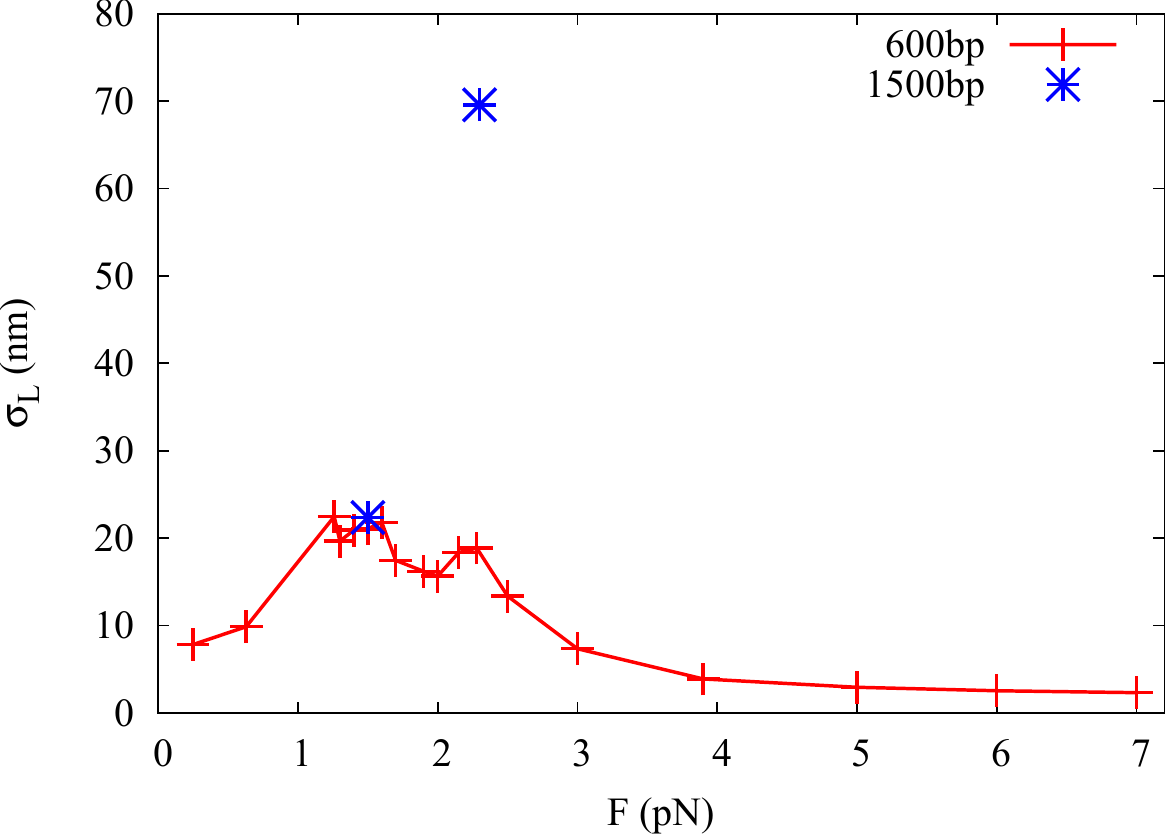}
 \caption{Fluctuation spectrum (standard deviations of the end-to-end distance) for $L=600$\,bp (red points, as in inset of Fig.~\ref{fig:hatcurves_torques}b in the main paper), together with two points obtained for a $L=1500$\,bp system (blue points). While the maximum at low force is approximately independent of strand length, the high-force maximum grows with system size in a roughly linear fashion.}
 \label{fig:fluct_spec_600_1500}
\end{figure}

Fluctuation results for both 600\,bp and 1500\,bp are plotted together in Fig.~\ref{fig:fluct_spec_600_1500}, corroborating the expectation 
that the first peak is largely independent of system size, while the second peak grows much more strongly, with an amplitude increase consistent with our expectation that it would grow roughly linearly with system size. 

In a series of pioneering experiments~\cite{Salerno2012a,Tempestini2013}, the fluctuation spectrum in $L/L_0$ was measured for the first time for systems of effective length  $L \approx 6$\,kbp, around the region for negative supercoiling where there is a  crossover from plectonemes to a bubble state.  
By extrapolating the increase in $\sigma_L$ observed in our simulations,  we would predict fluctuations to have a maximum magnitude of roughly $200$\, nm at the higher force peak  which is consistent with what was observed in experiment.  
We note that these experiments probably could not resolve the first peak, as this is expected to remain very small. 
The overall  width of the  peak in the fluctuation spectrum observed in experiments is somewhat more narrow than it is in our simulations for $L=600$.  
One reason is that the experiments would mainly resolve the higher force peak, for which the width at half maximum is smaller than the full spectrum we observe at this shorter length.
A second reason is that generic finite size effects should lead to a narrowing of the fluctuation peak at a transition for increasing system size. 
Nevertheless, as also discussed in the main text, the experiments discussed above observed considerably wider peaks in the fluctuation spectrum than they predicted based on a simple model that only includes plectonemes and bubbles. 
We argue that the bubbles and tip-bubble plectonemes each lower the nucleation barrier for the formation of the other, leading to a broader fluctuation spectrum than one would observe for two states separated by a large nucleation barrier.

We also note that for a similar salt concentration, the experiments of Ref.~\onlinecite{Tempestini2013} find the fluctuation peak at $F \approx 1$\,pN, a position about one pN lower than our large force peak. 
There may be a number of reasons for this.  Firstly, there are still unexplored generic finite size effects that may lower the transition force for longer strands.  
Secondly,  simulations for both strand lengths reported here have been performed for the average base parameterisation of oxDNA.  
As  discussed in Supplementary Sec.~\ref{sec:sequences}, when using a sequence-dependent model, the crossover between bubble and plectoneme states takes place at lower forces because bubbles preferentially form at AT rich regions. 
Moreover, for a random sequence, the longer the strand, the higher the probability for finding larger AT rich regions, which may also further enhance the probability of bubble formation. 
Taken together, these effects are expected to lower the forces at which the fluctuation peak occurs, bringing better agreement with the force at which a fluctuation maximum is observed in the experiments of Ref.~\onlinecite{Tempestini2013}. 
Finally, we note that it may also be the case that oxDNA overestimates the cost of forming twist-induced bubbles.  
It is hard to find direct comparisons to experiment for this phenomenon that would allow for an independent check.  
As previously mentioned, we  we do find good agreement with experiment for force induced melting~\cite{Romano2013} or for simple duplex melting~\cite{Ouldridge2011}.  
Nevertheless,  even if it is the case that oxDNA overestimates the cost of forming bubbles this is unlikely to change our qualitative conclusions.

\begin{figure}[h!]
 \includegraphics[width = 1.\columnwidth]{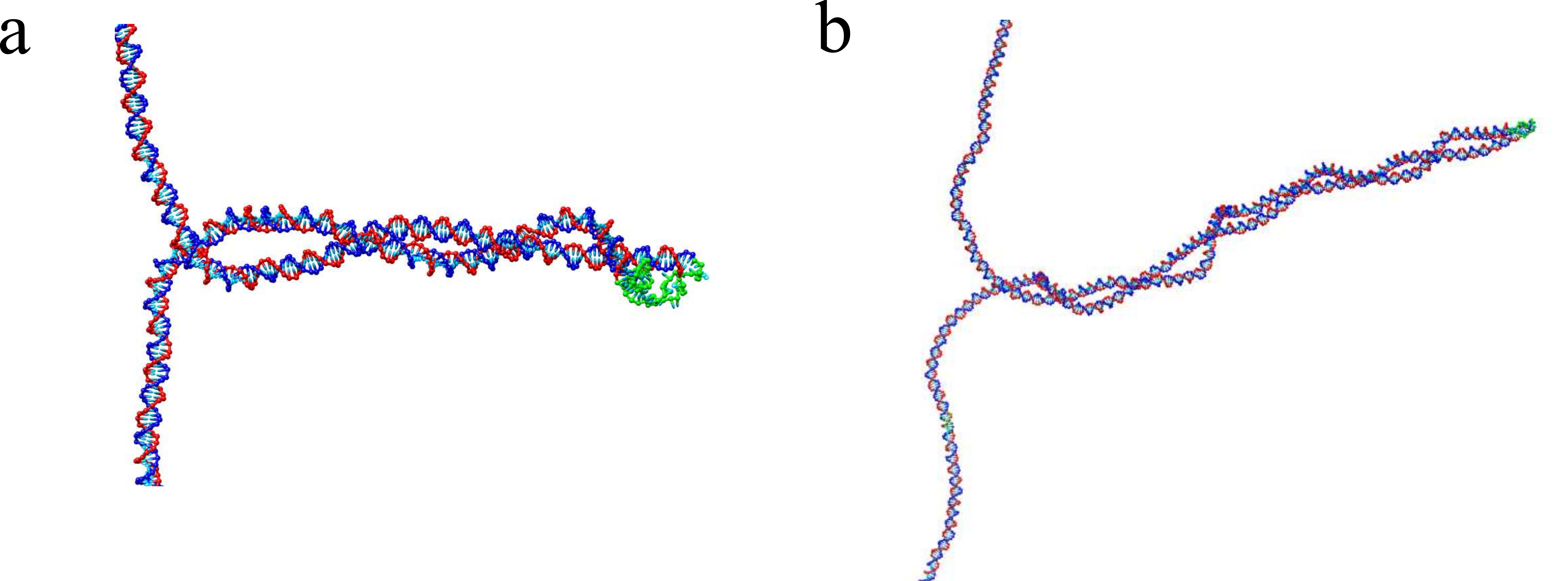}
 \caption{(a) Tip-bubble plectoneme structure at $L=1500$\,bp, for $\sigma=-0.08$ and $F=2.3$\,pN, with plectoneme size 211\,bp and bubble size 33\,bp. (b) Tip-bubble plectoneme structure at $L=1500$\,bp, for $\sigma=-0.08$ and $F=2.3$\,pN, with plectoneme size 576\,bp and bubble size 10\,bp. Note that both structures are obtained at the same values of $\sigma$ and $F$.}
 \label{fig:1500bp_plect_figs}
\end{figure}

\begin{figure}[h!]
 \includegraphics[width = 1.\columnwidth]{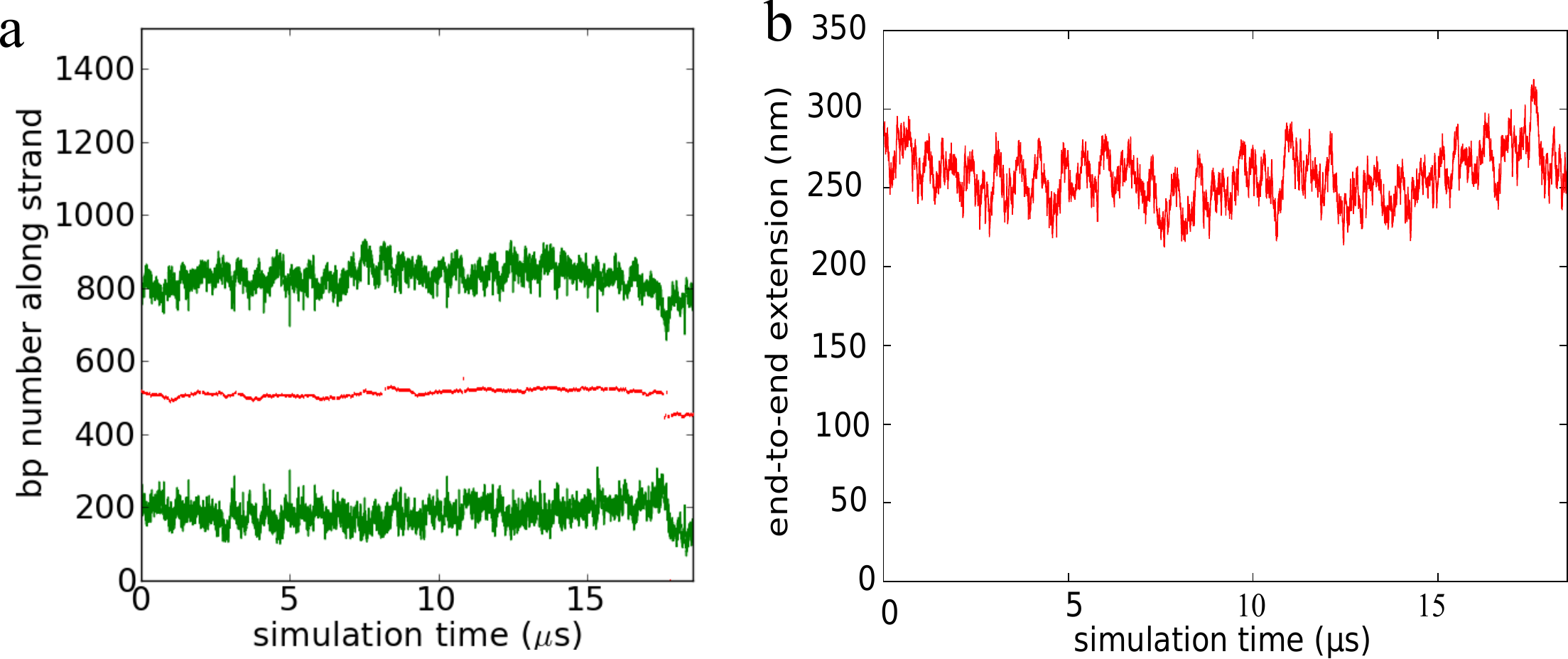}
 \caption{(a) Position kymograph of a tip-bubble plectoneme at $L=1500$\,bp, for $\sigma=-0.08$ and $F=2.3$\,pN, showing stability of the structure on a $\mu$s time scale. (b) End-to-end extension kymograph for the run shown in (a).}
 \label{fig:1500_plectokymo}
\end{figure}

Finally, in Fig.~\ref{fig:1500bp_plect_figs}, typical tip-bubble plectoneme configurations at $L=1500$\,bp observed for $F=2.3$\,pN and $\sigma=-0.08$ are shown.
We note that these structures are remarkably long-lived once they form. A kymograph for a tip-bubble plectoneme in the $L=1500$\,bp system at $F=2.3$\,pN and $\sigma=-0.08$ is shown in Fig.~\ref{fig:1500_plectokymo}, demonstrating that these structures can be stable on a $\mu$s timescale.

\clearpage
\section{Determination of plectoneme position, size and tip bubbles} \label{sec:plecto_pos_size_pinn}
\begin{figure}
 \includegraphics[width = 0.5\columnwidth]{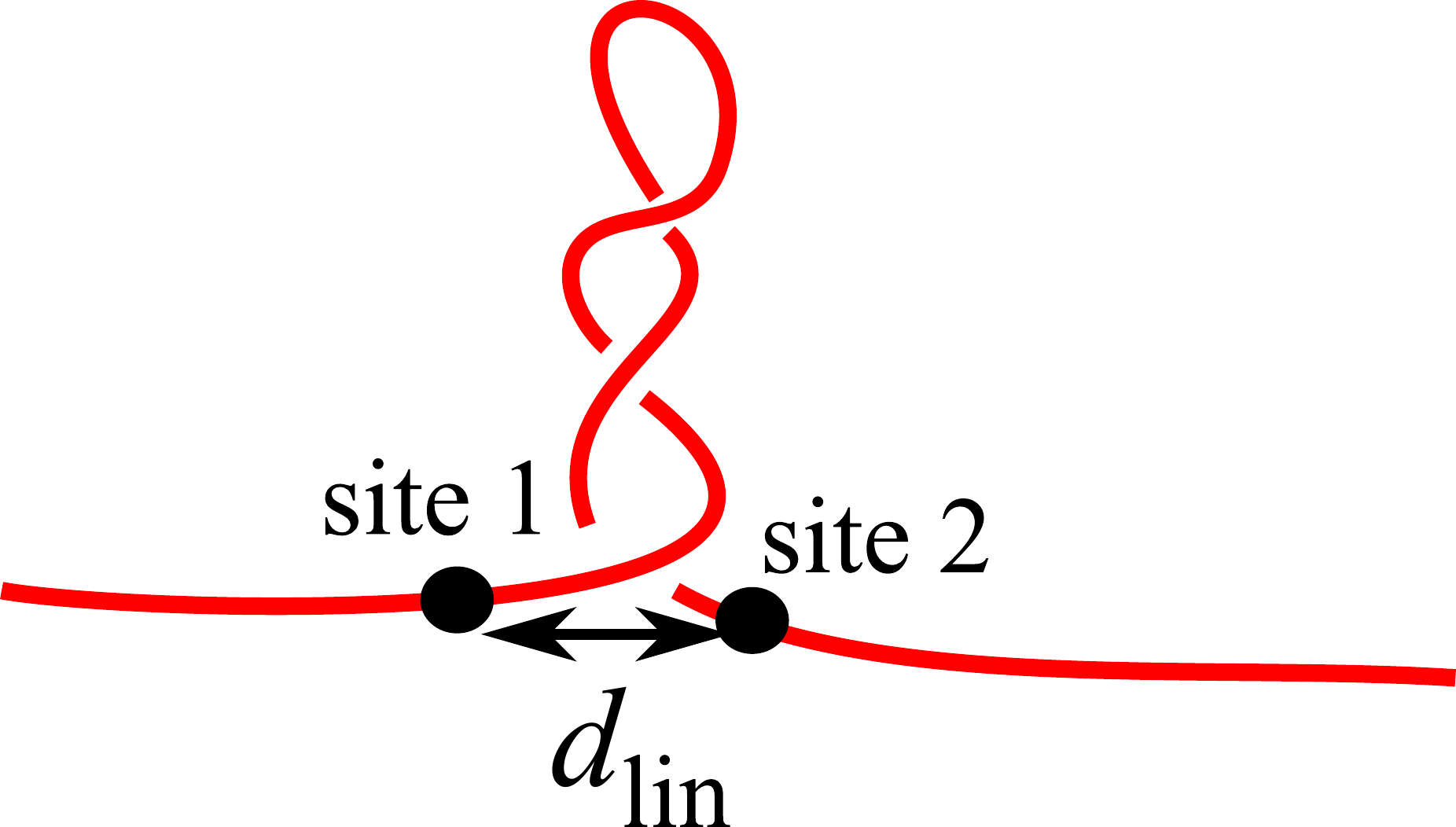}
 \caption{In plectonemic regions, the direct spatial distance $d_{\rm lin}$ between two sites is shorter than the distance along the double-strand contour.}
 \label{fig:plecto_detection}
\end{figure}

\noindent \underline{Plectoneme position and size}\\\\
Plectoneme position and size are determined using an algorithm which relies on the fact that in plectonemes, the spatial distance $d_{\rm lin}$ between two sites on the molecule is smaller than their separation along the strand (Fig. \ref{fig:plecto_detection}).
The algorithm only considers the midpoints between the centers of mass of corresponding bases on the single strands. It proceeds as follows:
\begin{itemize}
 \item Start from strand end, loop over all midpoints
 \begin{itemize}
  \item If any part of the remaining strand with a distance of more than $N_c$ bp along the contour has a distance $d_{\rm lin}<d_{\rm lin}^{0}$, record the current bp index as the beginning of a plectonemic region, if the beginning of a plectoneme has not yet been detected.
  \item If $d_{\rm lin}>d_{\rm lin}^{0}$ and a plectoneme beginning has been detected before, record the current bp index as the end of a plectonemic region and continue searching with the next bp
 \end{itemize}
 \item The plectoneme position is the mean between the bp indices of the beginning and end of a plectonemic region
 \item The plectoneme size is the difference between the bp indices of the beginning and end of a plectonemic region
\end{itemize}
We performed plectoneme detection using $d_{\rm lin}^0=7.24$\,nm and $N_c=40$\,bp.
The results are not very sensitive to the precise choice of these parameters, as long as $d_{\rm lin}^0 < N_c r_{\rm bp}$, where $r_{\rm bp}\approx 0.34$\,nm is the approximate rise of one base pair.
$N_c$ hence imposes a cutoff on the minimum size of plectonemes that can be detected with the search algorithm. The parameter choice made here ensures that writhed bubbles are reliably not counted as a plectonemic state, thus avoiding false positive detections.
The algorithm is able to detect multiple plectonemes along the DNA double strand. However, for the salt conditions and strand length used in this work, only one plectoneme occurs in the simulated system. Two simultaneous plectonemic regions were only detected transiently during initial formation of the plectoneme.
In order to obtain a single-valued plectoneme coordinate in these rare cases, we only consider the largest plectoneme structure.
\newpage
\noindent \underline{The plectoneme tip-bubble state}\\\\
For all sequences used in this work, we observed co-localization of plectonemes and bubbles. 
In these configurations the plectoneme position and the position of the midpoint of the largest denaturation bubble in the system coincide within a margin $M_c$.
\\A plectoneme with a tip bubble is defined as follows:
\begin{itemize}
 \item A denaturation bubble with size $l_b \geq 2$\,bp exists in the system.
 \item A plectoneme is detected in the system using the algorithm described in Supplementary Sec. \ref{sec:plecto_pos_size_pinn}.
 \item The bubble midpoint and the plectoneme position are separated by less than $M_{\rm c}$ bp
\end{itemize}
The restriction to bubbles with size $l_b \geq 2$\,bp was introduced in order to discard short-lived single base-pair denaturations, which can occur in the strand due to thermal noise.

\begin{figure}[bh!]
 \includegraphics[width = 0.7\columnwidth]{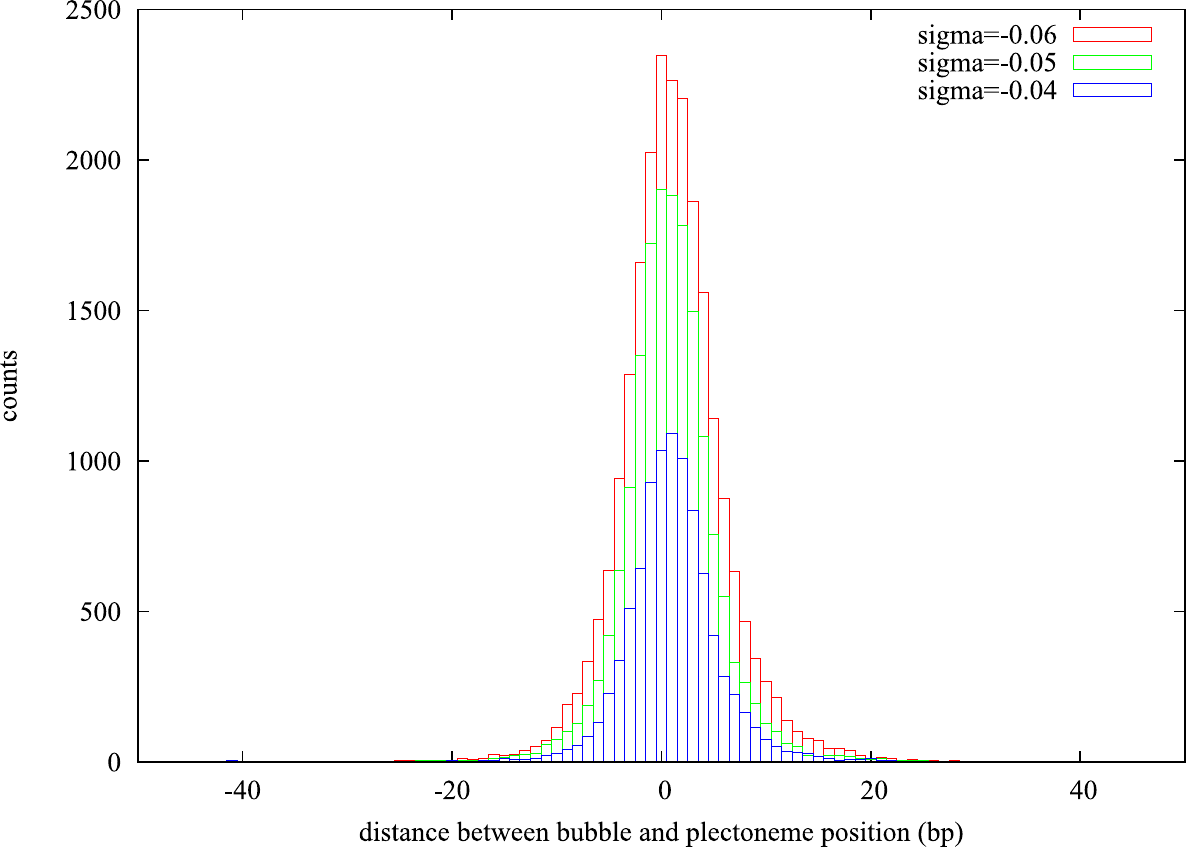}
 \caption{Distribution for the distance between bubbles and plectoneme midpoints for simulations of a 600-bp system with the random sequence defined in Supplementary Sec. \ref{sec:sequences}, at different values of $\sigma$ for $F=1.27$\,pN. Strong co-localization is observed.}
 \label{fig:plecto_bubb_colocalization}
\end{figure}

Fig. \ref{fig:plecto_bubb_colocalization} shows distributions exhibiting strong co-localization of bubble and plectoneme. 
For a large majority of configurations, the bubble-plectoneme distance  is less than 20\,bp. We therefore chose $M_c=20$\,bp for the detection of tip-bubble plectonemes.
\newpage
\section{Interconversion between bubble and plectoneme size}\label{sec:bubble_plectoneme_conversion}
\begin{figure}
 \includegraphics[width = .8\linewidth]{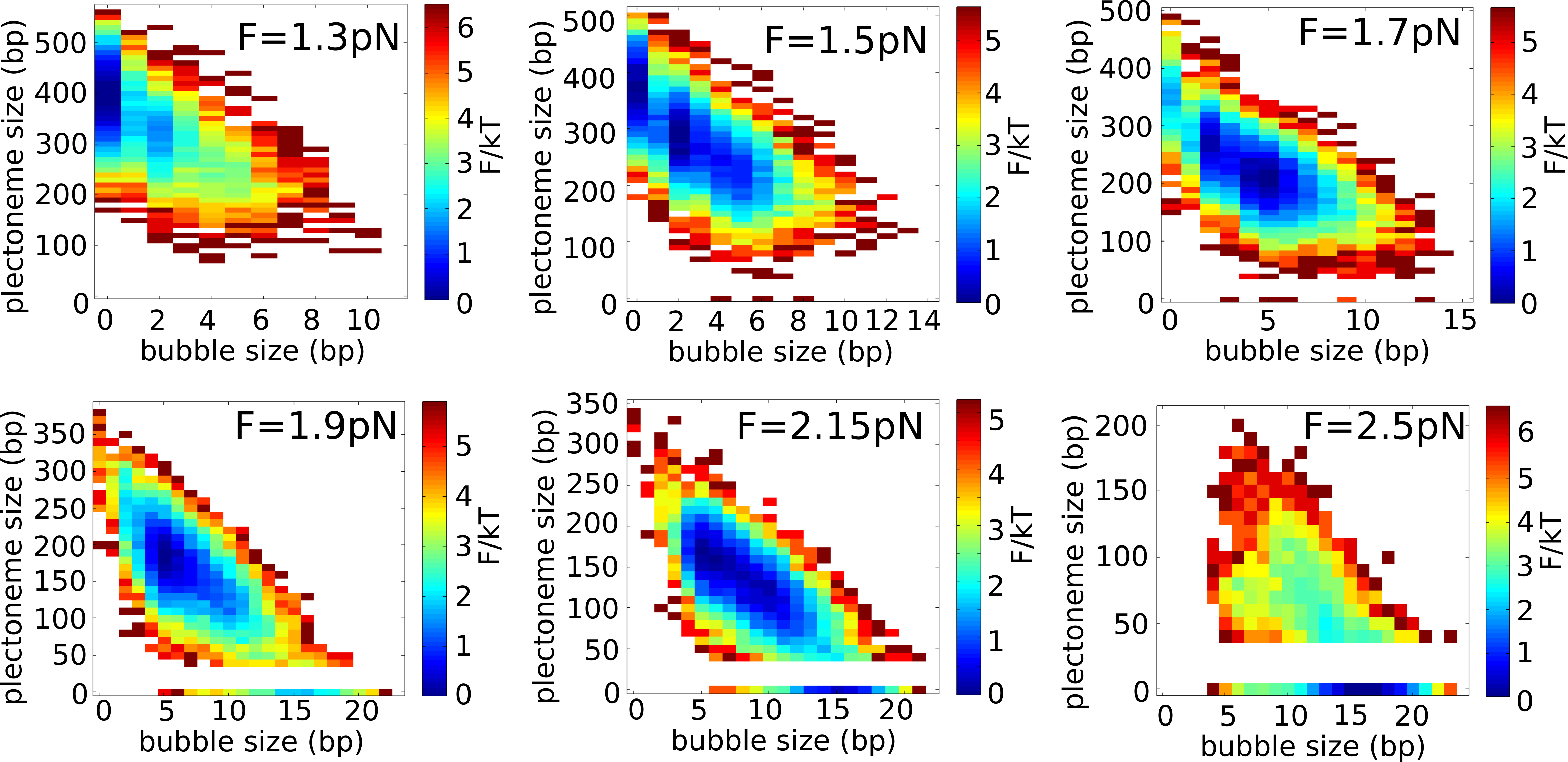}
 \caption{Free-energy landscapes as a function of bubble and plectoneme size at $\sigma=-0.08$ and different forces. A negative linear correlation of bubble and plectoneme size is observed after forming a kink-like tip-bubble defect of 1 to 2\,bp. The landscapes are obtained using the average-base parametrization of oxDNA.}
 \label{fig:plecto_bubb_size}
\end{figure}

Bubbles in the end loop tend to shorten plectonemes due to three effects. 
Firstly, a denaturation in the end loop constitutes a defect which allows the formation of a smaller, more tightly wound tip of the plectoneme structure, as shown in Fig.~\ref{fig:hatcurves_torques}d of the main text.
This causes an extension $\Delta L$ of the overall system.
Secondly, denaturation bubbles naturally possess very small twist. Thirdly, denatured single strands can twist back on themselves in a negative way, thus even leading to a negative twist contribution. 
As the overall linking difference $\Delta Lk$ of the system is conserved according to the relation $\Delta Lk= \Delta Tw+ \Delta Wr$ \cite{Calufareanu1959,White1969,Fuller1971o} (see Supplementary Sec.~\ref{sec:boundaries}), this increase in $\Delta Tw$ changes the system's writhe component and thus the average plectoneme size.
Fig.~\ref{fig:plecto_bubb_size} shows free-energy landscapes of the system as a function of bubble and plectoneme size for different values of $F$ (see also Figs.~\ref{fig:hatcurves_torques}e and \ref{fig:hatcurves_torques}f of the main text). 
In a 600-bp strand, due to thermal noise, very short-lived denaturations of small size can occur in a duplex at $T=300$\,K.
In order to separate out this contribution, bubbles with a size up of 1 or 2\,bp were only taken into account if they were co-localized with a plectoneme.

Fig.~\ref{fig:plecto_bubb_size} shows that after denaturation of the first two base pairs and consequent shape change of the end-loop, the size of plectonemes and bubbles can be interconverted in an approximately linear fashion, as seen also in Figs.~\ref{fig:hatcurves_torques}e and~\ref{fig:hatcurves_torques}f of the main text.
For this system, the initial formation of a 2-bp tip bubble causes a decrease in plectoneme size by roughly 100\,bp.
Assuming that the equivalent plectoneme size is converted into extended strand length aligned with $F$ and neglecting changes in bending energy, the free energy gain due to this extension is
$\Delta G = F \Delta L \approx 1.27$\,pN $\cdot 100 \cdot 0.34$\,nm$ \approx 10 k_B T $ for $T=300$\,K. 
This free energy gain is on the same order of magnitude as the cost of forming a small double strand denaturation in oxDNA.

\begin{figure}
 \centering
  \includegraphics[width=.8\textwidth]{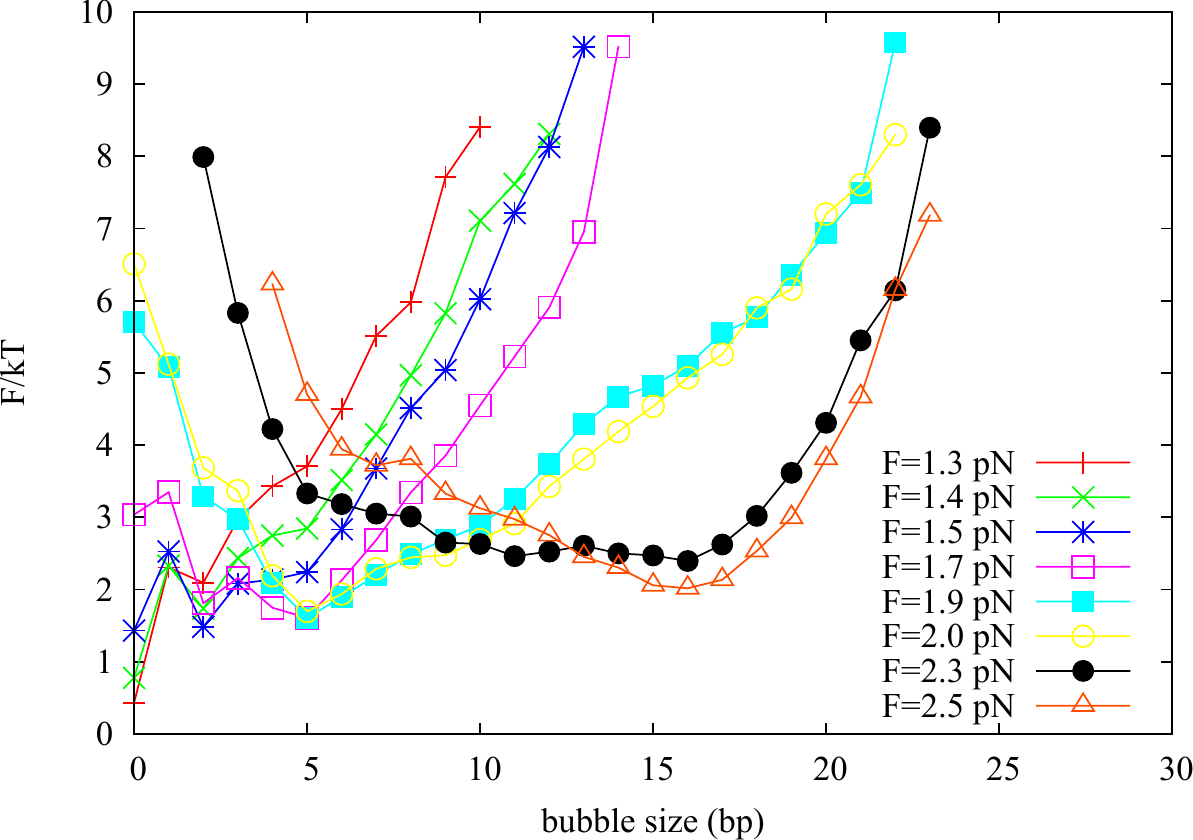}
  \caption{Free-energy profiles at $\sigma=-0.08$ and different values of $F$ as a function of bubble size, corresponding to projections of the 2-dimensional free energy landscapes of Fig.~\ref{fig:plecto_bubb_size} onto the x-axis. Note the stability of 2-bp bubbles at low force, due to free energy gains from initial end-loop rearrangement. For forces $F \gtrsim 2.3$\,pN, extended bubble states are more stable than plectonemic states, see also Fig.~\ref{fig:tipbubble-bubble-energetics}.}
  \label{fig:bubble-free-energy-lanscapes}
\end{figure}

Fig.~\ref{fig:bubble-free-energy-lanscapes} shows free energy profiles as a function of bubble size, corresponding to projections of the 2-dimensional free energy landscapes of Fig.~\ref{fig:plecto_bubb_size} onto the x-axis.
At $F \approx 1.5$\,pN, a 2-\,bp tip bubble denaturation starts to become favourable. 
This small initial denaturation is particularly stable, as it enables kinking and therefore tighter winding of the end-loop, as shown in Fig.~\ref{fig:hatcurves_torques}d of the main text.
As the force increases, it becomes more favourable to grow larger bubbles, as can be seen in the free-energy plots in Figs.~\ref{fig:plecto_bubb_size} and ~\ref{fig:bubble-free-energy-lanscapes}.
For large enough force, the plectoneme disappears, and the dominant states are extended bubble states. 

\begin{figure}
 \centering
  \includegraphics[width=.8\textwidth]{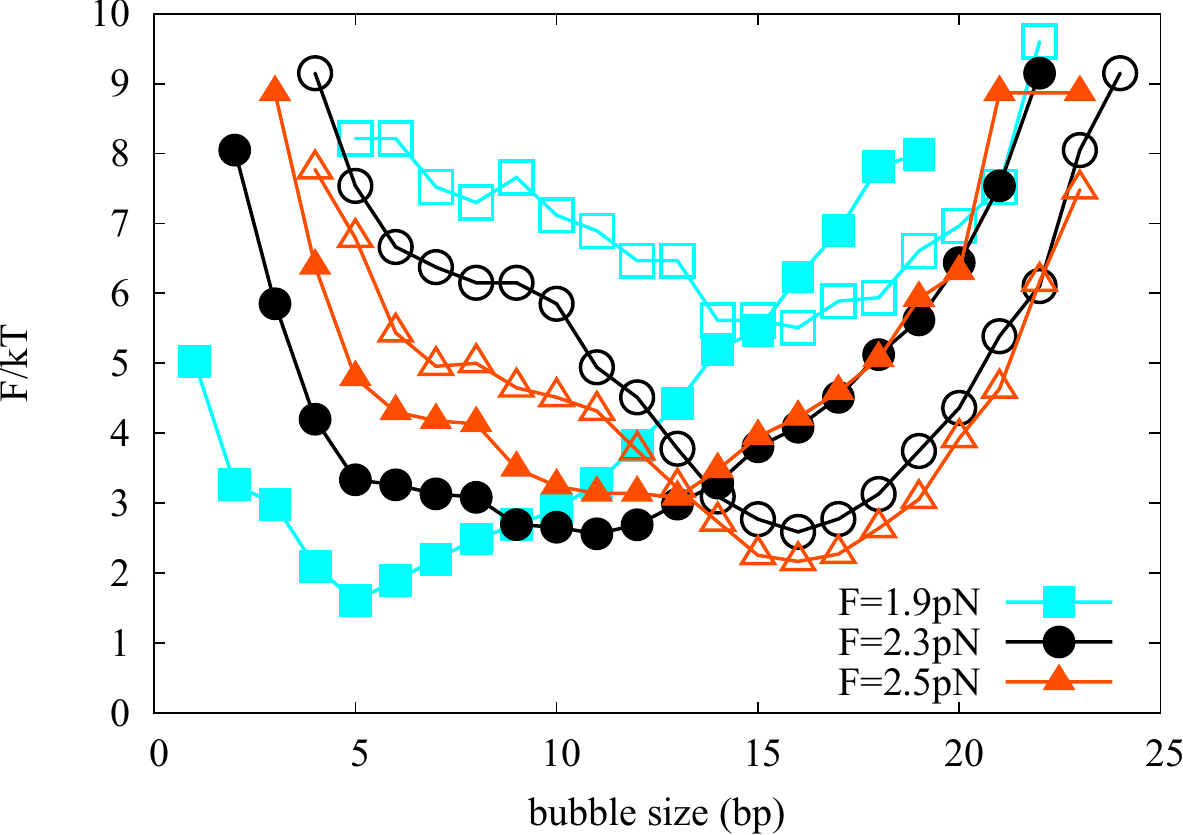}
  \caption{Free-energy plots that illustrate the thermodynamics of the crossover between tip bubbles and extended bubbles. The left curves (solid symbols) show the free energy of tip-bubble plectonemes at a particular size, while the right curves show the free energy of extended bubbles (open symbols) without a plectoneme. At small values of $F$, the tip bubble is more stable than the extended bubble. At $F \approx 2.3$\,pN, the extended bubble population has a free energy roughly equal to the tip-bubble plectoneme, indicating that above this force, extended bubbles become more favourable, as can be seen for the $F=2.5$\,pN curves.}
  \label{fig:tipbubble-bubble-energetics}
\end{figure}

The thermodynamics of the crossover between tip-bubble plectonemes and extended bubbles is illustrated in Fig.~\ref{fig:tipbubble-bubble-energetics}, where the projection of Fig.~\ref{fig:bubble-free-energy-lanscapes} is separated into contributions due to plectonemes, and contributions due to bubble states without plectonemes. 
As discussed in Supplementary Sec.~\ref{sec:plecto_pos_size_pinn}, the classification of tip-bubble plectonemes of small size somewhat depends on the cutoff of our plectoneme detection algorithm.
However, as the transition from tip-bubble plectonemes to bubbles is fairly narrow (cf. Fig.~\ref{fig:hatcurves_torques}b of the main text and Supplementary Sec.~\ref{sec:pop_frequencies}), this is not expected to significantly affect the force at which the crossover occurs.
Fig.~\ref{fig:tipbubble-bubble-energetics} clearly shows how, with increasing force, the system transitions from a regime with mainly
tip bubbles, to a regime with mainly extended bubble states. 

\begin{figure}
 \centering
  \includegraphics[width=.8\textwidth]{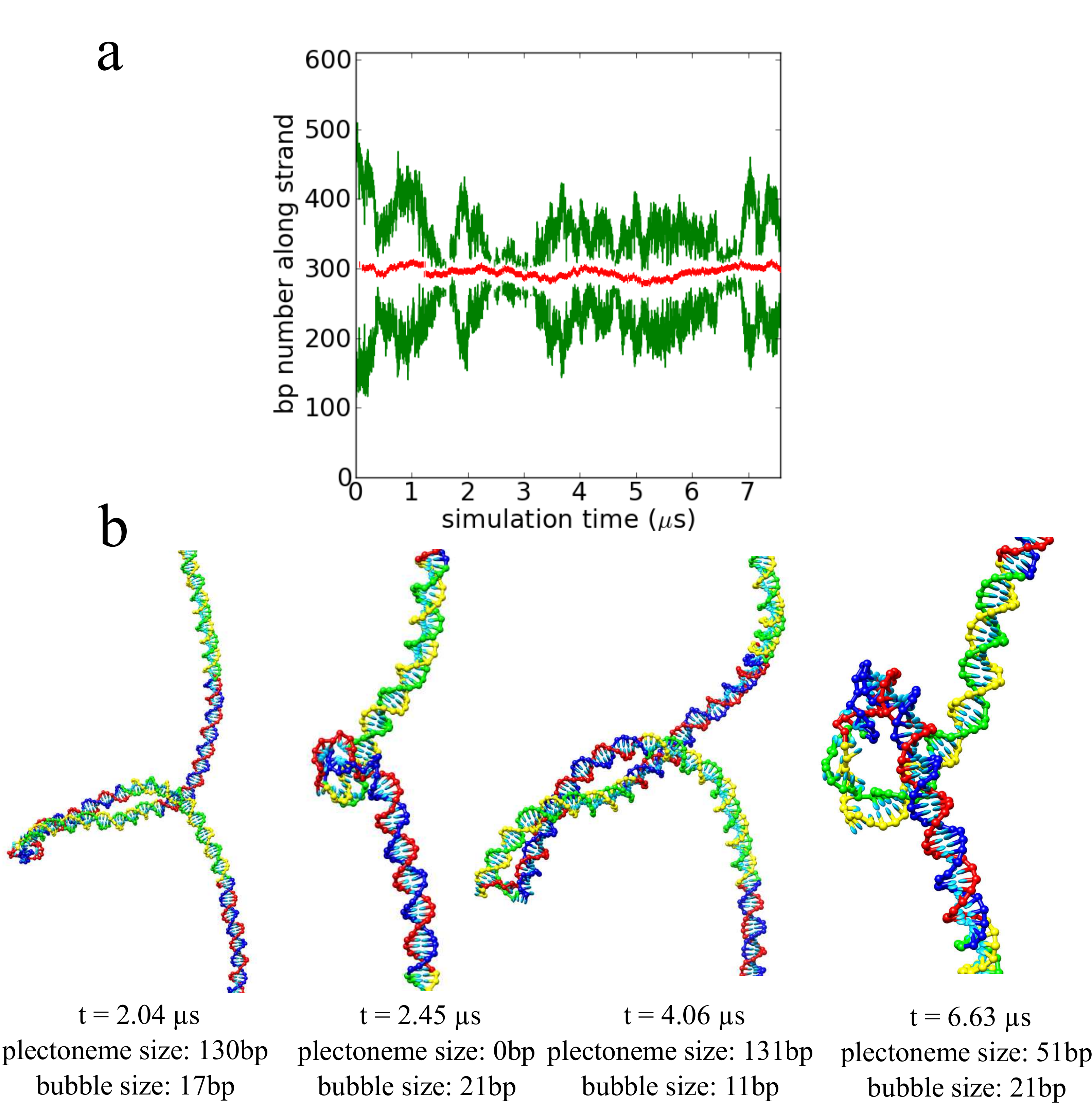}
  \caption{Interconversion between bubbles and plectonemes at $\sigma=-0.09$ and $F=2.3$\,pN. Different colouring of parts of the double strand is to facilitate comparison between structures. (a) Position kymograph of plectoneme boundaries (green) and bubble centre (red). (b) Example configurations taken from trajectory shown in (a) at the times indicated.}
  \label{fig:plecto-structures}
\end{figure}

\begin{figure}
 \centering
  \includegraphics[width=.6\textwidth]{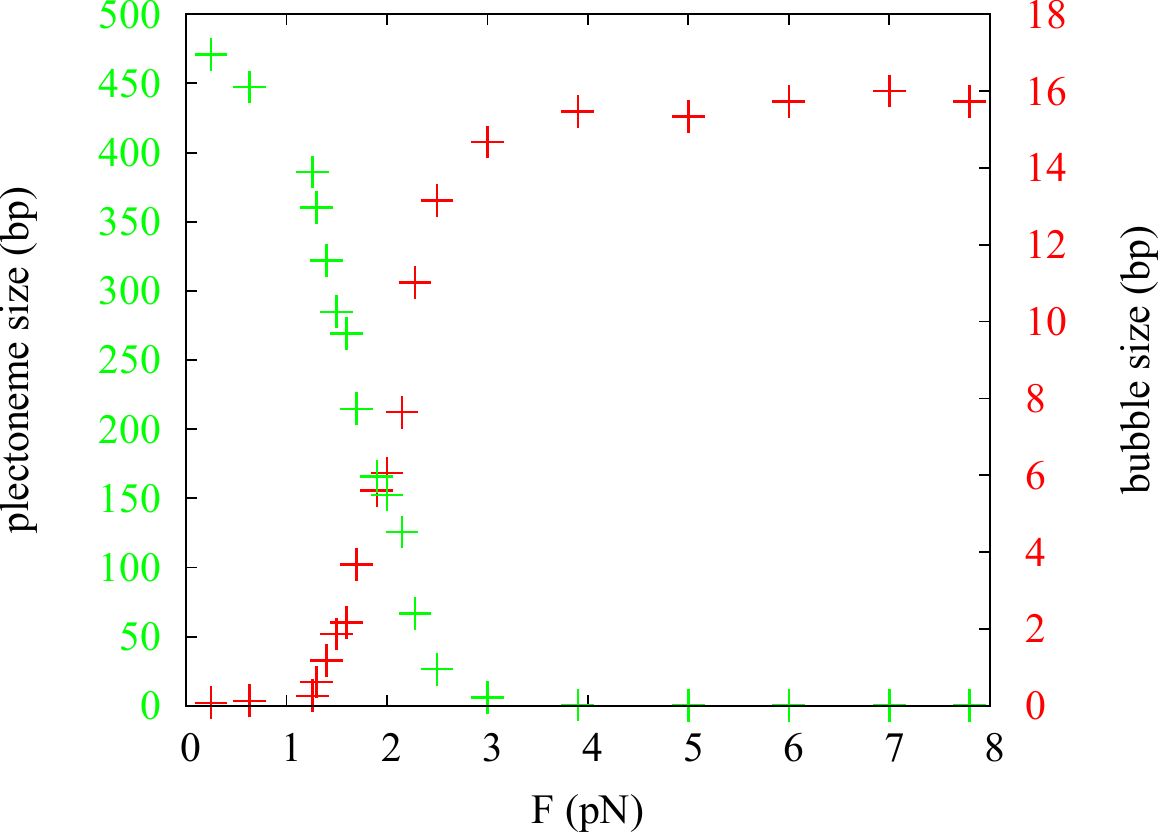}
  \caption{Mean sizes of plectoneme (green) and denaturation bubble (red) as a function of applied stretching force $F$ at $\sigma=-0.08$.}
  \label{fig:bubble_plecto_size_s-0p08}
\end{figure}

At parameter values where extended bubbles and tip-bubble plectonemes can coexist, local kinks induced by bubbles provide a preferred site for re-nucleation of a plectoneme. 
Structures and a corresponding kymograph are shown in Fig.~\ref{fig:plecto-structures} to illustrate the crossover dynamics between bubbles and plectonemes for $\sigma=-0.09$ and $F=2.3$pN.
The average size of both bubbles and plectoneme structures as a function of $F$ at $\sigma=-0.08$ is shown in Fig. \ref{fig:bubble_plecto_size_s-0p08}.



\clearpage
\newpage
\section{Displacement behaviour of plectonemes}\label{sec:plectoneme_displacement}
\begin{figure}[h!]
 \includegraphics[width = .7\columnwidth]{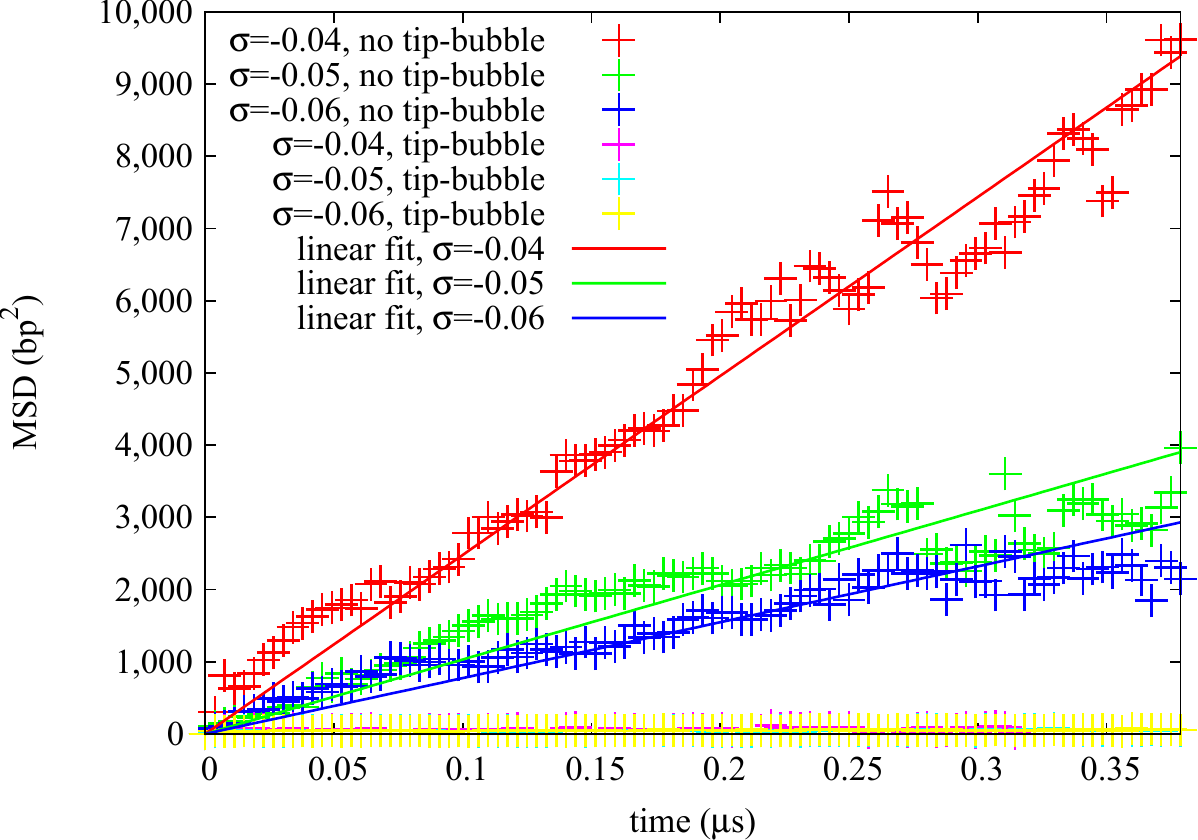}
 \caption{Mean square displacement of pinned and unpinned plectonemes at different superhelical densities and $F=1.27$\,pN. While tip-bubble plectonemes are pinned on a $\mu$s timescale, plectonemes without tip bubble exhibit significant diffusion.}
 \label{fig:plecto_diffusion}
\end{figure}

\noindent
\underline{Diffusion}\\\\
Fig.~\ref{fig:plecto_diffusion} shows the mean square displacement of plectoneme structures with and without tip bubbles in a 600-bp system, using the random sequence described in Supplementary Sec.~\ref{sec:sequences} for $\sigma=-0.04$, $\sigma=-0.05$ and $\sigma=-0.06$, and a stretching force $F=1.27$\,pN.
The data was extracted from sections of trajectories of a total simulated time of 140\,$\mu$s for each value of $\sigma$. On a $\mu$s time scale, the mean-square displacements (MSD) $\langle (d(t))^2 \rangle$ of the positions of plectonemes without tip bubbles show an approximately linear behaviour in time, indicating diffusive motion of these structures.
In contrast, tip-bubble plectonemes do not show any signature of MSD on that time scale. The diffusion coefficient $D$ and the MSD are related by $\langle \left( d(t) \right)^2 \rangle = 2Dt$, allowing determination of $D$ from linear fits to the data shown in Fig. \ref{fig:plecto_diffusion}.
The determined diffusion constant for unpinned structures decreases as $\sigma$ becomes more negative because of the larger plectonemes that form at larger $|\sigma|$.
The numerical values of $D$ determined from the fitting procedure are given in Table~\ref{tab:Diff_c}.

The diffusion constants obtained in this section are much higher than those measured in Ref.~\onlinecite{Loenhout2012}.
There are several reasons for this difference.
Firstly, at a strand length of 600\,bp, even the highest superhelical density used to study diffusive motion, $|\sigma| = 0.06$, corresponds to a linking difference of $\Delta Lk\approx3.5$. 
Thus, plectonemes of at most 3-4 double strand self-crossings are expected. In practice, plectoneme structures are slightly shorter, due to positive values of $\langle Tw \rangle$. 
In contrast to this, linking differences $\Delta Lk > 40$ were imposed in Ref.~\onlinecite{Loenhout2012}, leading to much larger plectoneme structures which are expected to diffuse more slowly than the simulated structures.
Secondly, a high effective monomer diffusion coefficient $D_{\rm sim}=6 \times 10^{-7}$\,m$^{2}$s$^{-1}$ was chosen in simulations for this work. 
This is a common choice in coarse-grained models, and increases the efficiency of sampling slow processes. From measurements of diffusive motion of DNA single strands, we estimate these effects to speed up the simulations by up to two orders of magnitude in simulated time~\cite{Ouldridge2013a}.
Finally, underlying free-energy landscapes tend to be smoothed out in coarse-grained models, leading to accelerated motion of the simulated diffusion over barriers compared to the experimental system (see e.g. Ref.~\onlinecite{Hills2010}).
The time scale used in our integrator is set by the mass, length and energy scales used, which determine the frequencies of intra-molecular vibration modes.
Coarse-graining may affect differently processes such as the overall bending mode of the double strand, which are important for strand reptation, making it difficult to define a homogeneous time scale for all these processes.
Therefore, it is generally safer to compare the relative time scale of two processes rather than their absolute duration.

In Ref.~\onlinecite{Loenhout2012}, experimental images were obtained with a 20\,ms time resolution.
This time-scale may be much larger than the lifetime of tip bubbles in a regime in which there is a finite population for both plectonemes with and without tip bubbles.
Thus, diffusion constants measured in experiment determine an average over pinned and unpinned states, and provide an effective diffusion coefficient $D_{\rm eff}$, as described in the main text.

\begin{figure}
 \includegraphics[width = 1.\columnwidth]{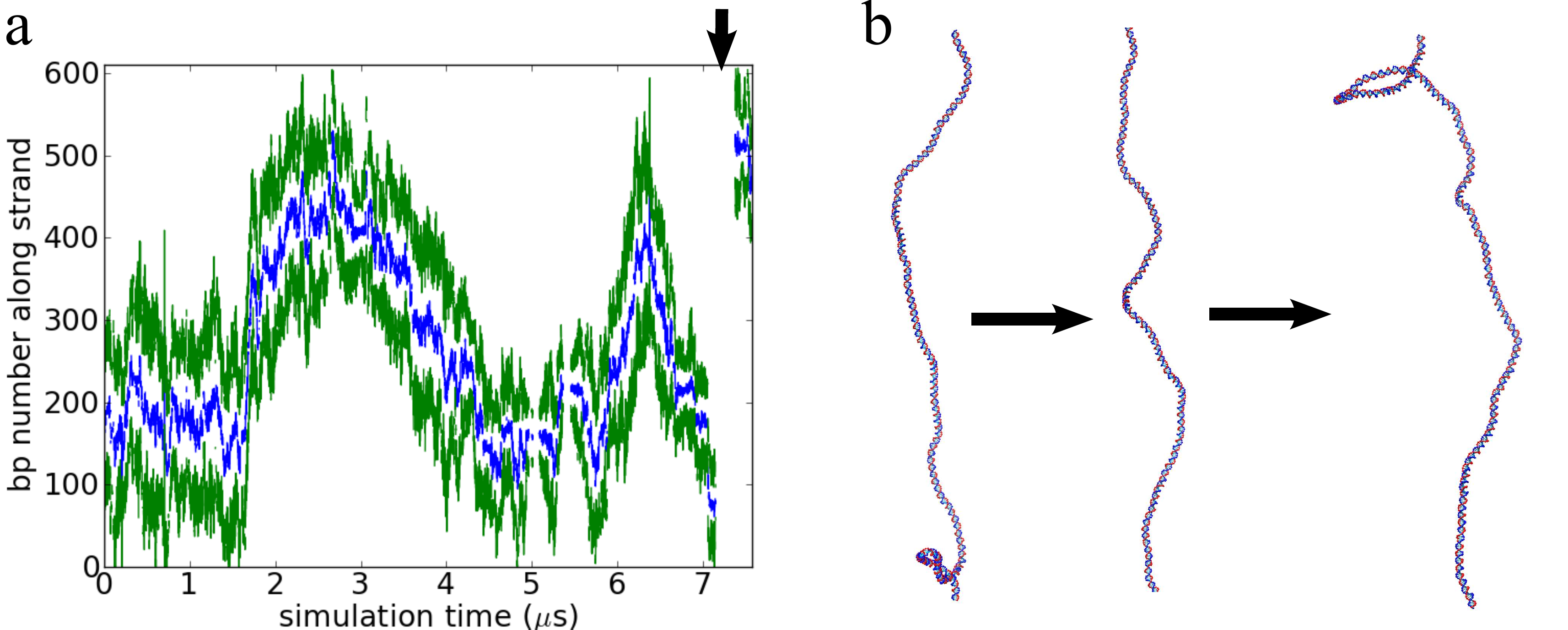}
 \caption{Hopping event observed in a simulation at $\sigma = -0.04$ and $F=1.27$\,pN: (a) Kymograph of plectoneme boundaries (green lines) and centre position (blue line) showing a hopping event at the time marked by a black arrow. (b) Structure of the DNA strand close to the hopping event at $t=7.09$\,$\mu$s (left), $t=7.34$\,$\mu$s (middle) and $t=7.39$\,$\mu$s (right).}
 \label{fig:plecto_hopping}
\end{figure}

\begin{table}[h!]
\begin{tabular}{c|cc}
\hline
$\sigma$ & $D$ (kbp$^2$s$^{-1}$) & mean plectoneme size (bp)\\
\hline
$-0.04$ &  $1.2 \times 10^4$ & 161\\ 
$-0.05$ & $5.1 \times 10^3$ & 218\\ 
$-0.06$ & $3.9 \times 10^3$ & 272\\ 
\hline
\end{tabular}
\caption{Unpinned Diffusion constants obtained for different superhelical densities at $F=1.27$\,pN from the linear fits shown in Fig. \ref{fig:plecto_diffusion}.}
\label{tab:Diff_c}
\end{table}
\newpage \noindent
\underline{Hopping}\\\\
We observed long-range displacement of plectonemes by ``hopping''. 
In 21 independent simulations of a 600-bp system at $F=1.27$\,pN and $\sigma = -0.04$ run for a total simulation time of 148\,$\mu$s, we observed 4 hopping events which involved rapid displacement of the plectoneme center position over a distance of more than 100\,bp.
In all cases, the plectoneme in the initial location fully unformed, and then quickly re-formed at a distant site (see Fig. \ref{fig:plecto_hopping} for a detailed plectoneme position kymograph and strand structures).
Coexistence of two plectonemes was not observed in equilibrium, as expected at the high-salt conditions used in oxDNA \cite{Emanuel2012,Loenhout2012}.
Due to the absence of equilibrium multi-plectoneme states in the 600-bp system, we observe hopping only close to the critical buckling superhelical density $\sigma_b$, where the system has a non-zero probability to dissolve a plectoneme and return to the extended state.

\begin{figure}
\centering
\begin{minipage}{.5\textwidth}
  \centering
  \includegraphics[width=.4\columnwidth]{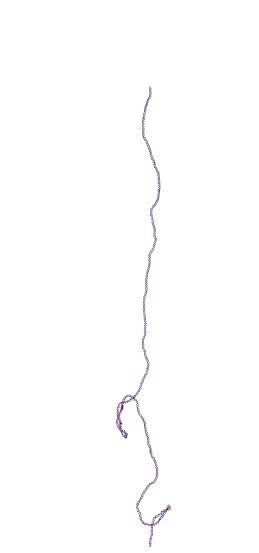}
\end{minipage}%
\begin{minipage}{.5\textwidth}
  \centering
  \includegraphics[width=1.\linewidth]{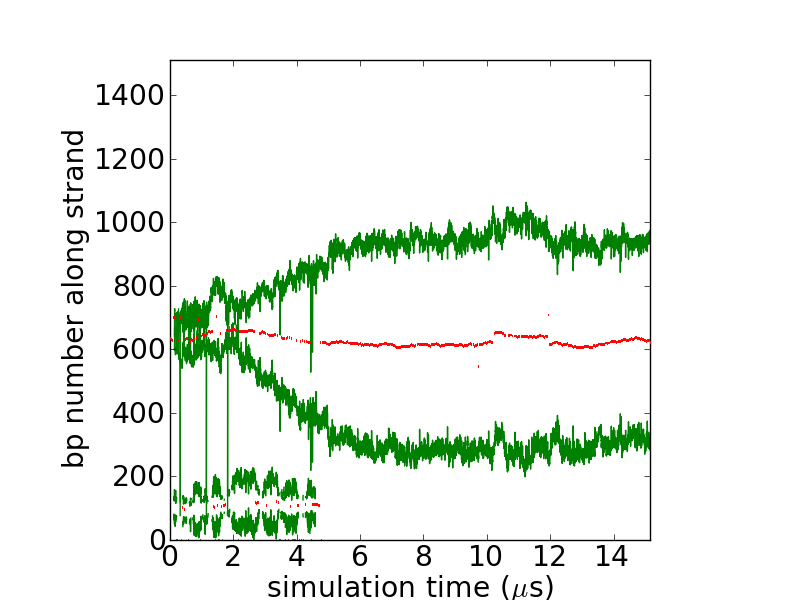}
\end{minipage}
 \caption{Coexisting plectonemes in an out-of-equilibrium run started from a homogeneously underwound linear double strand of length 1500\,bp at $\sigma=-0.08$ and $F=2.0$\,pN, as shown in the Supplementary Movie. left: Molecule configuration showing two coexisting plectonemes. right: Kymograph of two plectoneme sites. Ultimately, the smaller plectoneme dissolves and a single, large plectoneme is formed.}
 \label{fig:movie_conf}
\end{figure}

However, for DNA strands of length 1500\,bp, we did observe long-range writhe exchange between two simultaneously present plectonemes in an out-of-equilibrium situation.
Simulations were started from a linear, homogeneously undertwisted double strand at $\sigma=-0.08$ and $F=2.3$\,pN.
A typical example of such a run is shown in Fig.~\ref{fig:movie_conf}, where the kymograph initially shows the simultaneous presence of two plectonemes. 
Coexistence is followed by the disappearance of the smaller plectoneme, mediated by long-range transport of its writhe to the larger plectoneme,which then stays stable for a long time. 

Van Loenhout \textit{et al.}~\cite{Loenhout2012} found a similar long-range displacement behaviour over distances of up to 15\,kbp at $F=0.8$\,pN and $\sigma \approx +0.04$.
Some hopping events observed in the experiments of Ref.~\onlinecite{Loenhout2012} showed immediate dissolution of the initial plectoneme, and re-nucleation at a distant site, which is reminiscent of the equilibrium mechanism described above. 
However, the experiments are performed at ambient conditions that permit stable coexistence of multiple plectonemes, so that the mechanism reminiscent of the one we observe in non-equilibrium situations is likely to be present as well.

A more detailed treatment of plectoneme hopping in strands of different lengths is an interesting open problem for further study.
\clearpage
\section{Population frequencies of different DNA structures}\label{sec:pop_frequencies}
This section shows the population frequencies of bubbles, plectonemes and tip-bubble plectonemes as a function of $F$ for different fixed values of $\sigma$, as in Fig.~\ref{fig:hatcurves_torques}b of the main paper.
The pictures correspond to ``cuts'' along axes parallel to the y-axis of the state diagram (cf. Fig.~\ref{fig:hatcurves_torques}c of the main paper).
We used a population of $40\%$ in these plots to define the boundaries in the state-diagram shown in Fig.~\ref{fig:hatcurves_torques}c of the main paper.

For $\sigma<0$, there is a clear crossover from pure plectonemes to tip-bubble plectonemes and then to pure bubbles as a function of force.
For $\sigma>0$, a broader crossover to the tip-bubble regime occurs.
At low values of $|\sigma|$, the recognition of small plectonemes somewhat depends on the cutoff values chosen in the plectoneme detection algorithm, as described in Supplementary Sec.~\ref{sec:plecto_pos_size_pinn}.
As the transitions in this region of the state diagram are narrow as a function of $F$ and $\sigma$, this has only a small effect on the positions of state boundaries.

Note that as long as $\sigma$ is large enough to allow stable tip bubbles, the transition from plectonemes to tip bubbles appears to be at a very similar force for different $\sigma$. 
The reason for this is that the crossover is mainly determined by a change in the end-loop structure, rather than in the rest of the plectoneme, which grows for increasing $\sigma$. 
Similar arguments may explain the observation that the transition from tip bubbles to extended bubbles also happens at the same force for different negative values of $\sigma$.

\begin{figure}[!htb]
   \centering
   \includegraphics[width=\textwidth]{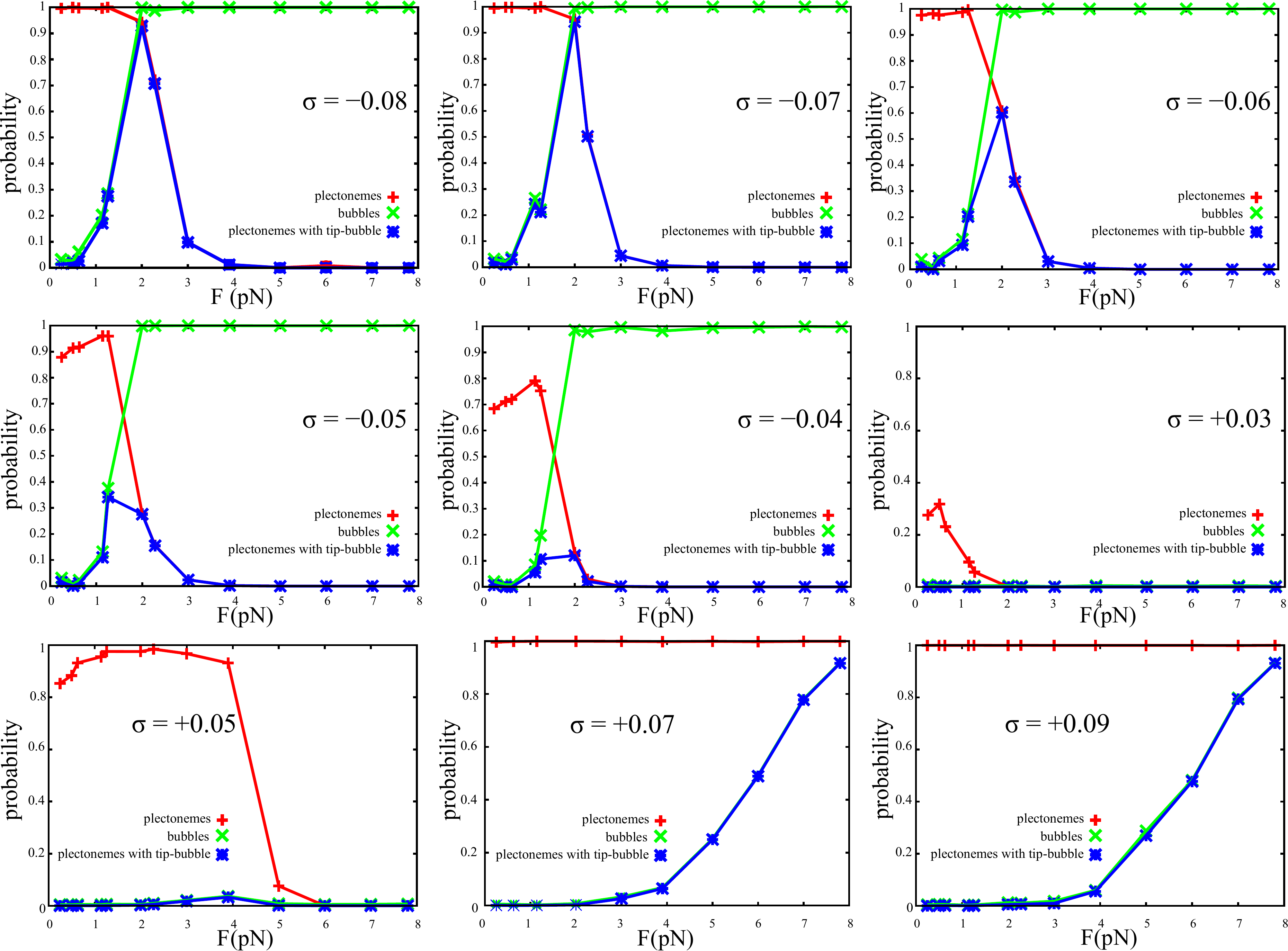}
   \caption{Population diagrams as a function of force for different imposed values of $\sigma$, analogous to Fig.~\ref{fig:hatcurves_torques}b of the main paper. For $\sigma<0$, note the occurrence of a tip-bubble regime for intermediate values of $F$. At $\sigma>0$, note the wide crossover to tip-bubble plectonemes for increasing $F$. No pure bubble regime exists for $\sigma>0$ at the force values used here. However, other forms of DNA may occur at sufficiently larger force.}
   \label{fig:state-diagram-cuts}
\end{figure}

\clearpage
\section{Sequence-dependent properties} \label{sec:sequences}
\begin{figure}[h!]
 \centering
 \includegraphics[width=.8\linewidth]{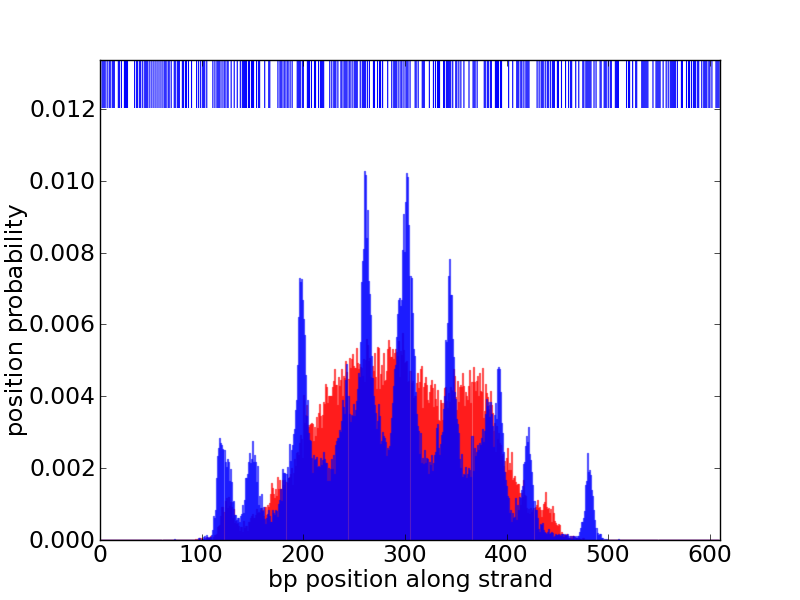}
 \caption{Plectoneme position distribution for the random sequence using the sequence-dependent (blue) and average-base (red) parametrization at $\sigma=-0.06$ and $F=1.27$\,pN. The sequence is given in the upper part of the figure, where a blue line indicates an AT base-pair. Mean plectoneme size is $260$\,bp for the average-base parametrization and 195\,bp for the random sequence.}
 \label{fig:plectopos_randseq_avgmod}
\end{figure}

\begin{figure}[h!]
 \centering
 \includegraphics[width=.8\linewidth]{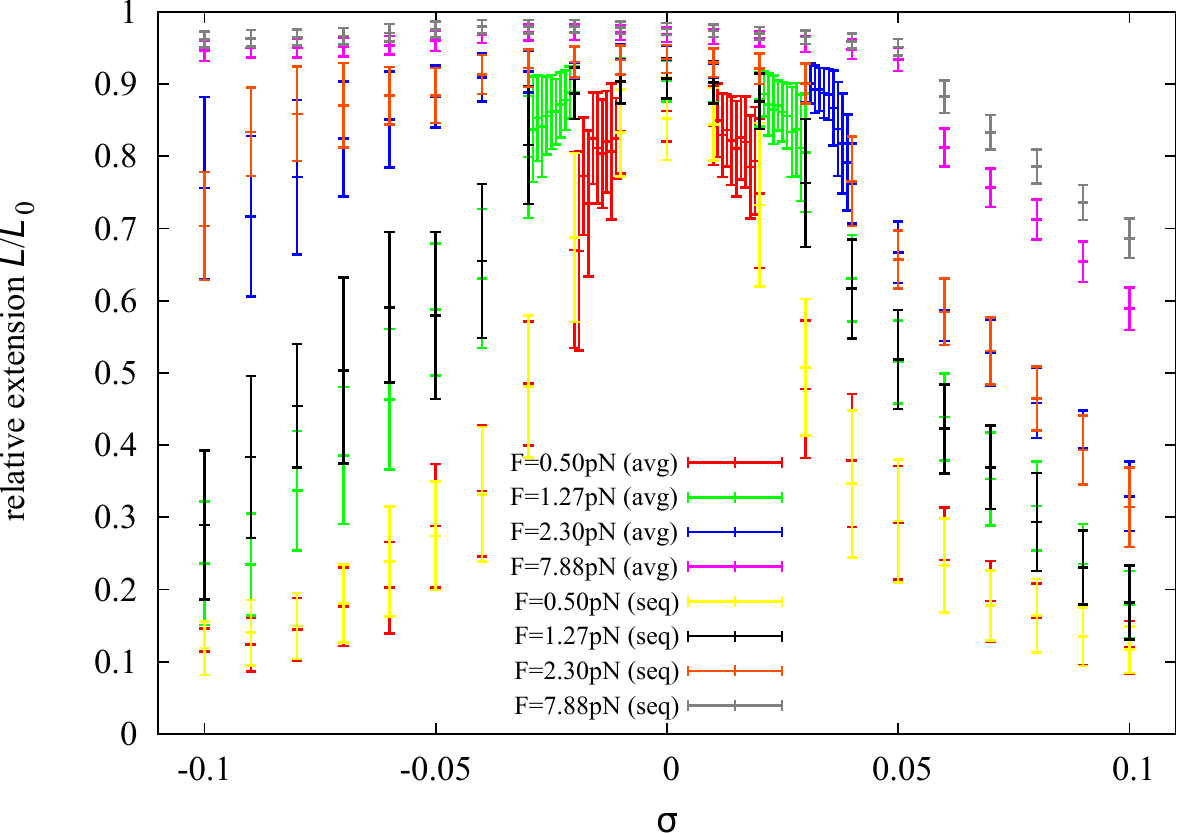}
 \caption{Comparison of extension curves for the random sequence given above in the average-base and sequence-dependent parametrization of oxDNA. For parameter values for which tip-bubble plectonemes occur, sequence-dependent simulations show a systematically higher extension, as weak regions in the sequence provide preferred sites of base-pair breaking, thus enhancing overall strand denaturation. Error bars indicate thermal fluctuations in the end-to-end distance, rather than sampling uncertainties.}
 \label{fig:hatcurves_avg_ran}
\end{figure}

\noindent \underline{Underlying base sequences} \\\\
In this work, we studied sequence-dependent properties of plectonemes using the sequence-dependent parametrization of oxDNA~\cite{Sulc2012}.
The sequences we used in the study of sequence-dependent behaviour are:
\\\\Random sequence:
\\ \texttt{5'-AGAGTACTTAGGCTTGACGATTTCGCGCCTGAACTTCTGATAACTCAGTCTGAGAGACTAAGTTGACGTTCTATCCATCATCA}
   \texttt{GGTGGGCTCAGAGATTGTGCGGCAGACTTAAGTGTAGTACCAGCTGCTGGTCAATTTGATCTATGCTGATCCGCTCGGAACGGGCC}\\
   \texttt{GTGAAAGAAGTACTCTCGCCTATAGAACGGTTAGTGCTACGACTTTTGCGCGACACAATGTGGTAGTTATCTTCTGTTTTCCTGAA}\\
   \texttt{TAGTGAGCCTACCAGAAGAGGCCACCGACAAATCTGATGAGATAGACGGGAACACGGTTTGCGGAGCCTCTGAAACGCTTGTTTAT}\\
   \texttt{GAGCAAGAGAGGTGCGGTGGGTATGACCGCCGTAGAAGTACCGTATTCTTCCGGGCTCGGTGGCAATGAACACTTAAGGGGCCGAC}\\
   \texttt{ACATTCTGAAGTCAATCGATGGACGGACCTCAACCGTGCACCCTTCTATATACGTGTGGCTAGGATACTCTAGCGTTTACCCGCCG}\\
   \texttt{TCTTCCACGATGCCGAATATAAGCCGAGGATAAAGGTGCAGACAAATATCAGGCTTCGCAGTTGTGTAACTTCCTGTATTGTTGTG}\\
   \texttt{C-3'}
\\This sequence was chosen at random, with a $25\%$ probability for each possible base identity.
\\\\Block-random sequence:
\\ \texttt{5'-\textcolor{blue}{AGACTCGACCGACCCGCGAGATCGGCTCCAGTCTCTGCGCCAAGTGCCGTTGCCCCGTCTCGTGGGCCGGTCGGTGAACCTTC}}\\
   \texttt{\textcolor{blue}{ATACGGTGGGAGTCGCTGAGGCCGTCCCATTTGTACC}\textcolor{red}{CATCGAACTCTTATTTTGTATTTTTTGGACATCCTCAGCTAACCACACG}}\\
   \texttt{\textcolor{red}{AGCCAAGCTATAGATCAGATTTGGGTATTCGGCGATCTTTCTAATCAACTGTATCCGATGCTATACAGATA}\textcolor{blue}{CTTTATTCTAAGGCG}}\\
   \texttt{\textcolor{blue}{GTCCGCGATGCGCCCAGTCCGTTGACCGGGCGAGTCATGTCAGAGTCGGCAATTATCGGGCACGTCGCCGGGGTGATACGTCCCTG}}\\
   \texttt{\textcolor{blue}{TGTCACTAGGCATAGGTCG}\textcolor{red}{TAACATATGATTATATATACTTTCCACTTTATGTATATCATTTGCAAGTTAGACATAATAAGGATAT}}\\
   \texttt{\textcolor{red}{ATAATATAAGAATCTCTTCACCTCTAAAGTGAGTGATTGGAATATAAGTATTT}\textcolor{blue}{GCGCCACTACCCGGCCGAAAGCCCGCGGCTCCT}}\\
   \texttt{\textcolor{blue}{CGCGGGTAGGTTGCCGGGGACCCGCGTGAAGAAAGGATGAAGCACCCGGACGCCCGCCTGCGAGTTGGCCACGGGCCCATAACGGC}}\\
   \texttt{\textcolor{blue}{G}-3'}\\
Above, stretches of 120 bp length are colored, with blue regions having a high GC content and red regions a high AT content. 
In an abbreviated form, the block-random sequence can be written as 5'-${\textnormal S_1W_1S_2W_2S_3}$-3'.
The overall GC content of the block-random sequence is $52\%$. The GC contents of the individual 120 bp stretches is given in Table \ref{tab:GC_cont}.\\
\begin{table}[h!]
\caption{GC-contents of 120 bp stretches in the block-random sequence}
\begin{tabular}{ccc}
\hline
stretch & GC content & AT content\\
\hline
$S_1$ & $65.0\%$ & $35.0\%$\\
$W_1$ & $39.2\%$ & $60.8\%$\\
$S_2$ & $60.0\%$ & $40.0\%$\\
$W_2$ & $24.2\%$ & $75.8\%$\\
$S_3$ & $71.7\%$ & $28.3\%$\\
\hline
\end{tabular}
\label{tab:GC_cont}
\end{table}
\newpage
\noindent \underline{plectoneme position distribution in average-base model}\\\\
As reported in Fig.~\ref{fig:pinning-diffusion}c of the main paper, the distribution of plectonemes exhibits a marked sequence dependence which is induced by the sequence-dependent enthalpic cost of forming tip-bubble denaturations.
In order to distinguish this sequence-dependent effect from the generic localization behaviour, we also ran simulations under the same stretching force and undertwist ($\sigma = -0.06$ and $F=1.27$\,pN) for the average-base parametrization of oxDNA.
Comparison between the two sequences studied and the average-base parametrization are shown in Figs.~\ref{fig:plectopos_randseq_avgmod} and~\ref{fig:plectopos_blocseq_avgmod}. Due to the repulsion plane boundary described in Supplementary Sec.~\ref{sec:boundaries} and finite plectoneme size, the probability for the position of plectoneme midpoints decreases towards the strand ends.
A clear influence of the sequence on plectoneme position is shown in Figs.~\ref{fig:plectopos_randseq_avgmod} and~\ref{fig:plectopos_blocseq_avgmod}.
\begin{figure}[h!]
 \centering
 \includegraphics[width=.8\linewidth]{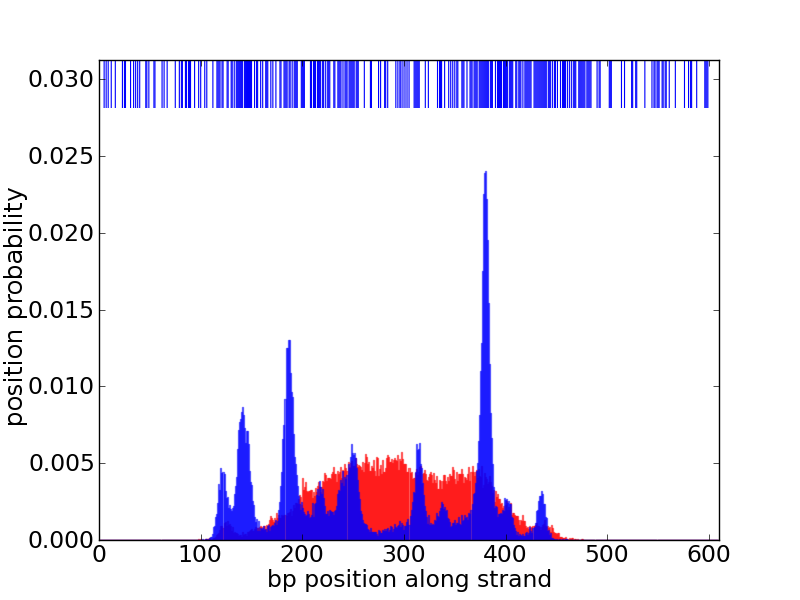}
 \caption{Plectoneme position distribution for the block-random sequence using the sequence-dependent (blue) and average-base (red) parametrization at $\sigma$=-0.06 and $F=1.27$\,pN. The sequence is given in the upper part of the figure, where a blue line indicates an AT base-pair. Mean plectoneme size is $260$\,bp for the average-base parametrization and 137\,bp for the block-random sequence, because tip-bubble plectonemes are more prevalent for the sequence-dependent model.}
 \label{fig:plectopos_blocseq_avgmod}
\end{figure}
\clearpage
\noindent \underline{Sequence effects in extension curves}\\\\
In Fig.~\ref{fig:hatcurves_avg_ran} we compare extension ``hat curves'' obtained for both the average-base and sequence-dependent parametrization of oxDNA.
Sequence-dependent simulations were performed for the random sequence given above.
While outside the bubble regime, results show no significant difference, the sequence-dependent model exhibits a systematically larger extension when forces become big enough to allow denaturation.
This indicates that denaturation occurs more easily in the sequence-dependent case than for the average-base parametrization.

Such behaviour might be expected, as weak AT base pairs represent preferred sites of base-pair breaking, making denaturation of the strand less enthalpically costly, while not affecting the bending energy cost much.
From simple considerations of energy scaling for the competition between denaturation and plectoneme formation, the crossover force can be estimated to depend on the free energy of base pair breaking $\alpha$ as $F_{\rm char} \propto \alpha^2 / B_0$ \cite{Salerno2012a}.
Hence, sequence-dependent differences in denaturation energy are expected to have a noticeable effect on strand extension properties. 
As denaturation bubbles can grow to significantly larger sizes than 1\,bp, not only the overall AT content of the strand, but also the base distribution may influence extension properties (see also the previous discussion on the block random sequence).
The detailed influence of sequence properties on strand extension behaviour represents an interesting problem for further study, which we are planning to address in the future.
\newpage

\section{Comparison to other modelling approaches} \label{sec:other_models}
DNA under superhelical stress has been modelled using various approaches on different scales of resolution.
Here, we briefly compare oxDNA to other models applicable to similar problems.

On a continuum level DNA is commonly described as a semi-flexible polymer. 
These models provide an efficient representation of DNA on large length scales, and have been widely used to study supercoiled DNA~\cite{Marko1994}.
Continuum models have also been used to study the phase behaviour of DNA plectonemes and other states deviating from the linear B-helical configuration~\cite{Marko2013,Marko2012,Neukirch2011}.
These models have been very successful at describing the effective mechanical behaviour of DNA on larger length scales. 
oxDNA captures the elastic behaviour of DNA, but also naturally has access to the physics of base-pair formation, which is important for the tip-bubble plectoneme regime we study in this work. 
It would be very interesting to refine continuum models by using oxDNA, to allow the direct calculation of tip-bubble plectonemes for positive and negative supercoiling.

To study the statistical properties of DNA denaturation, an important model has been developed by Benham and co-workers~\cite{Benham1992,Bauer1993,Fye1999}.
It is a thermodynamic model based on considering the effective free energy contributions of denatured and non-denatured strand regions.
Its simplicity allows rapid calculations of the statistical denaturation properties of supercoiled DNA on a genomic level~\cite{Strawbridge2010}.
Among other applications, it has been used successfully to describe sizes and locations of denaturation bubbles in circular DNA~\cite{Jeon2010}.
However, the model does not explicitly account for DNA structure beyond denaturation, or provide microscopic information on strand dynamics,
processes which are important for the tip-bubble plectoneme regime. 
It would be interesting to see if a similar thermodynamic model can be adapted to predict the distribution of locations and sizes of tip-bubble plectonemes.

On a much more structurally detailed level, DNA response to supercoiling has been studied in atomistic simulations~\cite{Randall2009,Kannan2009,Harris2006}. 
As these simulations are computationally expensive, they have only recently been used to access extended DNA structures under superhelical stress.
Atomistic simulations have been able to study local structural defects in small supercoiled minicircles, in particular microscopic properties of kinking at denaturations~\cite{Liverpool2008,Mitchell2011}.
Due to their computational expense, the timescales of these simulations are comparatively short, making the extraction of equilibrium properties difficult, in particular for writhed structures. 
For this reason, modelling of the properties of long DNA strands under twist and stretching force, as performed in this work, is currently outside the reach of fully atomistic models.
It may be interesting to prepare locally bent configurations in a plectonemic state, and use atomistic simulations to observe the local detailed configuration of a tip bubble.

Coarse-grained models, which combine structural information on a higher level of resolution with the possibility to measure thermodynamically averaged quantities, are needed to fill the gap between thermodynamic models and continuum models on the one hand and atomistic models on the other hand.
In Refs.~\onlinecite{Ouldridge2011}, \onlinecite{Sulc2012} and \onlinecite{Doye2013}, we review a number of other coarse-grained DNA models, although the interplay between twist, writhe and denaturation has not received much attention.
One model that has been used to study the interplay between denaturation and writhe is the coarse-grained model by Mielke {\it et al.}~\cite{Mielke2005}. 
In this approach, a single coarse-grained unit represents 3 nucleotides. The authors considered 141-bp minicircles, which they twisted in a dynamical fashion, up to $\sigma=-0.47$, a very high superhelical density.
For a few runs, the authors saw evidence that initial denaturation occurred at weak points in the sequence, near the loops of the writhed figure-of-8 structures that would form prior to denaturation.
As such, they saw tantalising evidence of coupled writhe and denaturation.
As the model of Ref.~\onlinecite{Mielke2005} does not allow reversible base-pair denaturation, its applicability is limited to non-equilibrium situations. 
It would be interesting to study the same system with oxDNA to see if this result for minicircles is robust in equilibrium.
One can imagine quite a number of other DNA configurations where the coupling of twist, writhe and denaturation may be interesting to study with oxDNA.

\bigskip 

\end{document}